\documentclass[aps,twocolumn,nofootinbib,groupedaddress,superscriptaddress,longbibliography]{revtex4-1}

\usepackage{amsmath,amssymb}
\usepackage{graphicx}
\usepackage{multirow}
\usepackage{bm}
\usepackage{mathtools}
\usepackage{dsfont}
\usepackage{amsfonts}
\usepackage{xcolor}
\usepackage{ulem}
\usepackage{url}
\usepackage{wasysym}

\usepackage[colorlinks=true,linkcolor=magenta,citecolor=magenta,hypertexnames=false]{hyperref}

\def\ba#1\ea{\begin{align}#1\end{align}}
\def\bg#1\eg{\begin{gather}#1\end{gather}}
\def\bpm{\begin{pmatrix}}
\def\epm{\end{pmatrix}}
\newcommand{\nn}{\nonumber}

\newcommand{\bb}[1]{{\boldsymbol #1}}
\newcommand{\bx}{\bb x}
\newcommand{\bk}{\bb k}
\newcommand{\bR}{\bb R}

\newcommand{\mc}[1]{\mathcal{#1}}

\newcommand{\dg}{\dagger}
\newcommand{\om}{\omega}
\newcommand{\sg}{\sigma}

\newcommand{\ep}{\epsilon}

\newcommand{\ket}[1]{|#1\rangle}
\newcommand{\bra}[1]{\langle#1|}
\newcommand{\brk}[2]{\langle#1|#2\rangle}

\newcommand{\Rf}[1]{Ref.~\onlinecite{#1}}

\newcommand{\eq}[1]{Eq.~\eqref{#1}}
\newcommand{\eqs}[1]{Eqs.~\eqref{#1}}
\newcommand{\fig}[1]{Fig.~\ref{#1}}
\newcommand{\figs}[1]{Figs.~\ref{#1}}

\newcommand{\cm}[1]{\overline{#1}}

\newcommand{\symeigv}{\xi_\sg}
\newcommand{\bchi}{\bb \chi}

\newcommand{\uef}[1]{\ket{\hat{u}_{#1}(\bk)}}
\newcommand{\uefL}[2]{\ket{\hat{u}#1_{#2}(\bk)}}

\newcommand{\magenta}[1]{\textcolor{magenta}{#1}}

\newcommand{\ourtitle}{
Flat bands with band crossings enforced by symmetry representation
}

\allowdisplaybreaks

\begin{document}
\title{\textbf{\ourtitle}}

\author{Yoonseok \surname{Hwang}}
\affiliation{Center for Correlated Electron Systems, Institute for Basic Science (IBS), Seoul 08826, Korea}
\affiliation{Department of Physics and Astronomy, Seoul National University, Seoul 08826, Korea}
\affiliation{Center for Theoretical Physics (CTP), Seoul National University, Seoul 08826, Korea}

\author{Jun-Won \surname{Rhim}}
\affiliation{Center for Correlated Electron Systems, Institute for Basic Science (IBS), Seoul 08826, Korea}
\affiliation{Department of Physics and Astronomy, Seoul National University, Seoul 08826, Korea}
\affiliation{Department of Physics, Ajou University, Suwon 16499, Korea}

\author{Bohm-Jung \surname{Yang}}
\email{bjyang@snu.ac.kr}
\affiliation{Center for Correlated Electron Systems, Institute for Basic Science (IBS), Seoul 08826, Korea}
\affiliation{Department of Physics and Astronomy, Seoul National University, Seoul 08826, Korea}
\affiliation{Center for Theoretical Physics (CTP), Seoul National University, Seoul 08826, Korea}

\begin{abstract}
Flat bands have band crossing points with other dispersive bands in many systems including the canonical flat-band models in the Lieb and kagome lattices.
Here we show that some of such band degeneracy points of nondegenerate flat bands are unavoidable because of the symmetry representation (SR) of flat bands under unitary symmetry.
We refer to such a band degeneracy point of flat bands as a \textit{SR-enforced band crossing}.
SR-enforced band crossing is distinct from the conventional band degeneracy protected by symmetry eigenvalues or topological charges in that
its protection requires both specific symmetry representation and band flatness of the flat band, simultaneously.
Even $n$-fold rotation $C_n$ ($n=2,3,4,6$) symmetry, which cannot protect band degeneracy without additional symmetries due to its abelian nature, can protect SR-enforced band crossings in flat-band systems.
In two-dimensional flat-band systems with $C_n$ symmetry, when the degeneracy of a SR-enforced band crossing is lifted by a $C_n$ \textit{symmetry-preserving perturbation}, 
we obtain a nearly flat Chern band.
Our theory not only explains the origin of the band crossing points of FBs existing in various models, but also gives a strict no-go theorem for isolated FBs in a given lattice arising from the SR.
\end{abstract}

\maketitle

\magenta{\it Introduction.|}
Flat bands (FBs)~\cite{lieb1989two,aoki1996hofstadter,huber2010bose,weeks2012flat,peotta2015superfluidity,julku2016geometric,ramachandran2017chiral,misumi2017new,pal2018flat,bilitewski2018disordered,mizoguchi2019flat,mizoguchi2019molecular,mizoguchi2020systematic,mizoguchi2020type,chiu2020fragile,hwang2020geometric,kuno2020flat,xie2020topology,lin2020chiral,kheirkhah2020majorana,morfonios2021flat,peri2021fragile,maimaiti2021flat} have been considered an ideal playground to study strong correlation physics because of the dispersionless energy-band structure.
However, recent studies of twisted bilayer graphene~\cite{bistritzer2011moire,cao2018correlated,cao2018unconventional} and kagome materials~\cite{ye2018massive,li2018realization,kang2020topological} have shown that not only the flat energy dispersion itself but also the characteristics of FB wave functions are important to understand fundamental properties of FB systems.
For example, the FBs in kagome materials are promising candidates to observe nearly flat Chern bands with nontrivial topology~\cite{tang2011high,sun2011nearly,neupert2011fractional}.
Also, the FB of twisted bilayer graphene at magic angle has demonstrated the first material realization of fragile band topology~\cite{song2019all,po2019faithful,ahn2019failure}.

In many systems, FBs often accompany band crossing points with other dispersive bands~\cite{bergman2008band,ma2020spin,rhim2019classification,rhim2021singular}.
Representative examples include the FB in the kagome lattice with a quadratic band crossing at the Brillouin zone (BZ) center, and that in the Lieb lattice with triple degeneracy at a BZ corner.
A FB with a band crossing is called a singular FB (SFB)~\cite{rhim2019classification,rhim2021singular} when the relevant momentum-space eigenstate has zeros.
Because of the inherent interband coupling, SFBs exhibit nontrivial geometric responses.
For instance, it was recently shown that a SFB with a quadratic band crossing exhibits anomalous Landau level spreading whose energy width measures the quantum distance of its momentum-space eigenstate~\cite{rhim2020quantum,rhim2021singular,hwang2021wave}.

In this Letter, we show that there is a class of FBs with unavoidable band crossings arising from the symmetry representation and band flatness of FBs, focusing on nondegenerate FBs.
To demonstrate this, we systematically investigate the symmetry representation (SR) of a compact localized state (CLS) under unitary symmetry.
A CLS is a real-space localized eigenstate of the FB, which is confined within a finite region~\cite{sutherland1986localization,vidal1998aharonov,vidal2001disorder,mukherjee2015observation,read2017compactly,maimaiti2017compact,maimaiti2019universal,rontgen2018compact,ma2020direct}.
We examine the SR of the CLS by considering it as a hybrid molecular state of atomic orbitals in a given lattice,
and show that irremovable singular points appear when there is a mismatch in the SRs between the CLS and its constituent atomic orbitals at high-symmetry points.
We refer to such a band crossing of FBs enforced by the SR of the CLS as a \textit{SR-enforced band crossing}.

SR-enforced band crossing is distinct from the conventional band degeneracy protected by symmetry eigenvalues or topological charges in that
its protection requires both specific SR and band flatness of the FB, simultaneously.
For instance, even $n$-fold rotation $C_n$ ($n=2,3,4,6$) symmetry, which cannot protect band degeneracy without additional symmetries as it has only one-dimensional irreducible representations (IRs), can protect SR-enforced band crossings in FB systems.
When the atomic orbitals constituting the CLS are located at maximal Wyckoff positions, the existence of the SR-enforced band crossing can be determined via a determinant test we developed.
On the other hand, if the constituent atomic orbitals are located at generic Wyckoff positions, the constraints from multiple symmetries can also give SR-enforced band crossings, which is applicable to various line and split graphs~\cite{mielke1991ferromagnetic,mielke1991ferromagnetism,mielke1993ferromagnetism,kollar2019line,ma2020spin}.

Interestingly, in two-dimensional lattices with $C_n$ rotation symmetry, a nearly FB with nonzero Chern number must emerge when the degeneracy of a single SR-enforced band crossing is lifted.
This clearly shows that the FB with SR-enforced band crossing is a promising platform to achieve nearly flat Chern bands~\cite{tang2011high,sun2011nearly,neupert2011fractional,wang2011nearly,trescher2012flat,yang2012topological,kalesaki2014dirac,pal2018nontrivial,bhattacharya2019flat}.

\begin{figure}[t!]
\centering
\includegraphics[width=0.48\textwidth]{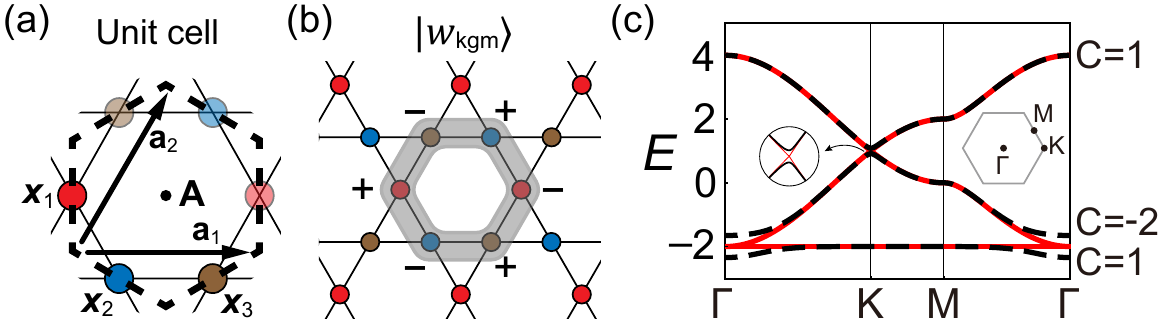}
\caption{
Kagome lattice.
(a) The unit cell composed of three sublattices located at $\bx_1=-\tfrac{1}{2}\bb a_1$, $\bx_2=-\tfrac{1}{2}\bb a_2$, and $\bx_3=\tfrac{1}{2}\bb a_1-\tfrac{1}{2}\bb a_2$, respectively, where $\bb a_1=(1,0)$ and $\bb a_2=(\tfrac{1}{2},\tfrac{\sqrt{3}}{2})$ are primitive lattice vectors.
The Wyckoff position $\bb A$ corresponds to the unit-cell center.
(b)  The CLS $\ket{w_{\rm kgm}(\bR)}$ whose shape is depicted by the gray region.
The sign ($\pm$) near each site denotes the amplitude $S_\alpha(\bR')$ of the CLS.
(c) The band structure (red solid lines) of $H_{\rm kgm}(\bk)$ having a FB with two-fold degeneracy at $\Gamma$.
The black dashed lines correspond to the case with
additional $C_6$-preserving perturbation $\delta H_{\rm kgm}(\bk)$.
$C$ near each gapped band denotes its Chern number.
}
\label{Fig1}
\end{figure}

\magenta{\it Kagome lattice.|}
We illustrate the general idea of SR-enforced band crossing by considering the FB in the kagome lattice composed of three sublattices as shown in \fig{Fig1}.
We place an $s$ orbital at each sublattice.
The kagome lattice belongs to the wallpaper group $p6mm$ generated by a six-fold rotation $C_6$ and two mirrors $M_x$ and $M_y$ that flip the sign of the $x$ and $y$ coordinates, respectively.
Below we focus on the role of $C_6$ symmetry on the band crossing of the FB.
The tight-binding Hamiltonian including only the nearest-neighbor hopping with the unit amplitude is
\ba
H_{\rm kgm}(\bk) = \bpm 0 & 1+Q_1 \cm{Q_2} & Q_1+Q_1\cm{Q_2} \\ c.c. & 0 & 1+Q_1 \\ c.c. & c.c. & 0 \epm,
\label{eq:kagome_H}
\ea
where $Q_i=e^{-i \bk \cdot \bb a_i}$ with lattice vectors $\bb a_i$ ($i=1,2$), and $\cm{x}$ denotes the complex conjugation (\textit{c.c.}) of $x$ (see the Supplemental Material (SM)~\cite{supple}).
The band structure of $H_{\rm kgm}(\bk)$ exhibits a FB at the energy $-2$ with a band crossing at the momentum $\Gamma=(0,0)$ as shown in \fig{Fig1}(c).
Conventionally, the degeneracy can be understood in terms of two-dimensional IRs of $C_3$ symmetry combined with time-reversal $T$ or a mirror symmetry. 
However, we find that the degeneracy survives even when all symmetries except $C_6$ are broken as long as the band flatness is maintained.
Degeneracy protection by an abelian symmetry is a hallmark of SR-enforced band crossing of FBs.
%

For any FB model with finite-ranged hoppings, an eigenstate of FB $\uef{}$ can always be expressed as Laurent polynomials in $Q_i$~\cite{read2017compactly,rhim2019classification}.
A relevant CLS is defined by a Fourier transform of $\uef{}$~\cite{supple}.
We refer to $\uef{}$ as a Fourier transform of CLS (FT-CLS).
In general, the center of a CLS is confined within a unit cell.
For a unit cell located at $\bR$, the relevant CLS is expressed as $\ket{w(\bR)}=\sum_{\bR',\alpha} S_\alpha(\bR')\ket{\bR+\bR',\alpha}$ where $\bR$ and $\bR'$ are lattice vectors, $\alpha$ is an orbital index, and $\ket{\bR,\alpha}$ denotes an electron located at $\bR+\bx_\alpha$ ($\alpha=1,\dots,n_{\rm tot}$) where $n_{\rm tot}$ indicates the total number of orbitals per unit cell.
$S_\alpha(\bR')$ is nonzero only inside a finite region, referred to as a $\textit{shape}$ here.
Note that we always require that there is no common divisor polynomial for all elements of $\uef{}$, which means that a CLS is \textit{elementary}, not expressed as a sum of other CLSs.
For $H_{\rm kgm}(\bk)$, FT-CLS and CLS are given by $\uef{\rm kgm}=(1-Q_1,Q_2-1,1-\cm{Q_1}Q_2)$ and $\ket{w_{\rm kgm}(\bR)} = \ket{\bR,1}-\ket{\bR+\bb a_1,1}+\ket{\bR+\bb a_2,2}-\ket{\bR,2}+\ket{\bR,3}-\ket{\bR-\bb a_1+\bb a_2,3}$, respectively.
The shape of $\ket{w_{\rm kgm}(\bR)}$ is illustrated in \fig{Fig1}(b).

Importantly, $\uef{\rm kgm}$ has a zero at $\Gamma$, i.e., $\ket{\hat{u}_{\rm kgm}(\Gamma)}=0$.
In general, when $\uef{}$ becomes zero at a momentum, the corresponding FB and its zero are called a singular FB (SFB) and a singular point, respectively.
A SFB must have a band crossing with other dispersive bands at the singular point~\cite{rhim2019classification}. 
A rigorous proof of the relation between zeros of FT-CLS and band crossing points is shown in SM~\cite{supple}. 
(Note that the number of degenerate states at the singular points depends on the symmetry of the model.)
Hence, the singularity of $\uef{\rm kgm}$ indicates the existence of band crossing points of FB in $H_{\rm kgm}(\bk)$.
Below we further show that the singularity originates from the SR of the CLS.

\magenta{\it Symmetry representation of CLS.|}
Now we discuss the symmetry property of CLS.
The hexagonal CLS $\ket{w_{\rm kgm}(\bR)}$ can be considered as a hybrid orbital formed by six atomic orbitals.
Interestingly, a hybrid orbital can have symmetry that its constituent atomic orbitals cannot have.
For instance, while each atomic orbital does not have $C_6$, the CLS $\ket{w_{\rm kgm}(\bR)}$ is symmetric under $C_6$.
Specifically, $\ket{w_{\rm kgm}(\bR)}$ is centered at the Wyckoff position $\bb A=(0,0)$ and has a $C_6$ eigenvalue $\xi_3=-1$ where $\xi_l=e^{i\pi l/3}$ ($l=0,\dots,5$), as readily inferred from its shape.
The center $\bb A$ and symmetry eigenvalue $\xi_3$ define the SR $\bb A_3$ of the CLS $\ket{w_{\rm kgm}(\bR)}$.

According to the SR, the CLS and FT-CLS transform under $C_6$ as 
\bg
\label{eq:kagome_w_Sym}
\hat{C}_6 \ket{w_{\rm kgm}(\bR)} = \ket{w_{\rm kgm}(O_{C_6}\bR)} \, \xi_3, \\
\label{eq:kagome_u_Sym}
U_{C_6}(\bk) \uef{\rm kgm} = 
\ket{\hat{u}_{\rm kgm}(O_{C_6}\bk)} \, \xi_3,
\eg
where $O_{C_6}$ and $U_{C_6}(\bk)$ indicate the matrix representations of $C_6$ in real and orbital spaces, respectively~\cite{supple}.
A CLS with its center at $\bb A$ and $C_6$ eigenvalue $\xi_3$, generally denoted as $\bb A_3$-CLS, follows the same symmetry transformation in \eqs{eq:kagome_w_Sym} and \eqref{eq:kagome_u_Sym}.

\begin{figure*}[t!]
\centering
\includegraphics[width=0.75\textwidth]{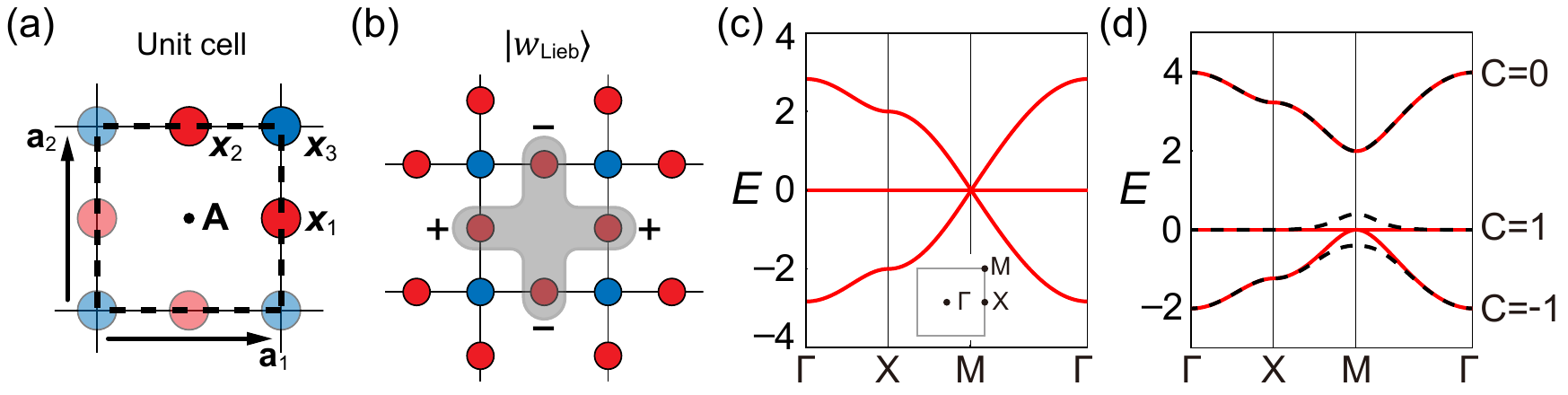}
\caption{
Lieb lattice.
(a) The unit cell composed of three sublattices at $\bb x_{1,2,3}$.
(b) The CLS $\ket{w_{\rm Lieb}(\bR)}$ composed of orbitals located at the sublattices $\bx_1$ and $\bx_2$.
(c) Band structure of $H^{(0)}_{\rm Lieb}(\bk)$ with a three-fold degeneracy at $M$.
(d) Band structure of $H_{\rm Lieb}(\bk)$ with on-site potential $\ep_3=2.0$ at the sublattice 3 (red lines).
There is a two-fold degeneracy at $M$.
Black dashed lines indicate bands when a $C_4$-preserving perturbation $\delta H_{\rm Lieb}(\bk)$ is added.
$C$ near each band denotes the corresponding Chern number.
}
\label{Fig2}
\end{figure*}

\magenta{\it SR-enforced band crossing.|}
Remarkably, the band crossing in $H_{\rm kgm}(\bk)$ is enforced by $\bb A_3$ SR of the CLS.
To elucidate this, let us consider $C_6$ eigenvalues at $\Gamma$, a unique $C_6$-invariant momentum.
$\bb A_3$-CLS always occupies all three sublattices.
At $\Gamma$ point in the BZ, the orbitals at these sublattices induce three IRs with $C_6$ eigenvalues $1$, $e^{2\pi i/3}$ and $e^{-2\pi i/3}$, respectively.
On the other hand, according to \eq{eq:kagome_u_Sym}, $C_6$ eigenvalue of the FT-CLS $\uef{\rm kgm}$ is $-1$, if the FT-CLS is nonsingular at $\Gamma$, i.e., $\ket{\hat{u}_{\rm kgm}(\Gamma)} \ne 0$.
We refer to such a symmetry eigenvalue of FT-CLS under the assumption of nonsingularity as a \textit{superficial} symmetry eigenvalue.
Clearly, the superficial $C_6$ eigenvalue of FT-CLS does not match any $C_6$ eigenvalue of the IRs induced from its constituent atomic orbitals: $-1 \notin \{1,e^{2\pi i/3},e^{-2\pi i/3}\}$.
The only way to circumvent this contradiction is that $\ket{\hat{u}_{\rm kgm}(\Gamma)}$ is singular, i.e., $\ket{\hat{u}_{\rm kgm}(\Gamma)}=0$.
Hence, a FB corresponding to $\bb A_3$-CLS cannot exist as a gapped isolated band, and it must have a SR-enforced band crossing at $\Gamma$.
We note that this \textit{no-go} statement for gapped FB with $\bb A_3$ SR holds for \textit{any} FB model defined in the kagome lattice
and any CLS with $\bb A_{1,3,5}$ SRs.

We further search SR-enforced band crossing at $K$ and $M$ in the BZ, which correspond to $C_3$ and $C_2$ invariant momenta, respectively.
The FB corresponding to $\bb A_3$-CLS has superficial $C_3$ ($C_2$) eigenvalue 1 ($-1$) at $K$ ($M$).
On the other hand, the IRs induced from constituent atomic orbitals have $C_3$ ($C_2$) eigenvalues $1$, $e^{2\pi i/3}$ and $e^{-2\pi i/3}$ ($-1$,$-1$, and 1).
Hence, the mismatch of symmetry eigenvalues does not happen, and there is no SR-enforced band crossing at $K$ and $M$.

\magenta{\it Determinant criterion for SR-enforced band crossing.|}
Based on the above discussion, we now propose a general condition for SR-enforced band crossings.
At $\Gamma$, \eq{eq:kagome_u_Sym} is reduced to $\mc{P}(\Gamma)\ket{\hat{u}(\Gamma)}=0$ where $\mc{P}(\Gamma)=U_{C_6}(\Gamma)-\xi_3 \mathds{1}_3$ and $\mathds{1}_3$ is the $3\times3$ identity matrix.
A condition for having SR-enforced band crossing is encapsulated in ${\rm Det} \mc{P}(\Gamma) \ne 0$.
This ensures a singular FT-CLS, i.e., $\ket{\hat{u}(\Gamma)}=0$.
Since $C_6$ eigenvalues of the atomic orbitals constituting CLS are equivalent to eigenvalues of $U_{C_6}(\Gamma)$, the condition ${\rm Det} \mc{P}(\Gamma) \ne 0$ implies that the superficial $C_6$ eigenvalue $\xi_3$ of FT-CLS does not match any of those of the constituent atomic orbitals.

In the kagome lattice, all atomic orbitals contribute to the hybrid orbital of the CLS.
However, in general, a CLS can also occupy only a part of atomic orbitals as in the Lieb lattice model discussed below.
We denote a set of such atomic orbitals unoccupied by CLS as $\{\alpha_\varnothing\}$.
Then a general condition for SR-enforced band crossing related to unitary symmetry $\sigma$ is as follows.
When a given SR under $\sg$ constrains FT-CLS such that $\mc{P}(\overline{\bk}_\sg) \ket{\hat{u}(\overline{\bk}_\sg)}=0$ at a high-symmetry point $\overline{\bk}_\sg$ for $\sg$,
\ba
{\rm Det}_{\{\alpha_\varnothing\}} \mc{P}(\overline{\bk}_\sg) \ne 0
\label{eq:condition_SFB}
\ea
indicates a SR-enforced band crossing at $\overline{\bk}_\sg$.
The detailed expression of $\mc{P}(\overline{\bk}_\sg)$ is determined by unitary symmetry $\sg$ and SR of a given CLS~\cite{supple}, and ${\rm Det}_{\{\alpha_\varnothing\}} \mc{O}$ is the determinant of $\mc{O}$ after removing the rows and columns corresponding to the orbital indices $\{\alpha_\varnothing\}$.

When only a single unitary symmetry is concerned, \eq{eq:condition_SFB} is the most general condition for SR-enforced band crossing.
However, when the constituent atomic orbitals are located at nonmaximal Wyckoff positions, they can support all possible symmetry eigenvalues including the superficial eigenvalue of FT-CLS, thus mismatch of symmetry eigenvalues does not occur.
Even in such cases, nevertheless, SR-enforced band crossing can occur when two or more symmetries are considered.
Suppose a SR of CLS under multiple symmetries imposes $N$ conditions on the FT-CLS: $\mc{P}_{I}(\overline{\bk}) \ket{\hat{u}(\overline{\bk})}=0$ ($I=1,\dots,N$).
When the constituent atomic orbitals of the CLS cannot span the FT-CLS at a high-symmetry point $\overline{\bk}$, these $N$ conditions lead to $\ket{\hat{u}(\overline{\bk})}=0$, hence the band crossing point at $\overline{\bk}$.
We illustrate this idea using the line graph of the Lieb lattice below.

\begin{figure*}[t!]
\centering
\includegraphics[width=0.75\textwidth]{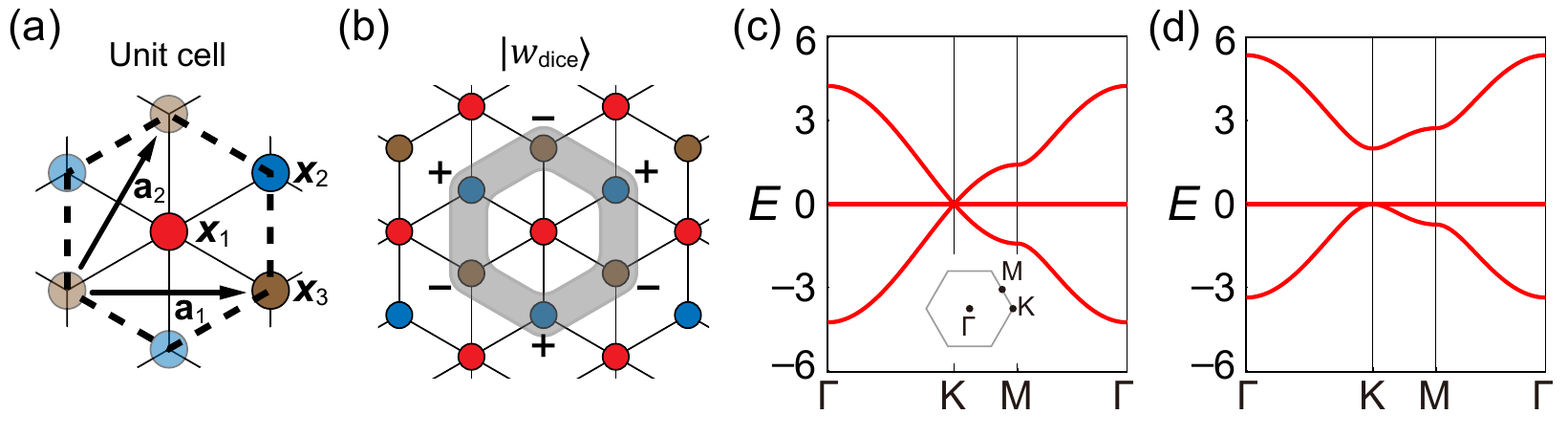}
\caption{
Dice lattice.
(a) The unit cell composed of three sublattices. 
(b) The CLS $\ket{w_{\rm dice}(\bR)}$ composed of orbitals located at the sublattices $\bx_2$ and $\bx_3$.
(c) Band structure of $H_{\rm dice}(\bk)$ with a three-fold degeneracy at $K$.
(d) Band structure of $H_{\rm dice}(\bk)$ with on-site potential $\ep_1=2.0$ at the sublattice 1.
There are two-fold degeneracies at $\pm K$.
}
\label{Fig3}
\end{figure*}

\magenta{\it Lieb lattice.|}
FB with band crossing enforced by SR for $C_4$ exists in the Lieb lattice composed of three sublattices, which belongs to the wallpaper group $p4mm$ (see \fig{Fig2}).
A tight-binding model $H_{\rm Lieb}(\bk)$ with nearest-neighbor hopping exhibits a FB having a three-fold degeneracy at the momentum $M=(\pi,\pi)$ [see \fig{Fig2}(b)].
The relevant CLS $\ket{w_{\rm Lieb}(\bR)}$ is centered at the Wyckoff position $\bb A=(0,0)$ and has a $C_4$ eigenvalue $-1$.
It is worth noting that the atomic orbitals constituting the CLS $\ket{w_{\rm Lieb}(\bR)}$ occupy only the sublattice 1 and 2, i.e., $\{\alpha_\varnothing\}=\{3\}$.
Therefore, the constituent atomic orbitals form IRs with $C_4$ eigenvalues $\pm i$ at $M$, while the FB has a superficial $C_4$ eigenvalue $-1$ at $M$.
This mismatch of $C_4$ eigenvalue at $M$ indicates a singular FT-CLS, i.e., $\ket{\hat{u}_{\rm Lieb}(M)}=0$, and a SR-enforced band crossing at $M$.
The same result can also be obtained by the determinant criterion in \eq{eq:condition_SFB}, which becomes ${\rm Det}_{\{3\}}\mc{P}(M)=2$ where $\mc{P}(M)=U_{C_4}(M)+\mathds{1}_3$.
The determinant criterion for $C_2$ SR-enforced band crossing applied at $\overline{\bk}=\Gamma,X,Y,M$ also predicts a SR-enforced band crossing only at $M$.

It is worth noting that the fact that $\ket{w_{\rm Lieb}(\bR)}$ does not occupy the sublattice 3 is critical for the presence of the band crossing at $M$.
For example, one can consider another CLS with its center at $\bb A$ and $C_4$ eigenvalue $-1$ whose constituent atomic orbitals occupy all three sublattices.
In this case, the corresponding FB is not enforced to have a band crossing point, hence it can be gapped~\cite{supple}.

\magenta{\it Dice lattice.|}
FB with band crossings enforced by SR for $C_3$ exists in the dice lattice composed of three sublattices, which belongs to the wallpaper group $p6mm$ (see \fig{Fig3}).
Considering only nearest-neighbor hoppings, a tight-binding model $H_{\rm dice}(\bk)$ exhibits a FB with three-fold degeneracy at $K$ and $-K$, respectively, as shown in \fig{Fig3}(c).
The CLS of the FB $\ket{w_{\rm dice}(\bR)}$ is composed of the atomic orbitals located at the sublattice 2 and 3, as shown in \fig{Fig3}(b), thus $\{\alpha_\varnothing\}=\{1\}$.
Also, it is centered at $\bb A=(0,0)$ and has $C_3$ eigenvalue 1.
Then the determinant criterion, ${\rm Det}_{\{1\}}[U_{C_3}(\pm K)-\mathds{1}_3]=3$, indicates SR-enforced band crossings at $\pm K$.

\begin{figure*}[t!]
\centering
\includegraphics[width=0.75\textwidth]{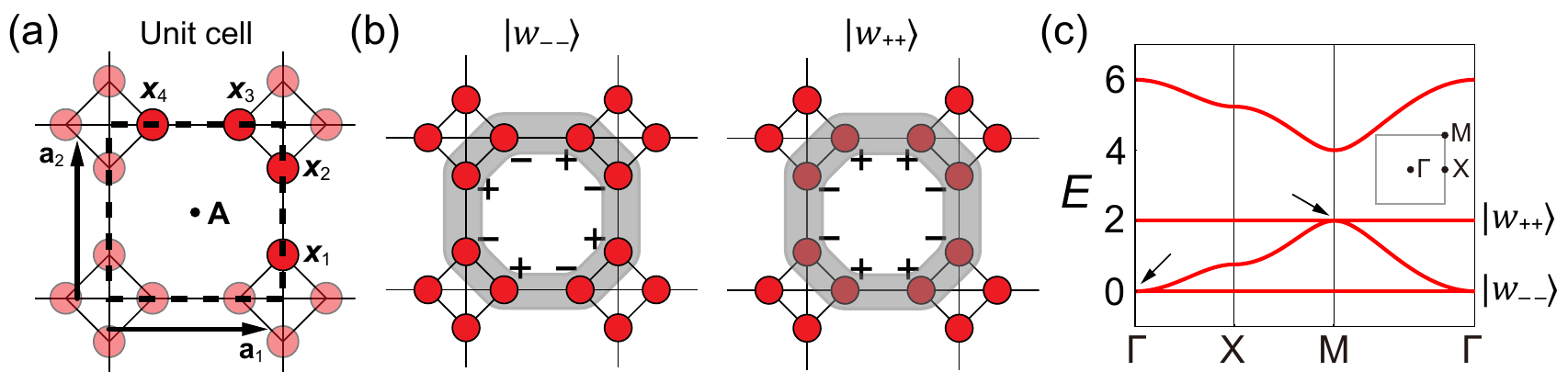}
\caption{
Line graph of the Lieb lattice.
(a) The unit cell composed of four sublattices located at $\bx_1=(1/2,-1/4)$, $\bx_2=(1/2,1/4)$, $\bx_3=(1/4,1/2)$ and $\bx_4=(-1/4,1/2)$, respectively.
The hoppings between the sites connected by black lines are considered.
(b) The CLSs $\ket{w_{++}(\bR)}$ and $\ket{w_{--}(\bR)}$ of $H_{\rm LG}(\bk)$.
The lower indices denote $M_x$ and $M_y$ eigenvalues, respectively.
(c) Band structure of $H_{\rm LG}(\bk)$.
The FB with $E=0$ ($E=2$), which corresponds to $\ket{w_{--}(\bR)}$ ($\ket{w_{++}(\bR)}$), has a band crossing point at $\Gamma$ ($M$).
}
\label{Fig4}
\end{figure*}

\magenta{\it FBs in lattices with two orthogonal mirrors.|}
When the atomic orbitals are located at generic Wyckoff positions, the determinant criterion in \eq{eq:condition_SFB} does not work. Even in this case, however, SR-enforced band crossings can appear because of multiple symmetries.
As a representative example, we consider a square lattice with four sublattices shown in \fig{Fig4}(a), belonging to the wallpaper group $p4mm$ as the Lieb lattice.
Here, we focus on two mirror symmetries $M_x$ and $M_y$.
When the hopping structure takes a particular configuration shown in \fig{Fig4}(a), the lattice is called a line graph~\cite{kollar2019line,ma2020spin,chiu2020fragile} of the Lieb lattice.
The relevant tight-binding Hamiltonian is denoted as $H_{\rm LG}(\bk)$ whose band structure possesses two FBs at the energy $E=0$ and 2, respectively [see \fig{Fig4}(c)].
The FB with $E=0$ (2) has a band crossing point at $\Gamma$ ($M$) and the corresponding CLS $\ket{w_{--}(\bR)}$ ($\ket{w_{++}(\bR)}$) is shown in \fig{Fig4}(b).
Note that the lower indices of CLSs denote the eigenvalues of $M_x$ and $M_y$ in order.
%

Now let us consider the SR of $\ket{w_{++}(\bR)}$ under mirror symmetries.
Because of its SR, the FT-CLS $\ket{\hat{u}_{++}(\bk)}$ satisfies $U_{M_{x,y}}(M) \ket{\hat{u}_{++}(M)}=\ket{\hat{u}_{++}(M)}$ where $4\times 4$ matrices $U_{M_x}(M)$ and $U_{M_y}(M)$ are given by the direct sum  $(-\tau_0) \oplus \tau_1$ and $\tau_1 \oplus (-\tau_0)$ in terms of the Pauli matrices $\tau_{0,1,2,3}$.
These two conditions necessarily lead to a singular FT-CLS satisfying $\ket{\hat{u}_{++}(M)}=0$.
This shows that a FB corresponding to the same SR as $\ket{w_{++}(\bR)}$ cannot be gapped in any model with arbitrary hopping structure including the line graph configuration.
Similarly, it can also be shown that the SR of $\ket{w_{--}(\bR)}$ leads to the singular FT-CLS at $\Gamma$.
It is worth noting that although two orthogonal commuting mirrors allow only one-dimensional IRs, they can induce SR-enforced band crossing of FBs.

\magenta{\it Nearly flat Chern bands.|}
What happens if the degeneracy of SR-enforced band crossing is lifted?
When a single band crossing with two-fold degeneracy is enforced by SR under $C_n$ symmetry alone and the degeneracy is lifted by a $C_n$-preserving perturbation, the resulting gapped nearly FB carries a nonzero Chern number.
This is because the $C_n$ eigenvalue of the nearly FB at the momentum where the degeneracy is lifted is different from the superficial $C_n$ eigenvalue of the FB so that the gapped FB is not band representable, thus topologically nontrivial~\cite{fang2012bulk}.

In the case of $H_{\rm kgm}(\bk)$, a FB has a single SR-enforced band crossing at $\Gamma$.
Hence, the FB becomes a nearly flat Chern band when a $C_6$-preserving perturbation $\delta H_{\rm kgm}(\bk)$, which lifts the degeneracy at $\Gamma$, is added [see \fig{Fig1}(c)].
The nearly flat Chern band induced by spin-orbit coupling can be understood in this way~\cite{tang2011high,beugeling2012topological,ye2018massive}.
The same mechanism can also be applied to magnon bands in the kagome ferromagnet~\cite{zhang2013topological,mook2014magnon,chisnell2015topological}.

In the case of $H_{\rm Lieb}(\bk)$, the FB has a three-fold degeneracy at $M$, which can be split into two-fold and nondegenerate ones, by introducing on-site potential $\ep_3$ at the sublattice 3.
Crucially, this procedure does not break the flat dispersion of FB because the CLS $\ket{w_{\rm Lieb}(\bR)}$ does not occupy the sublattice 3.
The SR-enforced degeneracy can be lifted and the FB becomes a nearly flat Chern band as shown in \fig{Fig2}(d), by introducing a perturbation $\delta H_{\rm Lieb}(\bk)$ that preserves $C_4$ but breaks other symmetries.

Finally, we note that a FB with multiple SR-enforced band crossing points, as in the model $H_{\rm dice}(\bk)$ [\figs{Fig3}(c) and (d)], is unnecessary to be a nearly flat Chern band, when the degeneracy is lifted by symmetry-preserving perturbation, as the Berry curvature from different crossing points can be canceled.

\magenta{\it Discussion.|}
We have shown that some FBs have unavoidable band crossings with dispersive bands, when there is a SR mismatch between the CLS and its constituent atomic orbitals under unitary symmetry.
This result can be applied to any lattices with arbitrary unitary symmetry representations of orbitals.
Especially, the notion of SR-enforced band crossing not only explains the origin of the band crossing points of FBs appearing in various lattice models, but also gives a stringent no-go theorem such that a class of FB can never be realized as an isolated FB in a given lattice due to the SR.

We also have shown that lifting the degeneracy of SR-enforced band crossing by symmetry-preserving perturbation gives nearly flat Chern bands.
Even the early models of the nearly flat Chern band~\cite{sun2011nearly,neupert2011fractional} can be adiabatically deformed to FB models with SR-enforced band crossing as shown in SM~\cite{supple}.

We expect that anti-unitary symmetries can provide more examples of FBs with unavoidable band crossings enforced by SR.
Further extending the notion of wave function singularity and SR of CLS to degenerate FBs is also an important topic for future research.

Finally, we anticipate that SR of CLS plays an important role in realizing symmetry-protected topological (SPT) phases.
In Ref.~\cite{yang2021symmetry}, interacting SPT phases with spinful bosons are obtained from FB models, when many-body ground state, given by a product of CLSs with specific SR, transforms nontrivially under point group symmetries.

\let\oldaddcontentsline\addcontentsline
\renewcommand{\addcontentsline}[3]{}
\begin{acknowledgments}
We thank Sungjoon Park and Sunje Kim for useful discussion.
Y.H. and B.J.Y. were supported by the Institute for Basic Science in Korea (Grant No. IBS-R009-D1), 
Samsung  Science and Technology Foundation under Project No. SSTF-BA2002-06,
the National Research Foundation of Korea (NRF) Grant funded by the Korea government (MSIT) (No. 2021R1A2C4002773, and No. NRF-2021R1A5A1032996).
J.W.R. was supported by IBS-R009-D1, the National Research Foundation of Korea (NRF) Grant funded by the Korea government (MSIT) (Grant No. 2021R1A2C1010572), and the New Faculty Research Fund of Ajou University.
\end{acknowledgments}

%

\let\addcontentsline\oldaddcontentsline
\clearpage
\onecolumngrid
\begin{center}
\textbf{\large Supplemental Material for ``\ourtitle"}
\end{center}
\setcounter{section}{0}
\setcounter{figure}{0}
\setcounter{equation}{0}
\renewcommand{\thefigure}{S\arabic{figure}}
\renewcommand{\theequation}{S\arabic{equation}}
\renewcommand{\thesection}{S\arabic{section}}
\tableofcontents
\hfill \\
\twocolumngrid

In this Supplementary Material, we provide detailed discussions on the results presented in the main text.
In Sec.~\ref{app:TB}, we review the tight-binding Hamiltonian and provide various formulas that are necessary to understand the main results.
Compact localized state (CLS) and its Fourier transform (FT-CLS) are defined in Sec.~\ref{app:FB_Def}.
These definitions must always be satisfied by a given CLS and FT-CLS, in order to define a singular flat band and prove the relation between band crossing points and singular points of FT-CLS in Sec.~\ref{app:SFB}.
In Sec.~\ref{app:symrep}, we define symmetry representation (SR) of CLS and FT-CLS and discuss SR-enforced band crossing.
Section~\ref{app:FB_models} provides detailed descriptions of the flat-band models discussed in the main text.
Finally, we demonstrate that the early models of nearly flat band with Chern number can be understood from SR-enforced band crossing of FB, in Sec.~\ref{app:NFBs}.

\section{Review of the tight-binding Hamiltonian \label{app:TB}}

\subsection{Tight-binding Hamiltonian}
We consider a periodic lattice system, composed of $N_{\rm cell}$ number of unit cells, in $d$ dimensions.
In each unit cell, a set of basis atomic orbitals is denoted as $\ket{\bR,\alpha}$ ($\alpha=1,\dots,n_{\rm tot}$) where $\bR$ is a (Bravais) lattice vector corresponding to the position of unit cell.
$\ket{\bR,\alpha}$ describes an electron localized at $\bR+\bx_\alpha$ with the sublattice site of $\alpha$-th orbital $\bx_\alpha$.
They are mutually orthonormal such that
\ba
\brk{\bR,\alpha}{\bR',\beta}=\delta_{\bR,\bR'}\delta_{\alpha\beta}.
\ea
A tight-binding Hamiltonian is spanned by $\ket{\bR,\alpha}$:
\ba
\hat{H}
&= \sum_{\bR,\bR'} \sum_{\alpha,\beta} \ket{\bR,\alpha} \, t_{\alpha \leftarrow \beta}(\bR-\bR') \, \bra{\bR',\beta},
\ea
where $t_{\alpha \leftarrow \beta}(\bR-\bR')$ is the hopping parameter between $\alpha$-th and $\beta$-th orbitals located at $\bR+\bx_\alpha$ and $\bR'+\bx_\beta$ respectively.

Two different conventions for Fourier transform of $\ket{\bR,\alpha}$ can be defined as
\ba
\label{eq:basis_def}
&\ket{\bk,\alpha}
\equiv \frac{1}{\sqrt{N_{\rm cell}}} \sum_{\bR} e^{i \bk \cdot \bR} \, \ket{\bR,\alpha}, \\
&\ket{\{\bk,\alpha\}}
\equiv \frac{1}{\sqrt{N_{\rm cell}}} \sum_{\bR} e^{i \bk \cdot (\bR + \bx_\alpha)} \, \ket{\bR,\alpha}.
\ea
Accordingly, the tight-binding Hamiltonian $\hat{H}$ is now expressed as
\ba
\hat{H} &= \sum_{\bk} \sum_{\alpha,\beta} \ket{\bk,\alpha} \, H(\bk)_{\alpha \beta} \, \bra{\bk,\beta} \\
&= \sum_{\bk} \sum_{\alpha,\beta} \ket{\{\bk,\alpha\}} \, \widetilde{H}(\bk)_{\alpha \beta} \, \bra{\{\bk,\beta\}},
\ea
in terms of $\ket{\bk,\alpha}$ and $\ket{\{\bk,\alpha\}}$.
Both $\ket{\bk,\alpha}$ and $\ket{\{\bk,\alpha\}}$ satisfy the orthonormality: $\brk{\bk,\alpha}{\bk',\beta} = \delta_{\bk,\bk'} \delta_{\alpha \beta}$ and $\brk{\{\bk,\alpha\}}{\{\bk',\beta\}} = \delta_{\bk,\bk'} \delta_{\alpha \beta}$.
Hence, $H(\bk)_{\alpha\beta} = \bra{\bk,\alpha} \hat{H} \ket{\bk,\beta}$ and $\widetilde{H}(\bk)_{\alpha\beta} = \bra{\{\bk,\alpha\}} \hat{H} \ket{\{\bk,\beta\}}$.
Explicitly, the tight-binding Hamiltonians in momentum space, $H(\bk)$ and $\widetilde{H}(\bk)$, are expressed as
\bg
H(\bk)_{\alpha\beta} = \sum_{\bR} \, t_{\alpha \leftarrow \beta}(\bR) \, e^{-i \bk \cdot \bR}, \nn \\
\widetilde{H}(\bk)_{\alpha\beta} = \sum_{\bR} \, t_{\alpha \leftarrow \beta}(\bR) \, e^{-i \bk \cdot (\bR+\bx_\alpha-\bx_\beta)}.
\eg
Note that $\widetilde{H}(\bk)$ and $\ket{\{\bk,\alpha\}}$ are not periodic in the Brillouin zone (BZ) unless all the orbitals are located at $\bx_\alpha=0$.

Such violation of periodicity is encapsulated in the sublattice embedding matrix $V(\bk)$: $
V(\bk)_{\alpha \beta} = e^{-i \bk \cdot \bx_\alpha} \delta_{\alpha \beta}$.
Using $V(\bk)$, one can easily change the basis through
\bg
\ket{\bk,\alpha} = \ket{\{\bk,\beta\}} V(\bk)_{\beta \alpha}, \nn \\
H(\bk) = V(\bk)^{-1} \widetilde{H}(\bk) V(\bk),
\label{eq:Htb_basis}
\eg
where we use the Einstein summation convention and matrix notation for $\alpha$ and $\beta$.
Since $\ket{\bk,\alpha}=\ket{\bk+\bb G,\alpha}$ and $\ket{\{\bk+\bb G,\alpha\}}=V(\bb G)^{-1}\ket{\{\bk,\alpha\}}$ for reciprocal vector $\bb G$, we obtain
\bg
\label{eq:Htb_Period1}
H(\bk+\bb G)=H(\bk), \\
\label{eq:Htb_Period2}
\widetilde{H}(\bk+\bb G) = V(\bb G) \widetilde{H}(\bk) V(\bb G)^{-1}.
\eg
We call $H(\bk)$ a tight-binding Hamiltonian in \textit{periodic} basis because of its periodicity in the BZ~\cite{shiozaki2017topological}.
On the other hand, we call $\widetilde{H}(\bk)$ a tight-binding Hamiltonian in \textit{non-periodic} basis.
Energy eigenstates $\ket{u_n(\bk)}$ and $\ket{\{u_n(\bk)\}}$ can be obtained by a diagonalization of  $H(\bk)$ and $\widetilde{H}(\bk)$:
\ba
H(\bk)_{\alpha \beta} \, \ket{u_n(\bk)}_\beta
=& E_n(\bk) \, \ket{u_n(\bk)}_\alpha, \\
\widetilde{H}(\bk)_{\alpha \beta} \, \ket{\{u_n(\bk)\}}_\beta
=& E_n(\bk) \, \ket{\{u_n(\bk)\}}_\alpha,
\ea
where we set $\ket{\{u_n(\bk)\}} = V(\bk) \ket{u_n(\bk)}$ according to \eq{eq:Htb_basis}.
The energy eigenstates $\ket{u_n(\bk)}$ satisfy
\bg
\brk{u_n(\bk)}{u_m(\bk)} = \delta_{n m}, \\
\sum_{n=1}^{n_{\rm tot}} \ket{u_n(\bk)}_\alpha \bra{u_n(\bk)}_\beta = \delta_{\alpha\beta},
\eg
and similar relations for $\ket{\{u_n(\bk)\}}$.
From Eqs.~\eqref{eq:Htb_Period1}-\eqref{eq:Htb_Period2}, we impose the periodic gauge:
\bg
\ket{u_n(\bk+\bb G)} = \ket{u_n(\bk)}, \nn \\
\ket{\{u_n(\bk+\bb G)\}} = V(\bb G) \ket{\{u_n(\bk)\}}.
\eg

It is worth noting that $\widetilde{H}(\bk)$ is proper to calculate physical quantities or study the Wilson loop spectrum.
Nevertheless, we use the periodic Hamiltonian $H(\bk)$ throughout this work, since the periodicity in \eq{eq:Htb_Period1} gives an algebraic simplification.
To be specific, each element of $H(\bk)$ is given by a Laurent polynomial in $d$ variables $Q_i=e^{-i\bk \cdot \bb a_i}=e^{-ik_i}$ ($i=1,\dots,d$) with $i$-th primitive lattice vector $\bb a_i$:
\ba
H(\bk)_{\alpha\beta} = \sum_{n_1 \dots n_d} \, t_{\alpha \leftarrow \beta}(\textstyle{\sum_i} n_i \bb a_i) \, \prod_i Q_i^{n_i}
\label{eq:H_Laurent}
\ea
for $n_1,\dots,n_d \in \mathbb{Z}$.
Hence, we only use $H(\bk)$ in the following sections and the main text, unless otherwise noted.
Note that the corresponding results for $\widetilde{H}(\bk)$ can be simply obtained by the basis change in \eq{eq:Htb_basis}.

\subsection{Symmetry transformation}
First, let us consider a unitary symmetry $\sg=\{O_\sg|\bb \delta_\sg\}$ whose action in real space is given by $\sg: \bb r \rightarrow O_\sg \bb r + \bb \delta_\sg$.
Here, $O_\sg$ is an orthogonal matrix.
The unitary symmetry $\sg$ also acts on the basis orbitals, $\ket{\bR,\alpha}$, such that
\ba
\hat{\sg} \, \ket{\bR,\alpha}
&= \ket{\bR_\sg(\alpha),\beta} \, U(\sg)_{\beta \alpha},
\label{eq:unisym_real}
\ea
where $\bR_\sg(\alpha)= O_\sg (\bR + \bx_\alpha) + \bb \delta_\sg - \bx_\beta$.
Combining Eqs.~\eqref{eq:basis_def} and \eqref{eq:unisym_real}, the symmetry transformation of $\ket{\bk,\alpha}$ under $\hat{\sg}$ is defined by
\ba
\hat{\sg} \, \ket{\bk,\alpha}
&= \frac{1}{\sqrt{N_{\rm cell}}} \sum_{\bR} e^{i \bk \cdot \bR} \, \hat{\sg} \, \ket{\bR,\alpha} \nn \\
&= \sum_\beta e^{i O_\sg \bk \cdot (\bx_\beta - O_\sg \bx_\alpha -\bb \delta_\sg)} \, \ket{O_\sg \bk,\beta} \, U(\sg)_{\beta \alpha} \nn \\
&\equiv \ket{O_\sg \bk,\beta} \, U_\sg(\bk)_{\beta \alpha}.
\label{eq:unisym_BZ}
\ea
Here, we define the matrix representation of $\sg$ on the basis orbitals in momentum space as $U_\sg(\bk) = e^{i O_\sg \bk \cdot (\bx_\beta - O_\sg \bx_\alpha -\bb \delta_\sg)} \, U(\sg)$.
We simply call $U_\sg(\bk)$ a symmetry operator of $\sg$.
Note that each element of $U_\sg(\bk)$ is proportional to the products of $Q_i$ with integer powers.
As an example, symmetry operator of $C_6$ in kagome lattice has three nonzero elements as $U_{C_6}(\bk)_{13}=Q_1\cm{Q_2}$, $U_{C_6}(\bk)_{21}=1$ and $U_{C_6}(\bk)_{32}=1$. 
It follows because $(\bx_\beta - O_\sg \bx_\alpha -\bb \delta_\sg)$ belongs to the lattice vector for $\alpha$ and $\beta$ such that $U(\sg)_{\beta\alpha}\ne0$.
Using $V(\bk)$, $U_\sg(\bk)$ is expressed as
\ba
U_\sg(\bk) = V(O_\sg \bk)^\dg U(\sg) V(\bk).
\label{eq:Uk_def}
\ea

For unitary symmetry $\sg$, \eq{eq:unisym_BZ} and $\hat{H}=\hat{\sg} \hat{H} \hat{\sg}^{-1}$ give a symmetry relation,
\ba
\label{eq:unisym_H}
H(O_\sg \bk) = U_\sg (\bk) \, H(\bk) \, U_\sg(\bk)^\dg.
\ea
For anti-unitary symmetry, the symmetry relations can be simply obtained by combining unitary symmetry and the complex conjugation $\mc{K}$.

Finally, we remark on the group multiplication.
Consider $\hat{\sg}_{1,2}$ and their multiplication $\hat{\sg}=\hat{\sg}_2 \hat{\sg}_1$.
For $\hat{\sg}$, $\ket{\bk,\alpha}$ transforms as
\ba
\hat{\sg} \ket{\bk, \alpha} &= \hat{\sg}_2 \ket{O_{\sg_1} \bk, \beta} \, U_{\sg_1}(\bk)_{\beta \alpha} \\
&= \ket{O_{\sg_2} O_{\sg_1} \bk, \gamma} \, U_{\sg_2}(O_{\sg_1}\bk)_{\gamma \beta} U_{\sg_1}(\bk)_{\beta \alpha} \\
&= \ket{O_\sg \bk, \beta} \, \left[ U_{\sg_2}(O_{\sg_1}\bk) U_{\sg_1}(\bk) \right]_{\beta \alpha},
\label{eq:group_mul}
\ea
thus $U_{\sg_2 \sg_1}(\bk) = U_{\sg_2}(O_{\sg_1}\bk) U_{\sg_1}(\bk)$.
%

\section{Eigenstate of flat band in real and momentum spaces \label{app:FB_Def}}
In this section, we define eigenstates of flat band (FB) in both real and momentum spaces.
First, consider a $d$-dimensional lattice system composed of $N_{\rm cell}$ unit cells and $n_{\rm tot}$ orbitals in each unit cell.
In this lattice system, a tight-binding Hamiltonian exhibits a nondegenerate FB in the band structure.
Without loss of generality, we assume that the flat band has a zero energy eigenvalue.
Then, the corresponding Hamiltonian $H(\bk)$ and an eigenstate of flat band in momentum space $\uef{}$ satisfy:
\ba
\sum_{\beta=1}^{n_{\rm tot}} H(\bk)_{\alpha \beta} \uef{}_\beta = 0
\label{eq:Heq}
\ea
where $\alpha,\beta=1,\dots,n_{\rm tot}$ and $\bk$ denotes a point in the BZ.
In the periodic basis (Appendix~\ref{app:TB}), each element of the Hamiltonian $H(\bk)_{\alpha\beta}$ and eigenstate $\uef{}_\alpha$ is given by a Laurent polynomial in $d$ variables $Q_i$ ($i=1,\dots,d$) with complex coefficient, when the hopping range is finite~\cite{read2017compactly,rhim2019classification}:
\ba
H(\bk)_{\alpha\beta}, \, \uef{}_\alpha \in \mathbb{C}[\bb Q,\cm{\bb Q}]
\ea
where $(\bb Q)_i=Q_i=e^{-i \bk \cdot \bb a_i}=e^{-ik_i}$, $(\cm{\bb Q})_i=\cm{Q}_i=Q_i^{-1}$, and $\bb a_i$ denotes $i$-th primitive lattice vector.
Here, $\mathbb{C}[\bb Q,\cm{\bb Q}]$ is a \textit{Laurent polynomial ring} over field $\mathbb{C}$, a set of all Laurent polynomials with coefficients from $\mathbb{C}$ in the variables $Q_1,\dots,Q_d$.
For example, $H(\bk)_{12}=c_1+c_2 Q_1\cm{Q_2}$ and $\uef{}_3=c_3+c_4Q_1$ with $c_{1,2,3,4} \in \mathbb{C}$.

In general, the momentum-space eigenstate can be written as
\ba
\uef{}_\alpha = \sum_{n_1 \dots n_d} \, S_\alpha(\textstyle{\sum_i} n_i \bb a_i) \prod_i Q_i^{n_i},
\label{eq:app_u_def0}
\ea
where $S_\alpha(\textstyle{\sum_i} n_i \bb a_i) \in \mathbb{C}$ denotes the coefficient of $\prod_i Q_i^{n_i}$, or equivalently, \ba
\uef{}_\alpha=\sum_{\bR} \, S_\alpha(\bR) \, e^{-i\bk\cdot\bR},
\label{eq:app_u_def1}
\ea
by noticing that any lattice vector $\bR$ can be written as $\bR=\sum_i n_i \bb a_i$.
We note that when $\uef{}$ vanishes at some points in the BZ, $\uef{}$ has zero norm and hence it cannot be served as an eigenstate at there.
Nevertheless, we refer to $\uef{}$ as a momentum-space eigenstate for simplicity.

Now, a real-space eigenstate of FB $\ket{w(\bR)}$ is defined as a Fourier transform of $\uef{}$:
\ba
\ket{w(\bR)} = \frac{1}{N_{\rm cell}} \sum_{\bR',\bk,\alpha} \, e^{i \bk \cdot \bR'} \ket{\hat{u}(\bk)}_\alpha \, \ket{\bR+\bR',\alpha}.
\label{eq:app_CLS_def0}
\ea
Straightforwardly, it can be shown that
\ba
\ket{w(\bR)} = \sum_{\bR',\alpha} \, S_\alpha(\bR') \ket{\bR+\bR',\alpha}.
\label{eq:app_CLS_def1}
\ea
Since the momentum-space eigenstate is given by a (finite) Laurent polynomial, the amplitude $S_\alpha(\bR')$ in \eq{eq:app_u_def1} is nonzero only inside a finite region, which we call a \textit{shape}.
Outside the shape, the amplitude $S_\alpha(\bR')$ is exactly zero.
For this reason, the real-space eigenstate $\ket{w(\bR)}$ is called a compact localized state (CLS)~\cite{sutherland1986localization,read2017compactly}.

For example, consider a tight-binding model in the kagome lattice.
When only the nearest neighbor hopping is considered, a FB exists in the band structure.
In this model, the momentum-space eigenstate of FB is given by $\uef{\rm kgm}=(1-Q_1,Q_2-1,1-\cm{Q_1}Q_2)$.
Then, according to \eq{eq:app_CLS_def1}, we obtain the corresponding CLS, $\ket{w_{\rm kgm}(\bR)} = \ket{\bR,1}-\ket{\bR+\bb a_1,1}+\ket{\bR+\bb a_2,2}-\ket{\bR,2}+\ket{\bR,3}-\ket{\bR-\bb a_1+\bb a_2,3}$.

Most importantly, we impose two conditions on $\uef{}$.
First, while $H(\bk)_{\alpha\beta}$ and $\uef{}_\alpha$ can be identically zero for some $\alpha$ and $\beta$, all elements of $H(\bk)_{\alpha\beta}$ or $\uef{}_\alpha$ cannot be identically zero at the same time:
\begin{itemize}
\item \textbf{Condition 1}: At least one element of $\uef{}$ is not identically zero.
Otherwise, $\uef{}$ is a trivial solution of \eq{eq:Heq}.
\end{itemize}
Second, we require that there is no nontrivial common divisor polynomial $F(\bb Q)$ of all elements of $\uef{}$.
If such common divisor polynomial exists, we factor it out and redefine $\uef{}$.
We note that $\mathbb{C}[\bb Q,\cm{\bb Q}]$ is a unique factorization domain.
Consequently, any nonzero Laurent polynomial in $\mathbb{C}[\bb Q,\cm{\bb Q}]$ can be factored into a product of irreducible (Laurent) polynomials and a unit in $\mathbb{C}[\bb Q,\cm{\bb Q}]$, and the factorization is unique up to order of the irreducible polynomials and the multiplication by units.
Hence, a common divisor polynomial can be defined up to the multiplication by a unit.
In $\mathbb{C}[\bb Q,\cm{\bb Q}]$, the units are \textit{monomials}.
Namely, any unit has the form of $c \prod_{i=1}^d Q_i^{n_i}$ with $c \in \mathbb{C}$, $n_i \in \mathbb{Z}$.
For the details on the polynomial rings, see Refs.~\onlinecite{dummit2004abstract,read2017compactly}.
Hence, the second condition can be rephrased as follows:
\begin{itemize}
\item \textbf{Condition 2} If $F(\bb Q) \in \mathbb{C}[\bb Q,\cm{\bb Q}]$ exists such that $\uef{}=F(\bb Q) \, \ket{\hat{u}'(\bk)}$ where $\uef{}_\alpha, \ket{\hat{u}'(\bk)}_\alpha \in \mathbb{C}[\bb Q,\cm{\bb Q}]$ for all $\alpha$, then $F(\bb Q)$ must be a monomial, \textit{i.e.} $F(\bb Q)=c \prod_{i=1}^d Q_i^{n_i}$ with $c \in \mathbb{C}$, $n_1,\dots,n_d \in \mathbb{Z}$.
\end{itemize}
In fact, this condition means that $\ket{w(\bR)}$, the CLS  corresponding to $\uef{}$, is \textit{elementary}.
Otherwise, $\ket{w(\bR)}$ is a sum of other CLS $\ket{w'(\bR)}$, the CLS corresponding to $\ket{\hat{u}'(\bk)}$.
To see this, let us suppose that $\uef{}$ does not satisfy the \textbf{Condition 2} such that $\uef{}=F(\bb Q) \ket{\hat{u}'(\bk)}$ and $F(\bb Q)$ is given by
\ba
F(\bb Q) = \sum_{n_1 \dots n_d} Q_1^{n_1} \cdots Q_d^{n_d} F_{n_1 \dots n_d}.
\ea
Then, $\ket{\hat{u}(\bk)}=F(\bb Q) \ket{\hat{u}'(\bk)}$ implies that
\ba
\ket{w(\bR)}
=& \sum_{n_1 \dots n_d} \, F_{n_1 \dots n_d} \ket{w'(\bR+\textstyle{\sum_i} n_i \bb a_i)}.
\ea
For example of $F(\bb Q)=1+Q_1$, $\ket{w(\bR)} = \ket{w'(\bR)} + \ket{w'(\bR+\bb a_1)}$.
Note that when $F(\bb Q)$ is given by a monomial, $F(\bb Q) = c \, Q_1^{n_1} \cdots Q_d^{n_d}$, $\ket{w(\bR)}$ is nothing but a translated copy of $\ket{w'(\bR)}$: $\ket{w(\bR)}=c \ket{w'(\bR+\textstyle{\sum_i}n_i \bb a_i)}$.

From now on, we call $\uef{}$ a Fourier transform of CLS (FT-CLS).
In general, $\uef{}$ can be called a momentum-space eigenstate of FB, only when a tight-binding Hamiltonian of FB model is specified.
On the other hand, when we consider a CLS without specifying any particular FB model, $\uef{}$ can be obtained by a Fourier transform of CLS without referring a relevant Hamiltonian.
Given this case, it is natural to call $\uef{}$ a FT-CLS.

\section{Singular flat band and zeros of FT-CLS \label{app:SFB}}
In this section, a singular flat band (SFB) is introduced and we discuss its physical implications.
First, consider a FT-CLS $\uef{}$ satisfying the \textbf{Conditions 1} and \textbf{2} defined in Appendix~\ref{app:FB_Def}, and suppose that $\uef{}$ vanishes at $\bk_*$: $\ket{\hat{u}(\bk_*)}=0$. 
In this case, $\uef{}$ cannot be considered as an eigenstate at $\bk_*$ since it has a zero norm at $\bk_*$.
Hence, we refer to such $\bk_*$, where $\uef{}$ becomes zero, as a singular point.
Note that the number of singular points can be larger than 1.
When the FT-CLS $\uef{}$ has singular point(s), then the corresponding FB is called a SFB~\cite{rhim2019classification,rhim2021singular}.
In this section, we elaborate the notion of SFB further.
Particularly, we prove that a SFB must have band crossing points at the singular points by using the properties of Laurent polynomial ring.
\subsection{Singular point and linear dependence of CLS}
First, let us consider the case where a FT-CLS $\uef{}$ has no singular point.
Then, a normalized eigenstate of FB is well-defined everywhere in the BZ.
Hence, a set of the corresponding CLSs $\{\ket{w(\bR)}\}$ for all $\bR$ spans the whole eigenstates of FB completely, and the FB is gapped in general.
(The FB may be gapless when fine-tuned.)

However, when $\ket{\hat{u}(\bk)}$ is singular at $\bk_*$, a normalized eigenstate of flat band is well-defined only at $\bk \ne \bk_*$.
According to \eq{eq:app_u_def0}, $\ket{\hat{u}(\bk_*)}=0$ implies that
\ba
\sum_{n_1\dots n_d} \, q_1^{n_1} \cdots q_d^{n_d} \, S_\alpha (\textstyle{\sum_i} n_i \bb a_i) = 0
\label{eq:zerou}
\ea
where we define $(\bb q)_i=e^{-i\bk_* \cdot \bb a_i}$ ($i=1,\dots,d$).
Now, we show that $\{\ket{w(\bR)}\}$ is linearly dependent and hence cannot span the whole eigenstates of FB completely, by proving that $\sum_\bR \, e^{i \bk_* \cdot \bR} \, \ket{w(\bR)} = 0$~\cite{bergman2008band}.
From \eq{eq:app_CLS_def1},
\begin{widetext}
\ba
\sum_\bR \, e^{i \bk_* \cdot \bR} \, \ket{w(\bR)}
=& \sum_{n_1\dots n_d} \sum_{m_1\dots m_d} \sum_\alpha \, q_1^{-n_1} \cdots q_d^{-n_d} \, S_\alpha(\textstyle{\sum_i} m_i \bb a_i) \, \ket{\textstyle{\sum_i} (n_i+m_i) \bb a_i, \alpha} \nn \\
=& \sum_\alpha \left( \sum_{n_1\dots n_d} q_1^{-n_1} \cdots q_d^{-n_d} \, \ket{\textstyle{\sum_i} n_i \bb a_i, \alpha} \right) \, \left( \sum_{m_1\dots m_d} q_1^{m_1} \cdots q_d^{m_d} \, S_\alpha(\textstyle{\sum_i} m_i \bb a_i) \right)
= 0.
\label{eq:incompCLS}
\ea
\end{widetext}
In the last equality, \eq{eq:zerou} is used.
Hence, the number of independent CLSs is equal to or less than $N_{\rm cell}-N_c$ when the number of singular points is $N_c$.
Without any further constraints, the number of independent CLSs may be given by $N_{\rm cell}-N_c$.
In this case, there must be additional eigenstates to span the whole eigenstates of FB.
Such states, called a non-contractible loop states (NLSs) or non-contractible planar states (NPSs), are extended along at least one direction~\cite{bergman2008band,rhim2019classification,ma2020spin,chiu2020fragile}.
When the number of NLSs and NPSs are greater than $N_c$, then the total number of zero-energy eigenstates exceeds $N_{\rm cell}$.
Hence, there must be a band crossing between flat and other bands.
However, it is not straightforward to find an independent set of NLSs and NPSs, and to specify the number and locations of band crossing points for arbitrary FB models.
In the following subsection, we show that when a given $\uef{}$ is singular at $\bk_*$ then there must be a band crossing between flat and other bands at $\bk_*$.

\subsection{SFB has band crossings at singular points \label{app:SFB_Proof}}
In this section, we show that a SFB has band crossing points at singular points.
Consider a FB Hamiltonian $H(\bk)$ and a relevant FT-CLS $\uef{}$.
For simplicity, we assume that $\uef{}$ has a singular point at $\bk_*$ because the general case with multiple singular points follows from the result for a singular point.
First, let us assume that $(\alpha,\alpha)$ minor of $H(\bk)$ vanishes at $\bk_*$, \textit{i.e.} ${\rm Det} H_{(\alpha,\alpha)}(\bk_*)=0$ for all $\alpha=1,\dots,n_{\rm tot}$. 
Here, we define a $(n_{\rm tot}-1) \times (n_{\rm tot}-1)$ matrix $H_{(\alpha,\beta)}(\bk)$ that is obtained by removing $\alpha$-th row and $\beta$-th column of $H(\bk)$.
Then, $H(\bk_*)$ must have at least two zero eigenvalues.
This can be shown by looking at the characteristic polynomial of $H(\bk_*)$:
\ba
P_{H(\bk_*)}(\ep) = \sum_{\alpha=1}^{n_{\rm tot}} \, (-1)^{n_{\rm tot}-\alpha} \, p_\alpha \ep^\alpha
\label{eq:chPol}
\ea
where
\bg
p_0 = {\rm Det}H(\bk_*), \quad
p_1 = \sum_{\alpha=1}^{n_{\rm tot}} {\rm Det}H_{(\alpha,\alpha)}(\bk_*), \nn \\
\dots, \quad
p_{n_{\rm tot}-1} = {\rm Tr}[H(\bk_*)], \quad
p_{n_{\rm tot}} = 1.
\eg
By definition of FB system, ${\rm Det}H(\bk)=0$ for all $\bk$.
Hence, $p_0=0$.
Also, $p_1=0$ since every $(\alpha,\alpha)$ minor of $H(\bk_*)$ vanishes by assumption.
Hence, $H(\bk_*)$ has at least two zero eigenvalues and this implies that a band crossing between flat and other bands must occur at $\bk_*$.

Now, let us prove that $(\alpha,\alpha)$ minor of $H(\bk)$ vanishes at $\bk_*$ for all $\alpha$, which concludes the proof that a SFB with a singular point $\bk_*$ has a band crossing at $\bk_*$.
First, consider the Hamiltonian equations in \eq{eq:Heq} for all $\alpha \ne A$.
Here, $A$ is chosen arbitrarily.
These ($n_{\rm tot}-1$) number of equations are expressed as
\ba
H_{(A,A)}(\bk)
\bpm u_1 \\ \vdots \\ u_{A-1} \\ u_{A+1} \\ \vdots \\ u_{n_{\rm tot}} \epm
=-\bpm h_{1,A} \\ \vdots \\ h_{A-1,A} \\ h_{A+1,A} \\ \vdots \\ h_{n_{\rm tot},A} \epm u_A,
\label{eq:reducedEq}
\ea
where $h_{\alpha,\beta}=H(\bk)_{\alpha\beta}$ and $u_\alpha=\ket{\hat{u}(\bk)}_\alpha$.
Without loss of generality, we have two cases where $u_A$ is given by (i) the zero polynomial (ZP), \textit{i.e.} $u_A=0$, or (ii) not the ZP.

In the first case (i) where $u_A=0$, ${\rm Det} H_{(A,A)}(\bk)$ must be the ZP.
Otherwise, \eq{eq:reducedEq} implies that $u_\alpha=0$ for all $\alpha=1,\dots,n_{\rm tot}$ and hence $\uef{}=0$ for all $\bk$ which does not obey the \textbf{Condition 1} (Appendix~\ref{app:FB_Def}).
In order to show this, we use the following property of the adjugate matrix:
\ba
{\rm adj}[C] C = C {\rm adj}[C] = {\rm Det}C \, \mathds{1}
\ea
where $C$ and $\mathds{1}$ denote an arbitrary square matrix and the identity matrix respectively, and ${\rm adj}[C]$ is an adjugate matrix of $C$.
Multiplying ${\rm adj}[H_{(A,A)}(\bk)]$ to \eq{eq:reducedEq}, we obtain
\bg
{\rm Det} H_{(A,A)}(\bk) \,
(u_1,\dots,u_{A-1},u_{A+1},\dots,u_{n_{\rm tot}})^T=0
\label{eq:reducedEq1}
\eg
Since ${\rm Det} H_{(A,A)}(\bk)$ and $u_\alpha \in \mathbb{C}[\bb Q, \cm{\bb Q}]$, \eq{eq:reducedEq1} is reduced to $u_\alpha=0$ for all $\alpha \ne A$ when ${\rm Det} H_{(A,A)}(\bk)$ is not the ZP.
Then, all elements of $\uef{}$ is the ZP and the \textbf{Condition 1} is violated.
Hence, ${\rm Det} H_{(A,A)}(\bk)=0$ and hence ${\rm Det} H_{(A,A)}(\bk_*)=0$.

Now, let us consider the second case (ii) where $u_A$ is not the ZP.
Apply the Cramer's rule to \eq{eq:reducedEq}:
\ba
u_{\alpha \ne A} \, {\rm Det}H_{(A,A)}(\bk)
= (-1)^{A-\alpha} u_A \, {\rm Det}H_{(A,\alpha)}(\bk).
\label{eq:Cramer}
\ea
Note that there must be at least one $\alpha$ such that $u_\alpha$ is not the ZP.
Otherwise, $u_\alpha=0$ for all $\alpha \ne A$ and hence $\uef{}=(0,\dots,u_A,\dots,0)$.
Since $u_A$ cannot be a monomial by the assumption of singular FT-CLS at $\bk_*$, this violates the \textbf{Condition 2} since $u_A$ can be factored out of $\uef{}$.
Thus, we can choose $\alpha$ such that $u_\alpha$ is not the ZP.
Because of the unique factorization, $u_A$ is expressed as
\ba
u_A = \mc{M}_A \times (p_{A,1} p_{A,2} \cdots) \times (r_{A,1} r_{A,2} \cdots) 
\ea
where $\mc{M}_A$ is a monomial and $P_A=\{p_{A,1},p_{A,2},\dots\}$ ($R_A=\{r_{A,1},r_{A,2},\dots\}$) is a set of irreducible polynomials which vanish (do not vanish) at $\bk_*$.
Similarly, $u_\alpha = \mc{M}_\alpha \times (p_{\alpha,1} p_{\alpha,2} \cdots) \times (r_{\alpha,1} r_{\alpha,2} \cdots)$.

Crucially, the \textbf{Condition 2} ensures that there must be at least one $\alpha$ such that $P_A-P_\alpha$ is a nonempty set.
For such $\alpha$, \eq{eq:Cramer} implies that ${\rm Det} H_{(A,A)}(\bk)$ can be factored by at least one irreducible polynomial in $P_A-P_\alpha$.
Hence, we conclude that ${\rm Det} H_{(A,A)}(\bk_*)=0$.
As $A$ is chosen arbitrarily, $(\alpha,\alpha)$ minor of $H(\bk_*)$ vanishes for all $\alpha$, \textit{i.e.} ${\rm Det}H_{(\alpha,\alpha)}(\bk_*)=0$ for all $\alpha=1,\dots,n_{\rm tot}$.
(We note that every minor of $H(\bk_*)$ vanishes, in fact, \textit{i.e.} ${\rm Det}H_{(\alpha,\beta)}(\bk_*)=0$ for all $\alpha,\beta=1,\dots,n_{\rm tot}$, and it can be shown in a similar way.)

\subsection{Implications of the relation between band crossing points and singular points \label{app:SFB_discuss}}
We comment and discuss the implications of our result, the relation between a singular point and a band crossing point.
First, we stress that the converse of the relation is not true.
For example, consider a simple FB-model in one dimension,
\ba
H(k_1)=(2+Q_1+\cm{Q_1}) \bpm 1 & 1 \\ 1 & 1 \epm,
\ea
and the corresponding FT-CLS $\ket{\hat{u}(k_1)}=(1,-1)$.
While $H(k_1)=0$ at $Q_1=-1$ ($k_1=\pi$) implies a band crossing between flat band dispersive bands, $\ket{\hat{u}(k_1)}$ is nonzero everywhere in the BZ.
Second, the relation between a singular point and a band crossing point of FB also holds for a general set of singular points.
For example, when a FT-CLS is given by $\ket{\hat{u}_1(k_1,k_2)}=(1+\om \cm{Q_1}+\om^2 \cm{Q_2},1+\om \cm{Q_1}+\om^2 \cm{Q_1}Q_2)$ with $\om=e^{2\pi i/3}$~\cite{hwang2021wave}, the flat band must have at least two band crossing points, since $\ket{\hat{u}_1(k_1,k_2)}$ has zeros $(Q_1,Q_2)=(1,1)$ and $(\om^2,\om)$.
As another example, consider $\ket{\hat{u}_2(k_1,k_2,k_3)}=(Q_2-Q_3,Q_3-Q_1,Q_1-Q_2)$ which has a line of zeros $Q_1=Q_2=Q_3$, \textit{i.e.} $k_1=k_2=k_3$.
In this case, the corresponding flat band and other band must form a line degeneracy.
Third, our proof does not rely on neither the spin of (quasi-)particles nor the hermiticity of $H(\bk)$.
Hence, our result can be applied to both bosonic and fermionic systems.
We also expect a potential application of our result to non-Hermitian system.

Note that there are two cases when $\uef{}$ cannot be singular when the \textbf{Conditions 1} and \textbf{2} are satisfied.
First, in one dimension, $\uef{}$ is a univariate polynomial in a single variable $Q_1$ up to the multiplication of a monomial~\cite{rhim2019classification}.
For $\uef{}$ satisfying the \textbf{Condition 1}, the \textbf{Condition 2} and the fundamental theorem of algebra prevent $\uef{}$ from having possible zero(s).
For example, $\uef{}=(1+Q_1,2+Q_1+\cm{Q_1})$ vanishes at $Q_1=-1$.
According to the fundamental theorem of algebra, there must be a common divisor polynomial $F(Q_1)=1+Q_1$ up to the multiplication of a monomial.
Then, we define $\ket{\hat{u}'(\bk)}$ such that $\uef{}=F(Q_1) \ket{\hat{u}'(\bk)}$ where $F(Q_1)=1+Q_1$, and hence $\ket{\hat{u}'(\bk)}=(1,1+\cm{Q_1})$ is not singular at every $Q_1$.
According to the \textbf{Condition 2}, $\ket{\hat{u}'(\bk)}$ is the proper FT-CLS.
Second, when $\uef{}$ has only one element as a nonzero Laurent polynomial, \textbf{Condition 2} prevent $\uef{}$ from having possible zero(s).
For example, consider $\uef{}=(0,\dots,F(\bb Q)\dots,0)$.
Then, we choose $\ket{\hat{u}'(\bk)}=(0,\dots,1,\dots,0)$ as the proper FT-CLS by following the \textbf{Condition 2}.

Finally, we comment on the case of degenerate FBs.
Even in this case, all the momentum-space eigenstates corresponding to flat bands are always given by Laurent polynomials.
When we focus a single FT-CLS $\uef{}$ among them, our discussion so far holds in general.
Namely, if $\uef{}$ has singular points, the corresponding FB must have band crossings between other band at the singular points.
However, such band crossings can be formed with other FB rather than dispersive bands.
Then, FBs can be isolated from dispersive bands by a gap, as like the models discussed in Refs.~\onlinecite{bergman2008band,peri2021fragile,chiu2020fragile}.

\section{Symmetry representation of CLS and SR-enforced band crossing \label{app:symrep}}
In this section, we define symmetry representation (SR) of compact localized state and its Fourier transform.
Also, a generic condition for having SR-enforced band crossings is discussed.

\subsection{Symmetry representation of CLS \label{app:SR}}
In a periodic lattice system, a CLS is centered at the Wyckoff position in each unit cell.
As a nondegenerate FB is considered, a CLS must be centered at the Wyckoff position $\bchi$ of unit multiplicity.
(Hence, we consider a symmorphic space group.)
Then, the site symmetry group $G_\bchi$ is defined such that any element $\sg$ in $G_\bchi$ leaves $\bchi$ invariant.
Here, we focus on a single symmetry element $\sg=\{O_\sg|\bb \delta_\sg\}$ in $G_\bchi$, \textit{i.e.} $\sg: \bb r \rightarrow O_\sg \bb r + \bb \delta_\sg$.
Then, the CLS can be labelled by a symmetry eigenvalue $\symeigv$ with respect to $\sg$, and it transforms under $\sg$ as
\bg
\hat{\sg} \ket{w(\bR);\bchi,\symeigv} = \ket{w(\bR_\sg);\bb \bchi,\symeigv} \, \symeigv,
\label{eq:SR_w_def}
\eg
where $\bR_\sg=O_\sg (\bR+\bchi)+\bb \delta_\sg-\bchi$.
Correspondingly, the FT-CLS transforms as
\bg
\label{eq:SR_u_def}
U_\sg(\bk) \ket{\hat{u}(\bk);\bchi,\symeigv} = 
\ket{\hat{u}(O_\sg \bk);\bchi,\symeigv} \, \symeigv(\bk), \\
\label{eq:SR_xi_def}
\symeigv(\bk) = \symeigv e^{-i O_\sg \bk \cdot (O_\sg \bchi-\bchi+\bb \delta_\sg)}.
\eg
Hence, SR of CLS and FT-CLS (SR of $\sg$ on CLS and FT-CLS, more precisely) are defined by $\symeigv$ and $\symeigv(\bk)$ respectively.
For notational simplicity, we now include the Wyckoff position $\bchi$ of CLS in the definition of SR of CLS.
Then, the SR of FT-CLS $\symeigv(\bk)$ is induced from the SR of CLS $(\bchi,\symeigv)$ according to \eq{eq:SR_xi_def}.
We note that a Wannier function with Wyckoff position $\bchi$ and symmetry representation $\symeigv$ transforms as \eq{eq:SR_w_def} in the same way as the CLS $\ket{w(\bR);\bchi,\symeigv}$ does.
Hence, a representation theory on CLS and FT-CLS coincides the known results on Wannier function and band representation~\cite{zak1980symmetry,zak1981band,bradlyn2017topological,cano2018building,alexandradinata2018no,holler2018topological,alexandradinata2020crystallographic}.
However, there is a significant difference between CLS and Wannier function, which is the main theme of this work.
A set of symmetric and exponentially localized Wannier functions induces a band representation.
On the other hand, for certain SRs of a CLS, the corresponding FT-CLS must be singular and hence the FB cannot be band representable.
This point is discussed in detail in the next section.

Finally, we explain how \eqs{eq:SR_u_def} and \eqref{eq:SR_xi_def} are derived from \eq{eq:SR_w_def}.
To this end, let us introduce $\ket{\psi(\bk);\bchi,\symeigv}$:
\ba
\ket{\psi(\bk);\bchi,\symeigv}
&= \frac{1}{\sqrt{N_{\rm cell}}} \, \sum_\bR \, e^{i \bk \cdot \bR} \ket{w(\bR);\bchi,\symeigv}.
\label{eq:app_psi}
\ea
From \eqs{eq:basis_def} and \eqref{eq:app_CLS_def0}, \eq{eq:app_psi} is reduced to
\ba
\ket{\psi(\bk);\bchi,\symeigv}
= \sum_\alpha \ket{\hat{u}(\bk);\bchi,\symeigv}_\alpha \, \ket{\bk,\alpha},
\ea
and hence $\ket{\hat{u}(\bk);\bchi,\symeigv}_\alpha=\brk{\bk,\alpha}{\psi(\bk);\bchi,\symeigv}$.
Finally, $U_\sg(\bk)_{\alpha\beta} \ket{\hat{u}(\bk);\bchi,\symeigv}_\beta$ can be evaluated as follows:
\ba
& U_\sg(\bk)_{\alpha\beta} \ket{\hat{u}(\bk);\bchi,\symeigv}_\beta \nn \\
=& U_\sg(\bk)_{\alpha\beta} \bra{\bk,\beta} \hat{\sg}^{-1} \hat{\sg} \ket{\psi(\bk);\bchi,\symeigv} \nn \\
=& \bra{O_\sg \bk, \alpha} \hat{\sg} \ket{\psi(\bk);\bchi,\symeigv} \nn \\
=& \bra{O_\sg \bk, \alpha} \frac{1}{\sqrt{N_{\rm cell}}} \sum_\bR \, e^{i \bk \cdot \bR} \, \hat{\sg} \ket{w(\bR);\bchi,\symeigv} \nn \\
=& \symeigv(\bk) \, \bra{O_\sg \bk, \alpha} \frac{1}{\sqrt{N_{\rm cell}}} \sum_{\bR_\sg} \, e^{i O_\sg \bk \cdot \bR_\sg}  \, \ket{w(\bR_\sg);\bb \bchi,\symeigv} \nn \\
=& \ket{\hat{u}(O_\sg \bk);\bchi,\symeigv} \, \symeigv(\bk).
\ea

\subsection{SR-enforced band crossing \label{app:SR_enforced}}
For each element in $G_\bchi$, a FT-CLS transforms according to its SR,
\ba
U_{\sg_I}(\bk) \ket{\hat{u}(\bk)}=\ket{\hat{u}(O_{\sg_I} \bk)} \, \xi_{\sg_I}(\bk),
\label{eq:general_SR}
\ea
for $\sg_I \in G_\bchi$ with $I=1,2,\dots,|G_\bchi|$.
Now, let us consider a high-symmetry point $\bk$ which is left invariant under $N(\le|G_\bchi|)$ symmetry elements, say, $\sg_1,\sg_2,\dots,\sg_N$ without loss of generality: $O_{\sg_J} \overline{\bk} = \overline{\bk}$ (mod $\bb G$) with $J=1,2,\dots,N$.
Then, \eq{eq:general_SR} can be written as $\mc{P}_{J}(\overline{\bk}) \ket{\hat{u}(\overline{\bk})}=0$ where $\mc{P}_{J}(\overline{\bk})=U_{\sg_J}(\overline{\bk})-\xi_{\sg_J}(\overline{\bk}) \mathds{1}_{n_{\rm tot}}$.
When the atomic orbitals constituting the CLS cannot span the FT-CLS at $\overline{\bk}$, the FT-CLS must be singular at there, \textit{i.e.} $\ket{\hat{u}(\overline{\bk})}=0$, and hence a band crossing point is enforced by SR at $\overline{\bk}$.

Now, we discuss the case of a single unitary symmetry $\sg$ in more detail.
At high-symmetry point $\overline{\bk}_\sg$ with respect to $\sg$, \eq{eq:SR_u_def} is reduced to
\ba
\left[ U_\sg(\overline{\bk}_\sg) - \symeigv(\overline{\bk}_\sg) \mathds{1}_{n_{\rm tot}} \right] \, \ket{\hat{u}(\overline{\bk}_\sg);\bchi,\symeigv} = 0.
\label{eq:suffcond0}
\ea
Suppose that some elements of FT-CLS are identically zero: $\ket{\hat{u}(\bk);\bchi,\symeigv}_\alpha=0$ for $\alpha \in \{\alpha_\varnothing\}$.
This means that the CLS $\ket{w(\bR);\bchi,\symeigv}$ does not occupy sublattice sites corresponding to $\{\alpha_\varnothing\}$.
By the definition of symmetry, $\sg$ maps sublattice sites corresponding to $\{\alpha_\varnothing\}$ onto themselves.
Hence, we interchange rows and columns such that $U_\sg(\bk)$ can be written as a block diagonal matrix:
\ba
\bpm \left[U^{(1)}_\sg(\bk)\right]_{AB} & \\ & \left[U^{(2)}_\sg(\bk)\right]_{\mu\nu} \epm
\ea
where $A,B=1,\dots,n_1$ and $\mu,\nu=1,\dots,n_2$.
In the above equation, $U^{(1)}_\sg(\bk)$ acts only on the sublattice sites belonging to $\{\alpha_\varnothing\}$ while $U^{(2)}_\sg(\bk)$ acts only on the sublattice sites not belonging to $\{\alpha_\varnothing\}$.
Then, in this basis, \eq{eq:suffcond0} is expressed as
\bg
\sum_{B=1}^{n_1} \left[U^{(1)}_\sg(\overline{\bk}_\sg) - \symeigv(\overline{\bk}_\sg) \mathds{1}_{n_1}\right]_{AB} \bpm 0 \\ \vdots \\ 0 \epm_B = 0, \\
\sum_{\nu=1}^{n_2} \left[U^{(2)}_\sg(\overline{\bk}_\sg) - \symeigv(\overline{\bk}_\sg) \mathds{1}_{n_2}\right]_{\mu\nu} \ket{\hat{u}(\overline{\bk}_\sg);\bchi,\symeigv}_\nu=0
\eg
When ${\rm Det}\left[U^{(2)}_\sg(\overline{\bk}_\sg) - \symeigv(\overline{\bk}_\sg) \mathds{1}_{n_2}\right]$ is not zero, all the elements of $\ket{\hat{u}(\overline{\bk}_\sg);\bchi,\symeigv}$ must be zero.
Thus, we conclude that a FB is singular at $\overline{\bk}_\sg$, if its FT-CLS satisfies
\ba
{\rm Det}_{\{\alpha_\varnothing\}} \left[ U_\sg(\overline{\bk}_\sg)-\symeigv(\overline{\bk}_\sg) \mathds{1}_{n_{\rm tot}} \right] \ne 0,
\label{eq:criterion_app}
\ea
where ${\rm Det}_{\{\alpha_\varnothing\}}[M]$ denotes the determinant of $M$ after removing rows and columns belong to $\{\alpha_\varnothing\}$.

The interpretation of the determinant test \eq{eq:criterion_app} is as follows.
If a FB with a given SR can exists without a band crossing, the FB must have a symmetry eigenvalue $\symeigv(\overline{\bk}_\sg)$ at high-symmetry point $\overline{\bk}_\sg$.
On the other hand, symmetry eigenvalues of full set of bands are given by eigenvalues of $U_\sg(\overline{\bk}_\sg)$.
If the set of eigenvalues of $U_\sg(\overline{\bk}_\sg)$ does not contain $\symeigv(\overline{\bk}_\sg)$, our initial assumption that the FB is gapped must be broken.
Such mismatch of symmetry eigenvalues is captured by the determinant test.
Considering that some sublattice sites are not occupied by the CLS, the determinant is performed only for submatrix as in \eq{eq:criterion_app}.
Below, we make clear the role of unoccupied sublattice sites $\{\alpha_\varnothing\}$ by interpreting the SR-enforced band crossing in terms of band representation.

We note that not all unitary symmetries can give rise to a SR-enforced band crossing.
An example of such symmetries is a single mirror.
If a band crossing is enforced by SR under a single mirror, $\uef{}$ vanishes along a high-symmetry line (plane) in two (three) dimensions.
Then, $\uef{}$ has a overall common divisor polynomial and it violates our requirement that the CLS is elementary (Appendix~\ref{app:FB_Def}).
For instance, let us consider a mirror $\mc{R}$ which acts on momentum coordinates as $\mc{R}:$ $(k_1,k_2) \rightarrow (k_2,k_1)$, \textit{i.e.} $(Q_1,Q_2) \rightarrow (Q_2,Q_1)$. If $\ket{\hat{u}(k_1,k_1)}=0$ for all $k_1$, then $(Q_1-Q_2)$ divides $\uef{}$.

\begin{figure}[t!]
\centering
\includegraphics[width=0.45\textwidth]{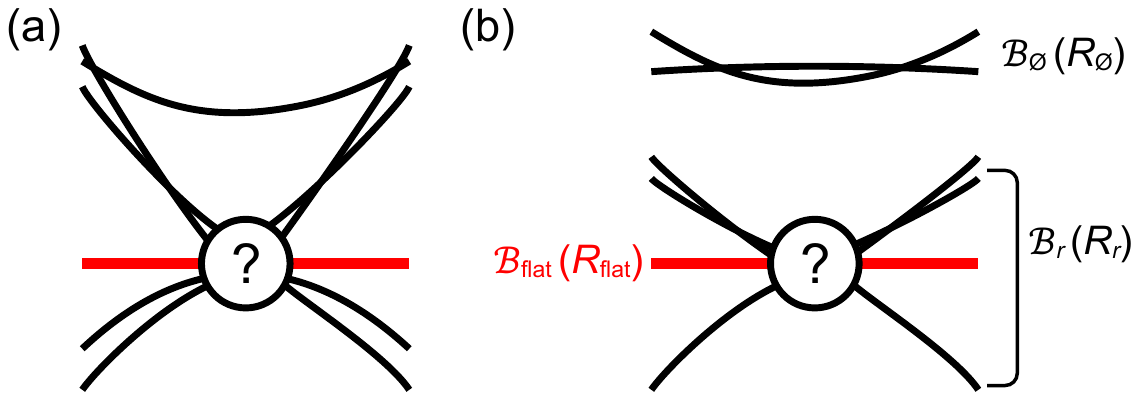}
\caption{
A schematic illustration of how energy bands are divided into a flat band $\mc{B}_{\rm flat}$, detachable band(s) $\mc{B}_\varnothing$, and the rest of bands $\mc{B}_r$.
Their representations ($R_{\rm flat}$, $R_\varnothing$ and $R_r$ in order) are expressed as the sums and differences of elementary band representations.
(a) A flat band exists in the band structure.
The flat band (red) may or may not have a band crossing with other bands (black).
(b) We increase the on-site energy of sublattice sites that the CLS does not occupy.
Then, some bands $\mc{B}_\varnothing$ are detached from the flat band $\mc{B}_{\rm flat}$ and the rest of bands $\mc{B}_r$.
Whether $\mc{B}_{\rm flat}$ can be isolated or not is determined by $R_r$.
}
\label{FigS1}
\end{figure}

\subsection{Interpretation of SR-enforced band crossing based on band representation \label{app:SR_BR}}
Let us consider a band structure in which a nondegenerate FB exists, as illustrated in \fig{FigS1}(a).
The SR of FB is denoted as $R_{\rm flat}$.
Now, assume that the band structure can be deformed so that the FB with $R_{\rm flat}$ SR can be gapped and isolated from other bands.
Since any single isolated FB is band representable~\cite{Chen2014impossibility,read2017compactly,alexandradinata2018no}, we note the band representation (BR) of FB as $R_{\rm flat}$.
Then, we divide the full set of bands except the FB into two set of bands, $\mc{B}_\varnothing$ and $\mc{B}_r$ with representations $R_\varnothing$ and $R_r$ respectively (see \fig{FigS1}).
While $\mc{B}_\varnothing$ is band representable and $R_\varnothing$ is a BR, $R_r$ does not have to be a BR necessarily, as we will now explain.
Nevertheless, $R_r$ can be expressed as the sums and differences of elementary BRs~\cite{bradlyn2017topological,elcoro2020magnetic}.
Note that the BR of the whole bands is induced by all basis atomic orbitals, and it can be expressed as $R_{\rm full}=R_{\rm flat} \oplus R_r \oplus R_\varnothing$.

We define $\mc{B}_\varnothing$ first, because $\mc{B}_r$ is determined by bands except the FB and $\mc{B}_\varnothing$.
When the CLS does not occupy some sublattice sites $\{\alpha_\varnothing\}$, some bands can be detached from other bands by reaching an atomic limit while maintaining the flat dispersion of FB.
Specifically, we can turn on on-site energy of sublattice sites $\{\alpha_\varnothing\}$.
For sufficiently large on-site energy, a set of bands $\mc{B}_\varnothing$ is detached from the rest of bands [\fig{FigS1}(b)].
Their representation $R_\varnothing$ is equal to a BR induced by basis atomic orbitals located at the sublattice sites $\{\alpha_\varnothing\}$.

After an exclusion of detachable bands $\mc{B}_\varnothing$ from the whole bands, the rest of bands, which can have a band crossing with the FB, corresponds to the representation $R_r = R_{\rm full} \ominus R_\varnothing \ominus R_{\rm flat}$.
Then, $R_r$ determines the presence or absence of band crossing with the FB and other bands.
If $R_r$ is incompatible with a set of bands isolated from others, our initial assumption that the FB with $R_{\rm flat}$ SR is isolated must be broken.
Therefore, a band crossing between the FB and other bands $\mc{B}_r$ must exist.
On the other hand, when $R_r$ is compatible with a set of bands isolated from others, there is no SR-enforced band crossing.

{
\renewcommand{\arraystretch}{1.4}
\begin{table}[t!]
\centering
\begin{minipage}{0.48\textwidth}
\caption{
Band representations (BRs) of $C_4$-symmetric lattice.
First column:
$\bchi_l$ indicates the BR induced from the Wyckoff position $\bchi=\bb A, \bb D$ with the $C_4$ eigenvalue $\xi_l=e^{i\pi l/2}$ where $l=0,1,2,3$.
$\bb B_{\pm}$ is the BR induced from the Wyckoff position $\bb B$ with the $C_2$ eigenvalue $\pm1$.
The maximal Wyckoff positions are $\bb A=(0,0)$, $\bb B=\{(1/2,0),(0,1/2)\}$, and $\bb D=(1/2,1/2)$.
Second to fourth columns:
Symmetry eigenvalues of irreducible representations at each high symmetry point are listed, where $\Gamma=(0,0)$, $X=(\pi,0)$ and $M=(\pi,\pi)$ in the BZ.
}
\label{Table1}
\end{minipage}
\begin{tabular*}{0.48\textwidth}{@{\extracolsep{\fill}}c c c c}
\hline \hline
\multirow{2}{*}{BR} & \multicolumn{2}{c}{$C_4$ eigenvalue} & $C_2$ eigenvalue \\
\cline{2-3} \cline{4-4}
& $\Gamma$ & $M$ & $X$ \\
\hline
$\bb A_l$ & $\xi_l$ & $\xi_l$ & $\xi_l^2$ \\
$\bb D_l$ & $\xi_l$ & $-\xi_l$ & $-\xi_l^2$ \\	
$\bb B_+$ & ($-1,1$) & ($-i,i$) & ($-1,1$) \\
$\bb B_-$ & ($-i,i$) & ($-1,1$) & ($-1,1$) \\ \hline \hline
\end{tabular*}
\end{table}
}

Based on the above discussion, let us discuss SR-enforced band crossing in the Lieb lattice.
The Lieb lattice has $C_4$ rotation and is composed of three sublattices [see \fig{FigS3}(a)].
The sublattices are located at the maximal Wyckoff positions $\bb B$ and $\bb D$.
Note that the maximal Wyckoff positions in $C_4$-symmetric lattice are $\bb A=(0,0)$, $\bb B=\{(1/2,0),(0,1/2)\}$, and $\bb D=(1/2,1/2)$.
The BRs of $C_4$-symmetric lattice are listed in Table~\ref{Table1}.
According to Table~\ref{Table1}, the BR of whole three bands is equivalent to $R_{\rm full}=\bb B_+ \oplus \bb D_0$.
Now, consider a CLS with $\bb A_2$ SR, which is centered at $\bb A$ and has $C_4$ eigenvalue $\xi_2=-1$.
When the CLS does not occupy the sublattice corresponding to the Wyckoff position $\bb D$, the relevant component of FT-CLS is identically zero: $\uef{}_3=0$ without loss of generality.
In this case, we can increase on-site potential of the third sublattice site by adding $\delta H(\ep_3)={\rm Diag}(0,0,\ep_3)$ to the unperturbed Hamiltonian.
For any $\ep_3$, the FB remains since $\uef{}_3=0$ ensures that $\delta H(\ep_3)\uef{}=0$.
Thus, a band with $\bb D_0$ BR is detachable.
This means that $R_\varnothing=\bb D_0$ and $R_r=\bb B_+ \ominus \bb A_2$.
However, any isolated band cannot have its representation as $R_r$.
This can be seen from $C_4$ eigenvalues at $M$ in Table~\ref{Table1}, because the irreducible representation of $\bb B_+$ has $C_4$ eigenvalue $\pm i$ at $M$ while that of $\bb A_2$ has $-1$.
This mismatch of eigenvalues implies that our initial assumption of gapped FB is invalid.
Hence, the FB must have a band crossing point at $M$.
On the other hand, when the CLS now occupies all three sublattices, $R_r= \bb D_0 \oplus \bb B_+ \ominus \bb A_2$.
In this case, $R_r$ is compatible with an isolated bands and thus there is no SR-enforced band crossing.
Interestingly, $\bb D_0 \oplus \bb B_+ \ominus \bb A_2$ corresponds to fragile topological bands~\cite{bradlyn2017topological,po2018fragile,cano2018topology,bradlyn2019disconnected,bouhon2019wilson,po2019faithful,else2019fragile,wieder2018axion,ahn2019failure,liu2019shift,hwang2019fragile,bouhon2020geometric,alexandradinata2020crystallographic,song2020fragile,song2020twisted,peri2020experimental,chiu2020fragile,zhang2021tunable}.
Finally, we comment on the case where the CLS with $\bb A_2$ SR occupies only the sublattice corresponding to the Wyckoff position $\bb D$.
In this case, the FT-CLS has the form of $\uef{}=(0,0,f(\bk))$.
If $f(\bk)$ is factored out, $\uef{}$ becomes a nonsingular FT-CLS, $\ket{\hat{u}'(\bk)}=(0,0,1)$.
This means that the CLS corresponding to $\uef{}$ is not elementary (Appendix~\ref{app:FB_Def}), we do not consider this case.
Also note that the CLS corresponding to $\ket{\hat{u}'(\bk)}$ has now $\bb D_0$ SR, not $\bb A_2$ SR.

\section{Flat-band models \label{app:FB_models}}
This section provides a detailed description of FB models, including those introduced in the main text.
Various features of each FB model are discussed to help understand SR-enforced band crossings.
Especially in the case of the kagome lattice, we demonstrate how the degeneracy at SR-enforced band crossing point and two-dimensional irreducible representations of point groups differ.
The hopping structure of each model can be read off from the corresponding tight-binding Hamiltonian by using
\ba
t_{\alpha \leftarrow \beta}(\Delta \bR) = \frac{1}{N_{\rm cell}} \, \sum_{\bk} e^{i \bk \cdot \Delta \bR} H(\bk)_{\alpha\beta}
\label{eq:hopping},
\ea
according to \eq{eq:H_Laurent}.
Also, we places an $s$-orbital at each sublattice, unless otherwise noted.

\begin{figure}[t!]
\centering
\includegraphics[width=0.48\textwidth]{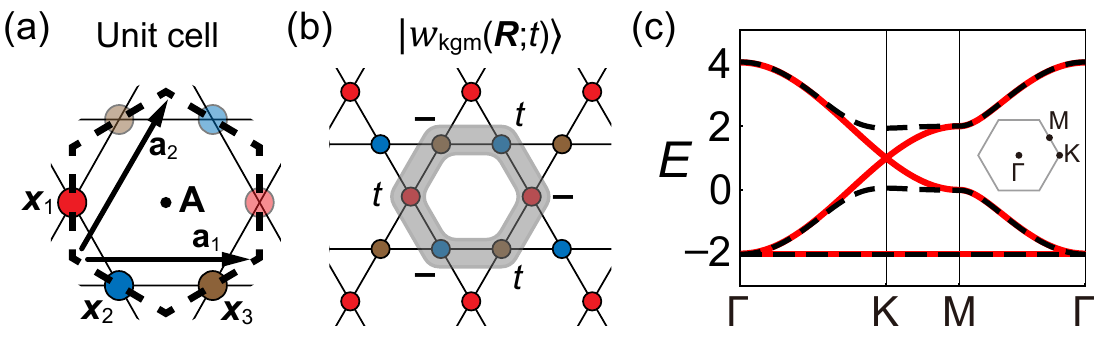}
\caption{
(a) Unit cell of the kagome lattice.
(b) Description of CLS $\ket{w_{\rm kgm}(\bR;t)}$.
For generic $t\in \mathbb{C}$, it is symmetric under $C_3$ and $M_y$.
When $t=1$, $\ket{w_{\rm kgm}(\bR;1)}$ has $C_6$, $M_{x,y}$ and $T$ symmetries.
(c) Band structure of $H_{\rm kgm}(\bk)$ ($H'_{\rm kgm}(\bk)$ with $t'=0.06$) is denoted by red lines (black dashed lines).
In both the FB models, the FB has a band crossing point at $\Gamma$.
In $H'_{\rm kgm}(\bk)$, $M_{x,y}$ and $T$ are broken while $C_6$ is preserved.
}
\label{FigS2}
\end{figure}

\subsection{Kagome lattice}
The kagome lattice is composed of three sublattices [\fig{FigS2}(a)].
The sublattices are located at $\bx_1=-\tfrac{1}{2} \bb a_1$, $\bx_2=-\tfrac{1}{2} \bb a_2$, and $\bx_3=\tfrac{1}{2} \bb a_1 - \tfrac{1}{2} \bb a_2$, where the primitive lattice vectors are $\bb a_1=(1,0)$ and $\bb a_2=(1/2,\sqrt{3}/2)$.
Hence, the sublattice embedding matrix $V(\bk)$ is given by $V(\bk)={\rm Diag}(Q_1^{-1/2},Q_2^{-1/2},Q_1^{1/2}Q_2^{-1/2})$.
The orbitals are permuted by $C_6$ rotation: $\hat{C}_6 \ket{\bR,1}=\ket{O_{C_6}\bR,2}$, $\hat{C}_6 \ket{\bR,2}=\ket{O_{C_6}\bR,3}$ and $\hat{C}_6 \ket{\bR,3}=\ket{O_{C_6}\bR+\bb a_1,1}$ for $C_6=\{O_{C_6}|\bb 0\}$.
Note that $C_6$ transforms $(k_1,k_2)$ and $(Q_1,Q_2)$ as $(k_1-k_2,k_1)$ and $(Q_1\cm{Q_2},Q_1)$ respectively.
From \eqs{eq:unisym_real}-\eqref{eq:unisym_BZ}, symmetry operator of $C_6$ is expressed as
\bg
U_{C_6}(\bk)=\bpm 0 & 0 & Q_1\cm{Q_2} \\ 1 & 0 & 0 \\ 0 & 1 & 0 \epm.
\eg

When only nearest-neighbor hoppings are considered, a tight-binding Hamiltonian $H_{\rm kgm}(\bk)$,
\ba
H_{\rm kgm}(\bk) = \bpm 0 & 1+Q_1 \cm{Q_2} & Q_1+Q_1\cm{Q_2} \\ c.c. & 0 & 1+Q_1 \\ c.c. & c.c. & 0 \epm,
\ea
exhibits a FB with a band crossing point at $\Gamma=(0,0)$ in the BZ [\fig{FigS2}(c)].
The CLS and FT-CLS are given by $\ket{w_{\rm kgm}(\bR)} = \ket{\bR,1}-\ket{\bR+\bb a_1,1}+\ket{\bR+\bb a_2,2}-\ket{\bR,2}+\ket{\bR,3}-\ket{\bR-\bb a_1+\bb a_2,3}$ and $\uef{\rm kgm}=(1-Q_1,Q_2-1,1-\cm{Q_1}Q_2)$, respectively.
The CLS $\ket{w_{\rm kgm}(\bR)}$ is centered at $\bb A=(0,0)$ and has $C_6$ eigenvalue $-1$ [Note that $\ket{w_{\rm kgm}(\bR)}$ is equal to $\ket{w_{\rm kgm}(\bR;t)}$ with $t=1$, which is illustrated in \fig{FigS2}(b)].
As discussed in the main text, the band crossing point of FB can be explained by SR of CLS under $C_6$.
That is, the SR imposes a condition $U_{C_6}(\Gamma) \ket{\hat{u}_{\rm kgm}(\Gamma)} = -\ket{\hat{u}_{\rm kgm}(\Gamma)}$ on the FT-CLS, and the only solution to the condition is $\ket{\hat{u}_{\rm kgm}(\Gamma)}=(0,0,0)$.
Thus, the singularity of $\uef{\rm kgm}$ ensures a band crossing point at the singular point $\Gamma$.

When a SR under $C_n$ rotation ($n=2,3,4,6$) alone enforces a single band crossing point of FB, any gap-opening and $C_n$-preserving perturbation leads to a nearly flat Chern band.
The FB must have a superficial $C_n$ eigenvalue according to its SR at the band crossing point, in order for the nearly FB to be topologically trivial after the gap is opened.
On the other hand, SR-enforced band crossing occurs since dispersive bands, which can have a possible band crossing with the FB, do not have the superficial eigenvalue as their eigenvalues.
Hence, the nearly FB must be topological.
In fact, it has nonzero Chern number~\cite{fang2012bulk,alexandradinata2018no}.
This generic result holds in the model $H_{\rm kgm}(\bk)$.
It is known that spin-orbit couplings lift the degeneracy of the FB and dispersive band at $\Gamma$~\cite{tang2011high,beugeling2012topological}.
Hence, we turn on the nearest and next-nearest neighbor spin-orbit couplings with the strengths $\lambda_1=0.05$ and $\lambda_2=0.05$ respectively.
The relevant Hamiltonian $\delta H_{\rm kgm}(\bk)$,
\bg
\delta H_{\rm kgm}(\bk)
=i \lambda_1 \bpm 0 & -1-Q_1\cm{Q_2} & Q_1+Q_1\cm{Q_2} \\ -c.c. & 0 & -1-Q_1 \\ -c.c. & -c.c. & 0 \epm \nn \\
+i \lambda_2 \bpm 0 & -Q_1-\cm{Q_2} & 1+Q_1^2\cm{Q_2} \\ -c.c. & 0 & -Q_2-Q_1\cm{Q_2} \\ -c.c. & -c.c. & 0 \epm,
\eg
preserves $C_6$ but breaks other irrelevant symmetries such as time-reversal and mirror symmetries.
Then, all the bands are gapped and the FB becomes a nearly FB with Chern number $C=1$ [Fig.~\magenta{1}(c) in the main text].
We confirmed this by calculating the Wilson loop spectrum.
For the calculation, the tight-binding in non-periodic basis (Appendix~\ref{app:TB}) is used.

Now, we compare how different the SR-enforced band crossing is from the understanding the degeneracy at the band crossing point as an irreducible representation (IR) of point groups.
The two-fold degeneracy at the $\Gamma$ is a two-dimensional IR of $C_3+T$ or $C_3+M_{x,y}$.
Here, $T$ is time-reversal and $M_x$ ($M_y$) denotes a mirror symmetry, which inverts the $x$ ($y$) coordinate.
Since $[C_3,T]=0$ and $M_{x,y} C_3 M_{x,y}^{-1} = C_3^{-1}$ for spinless electron, two eigenstates with $C_3$ eigenvalue $e^{\pm 2\pi i/3}$ must form two-dimensional IR.
However, in order to prove the degeneracy is protected by $C_3+T$ or $C_3+M_{x,y}$, the band structure and $C_3$ eigenvalues at the band crossing point must be known a priori.
Also note that the role of $M_x$ and $M_y$ is no different in protecting the degeneracy.

In contrast, we understand all FB models through CLS, and regard the models with the same SR of CLS as one universality class regardless of details such as hopping structure.
Therefore, for the models in the same universality class, SR-enforced band crossing can be defined without knowing the specific band structure.
As already discussed, SR under $C_6$ alone enforces the band crossing of FB at $\Gamma$.
For the sake of rigor, let us identify the role of $C_3$, $M_{x,y}$ and $T$ symmetries.
Their representations on the basis atomic orbitals and $\bb Q=(Q_1,Q_2)$ are defined as
\begin{widetext}
\bg
U_{C_3}(\bk)=\bpm 0 & \cm{Q_2} & 0 \\ 0 & 0 & Q_1\cm{Q_2} \\ 1 & 0 & 0 \epm, \quad
U_{M_x}(\bk)=\bpm \cm{Q_1} & 0 & 0 \\ 0 & 0 & 1 \\ 0 & 1 & 0 \epm, \quad
U_{M_y}(\bk)=\bpm 1 & 0 & 0 \\ 0 & 0 & Q_1\cm{Q_2} \\ 0 & \cm{Q_2} & 0 \epm, \quad U_{T}(\bk) = \mc{K}, \\
\hat{C}_3: \bb Q \rightarrow (\cm{Q_2},Q_1\cm{Q_2}), \quad
\hat{M}_x: \bb Q \rightarrow (\cm{Q_1},\cm{Q_1}Q_2), \quad
\hat{M}_y: \bb Q \rightarrow (Q_1,Q_1\cm{Q_2}), \quad
\hat{T}: \bb Q \rightarrow (\cm{Q_1},\cm{Q_2}),
\eg
\end{widetext}
where $\mc{K}$ denotes the complex conjugation operator and $O_\sg$ for $\sg=C_3,M_{x,y},T$ is defined by $\sg=\{O_\sg|\bb \delta_\sg\}$ for each symmetry.

The CLS $\ket{w_{\rm kgm}(\bR)}$ has $C_3$, $M_x$ and $M_y$ eigenvalues as 1, $-1$ and 1 respectively.
Also, it is symmetric under $T$.
Accordingly, the FT-CLS $\uef{\rm kgm}$ transforms as $U_{C_3}(\bk) \uef{\rm kgm} = \ket{\hat{u}_{\rm kgm}(O_{C_3}\bk)}$, $U_{M_x}(\bk) \uef{\rm kgm} = -\ket{\hat{u}_{\rm kgm}(O_{M_x}\bk)}$, $U_{M_y}(\bk) \uef{\rm kgm} = \ket{\hat{u}_{\rm kgm}(O_{M_y}\bk)}$ and $U_T(\bk) \uef{\rm kgm} = \ket{\hat{u}_{\rm kgm}(-\bk)}$.
Let us see how these conditions constrain the FT-CLS at $\Gamma$.
$C_3$ imposes that $\ket{\hat{u}_{\rm kgm}(\Gamma)}=(c_1,c_1,c_1)$ for $c_1 \in \mathbb{C}$.
And, $\ket{\hat{u}_{\rm kgm}(\Gamma)}=(0,c_2,-c_2)$, $(c_3,c_4,c_4)$ and $(r_1,r_1,r_1)$ by $M_x$, $M_y$ and $T$, respectively, where $c_{2,3,4} \in \mathbb{C}$ and $r_1 \in \mathbb{R}$.

First, we note that the SR under $C_3+T$ and $C_3+M_y$ cannot enforce a singular point at $\Gamma$: $\ket{\hat{u}_{\rm kgm}(\Gamma)}=(r_1,r_1,r_1)$ by $C_3+T$, and $\ket{\hat{u}_{\rm kgm}(\Gamma)}=(a_1,a_1,a_1)$ by $C_3+M_y$.
Hence, a FB with the same SR under $C_3,M_y,T$ as $\ket{w_{\rm kgm}(\bR)}$ is not singular in general, and it does not necessarily have to have a band crossing point.
For instance, consider a CLS $\ket{w_{\rm kgm}(\bR;t)} = t\ket{\bR,1}-\ket{\bR+\bb a_1,1}+t\ket{\bR+\bb a_2,2}-\ket{\bR,2}+t\ket{\bR,3}-\ket{\bR-\bb a_1+\bb a_2,3}$, illustrated in \fig{FigS2}(b).
For $t\ne1$, $\ket{w_{\rm kgm}(\bR;t)}$ is symmetric under $C_3$ and $M_y$, and has the same eigenvalues under these symmetry as $\ket{w_{\rm kgm}(\bR)}$.
Also, it is time-reversal symmetric when $t \in \mathbb{R}$.
The corresponding FT-CLS is not singular at $\Gamma$: $\ket{\hat{u}_{\rm kgm}(\bk;t)}=(t-Q_1,tQ_2-1,t-\cm{Q_1}Q_2)$ and $\ket{\hat{u}_{\rm kgm}(\Gamma;t)}=(t-1)(1,1,1)$.
While, the SR under $C_3+M_x$ gives a singular FT-CLS, $\ket{\hat{u}_{\rm kgm}(\Gamma)}=(0,0,0)$.
We note that $M_x$ and $M_y$ play different roles for SR-enforced band crossing.
So far, we only consider the case where a CLS has $C_3$ eigenvalue $1$.
When a CLS has a $C_3$ other eigenvalues $e^{\pm 2\pi i/3}$, $T$ and $M_{x,y}$ are broken.
As it can be diagnosed by the determinant test, there is no SR-enforced band crossing by $C_3$.

Finally, we present a FB model in the kagome lattice, which has $C_6$ only:
\bg
H'_{\rm kgm}(\bk) = H_{\rm kgm}(\bk) + f(\bk) h(\bk), \\
h(\bk)=\bpm 0 & 1-Q_1\cm{Q_2} & Q_1-Q_1\cm{Q_2} \\ c.c. & 0 & Q_1-1 \\ c.c. & c.c. & 0 \epm, \\
f(\bk)=it_2(1-Q_1)(1-\cm{Q_2})(1-\cm{Q_1}Q_2).
\eg
The band structure with $t_2=0.06$ is shown in \fig{FigS2}(c).
All IR of $C_6$ are one-dimensional.
Nevertheless, if a FB exists in $C_6$-symmetric lattice and its CLS has $C_6$ eigenvalue $-1$, then it cannot be gapped.

\subsection{Lieb lattice}
\begin{figure*}[t!]
\centering
\includegraphics[width=0.95\textwidth]{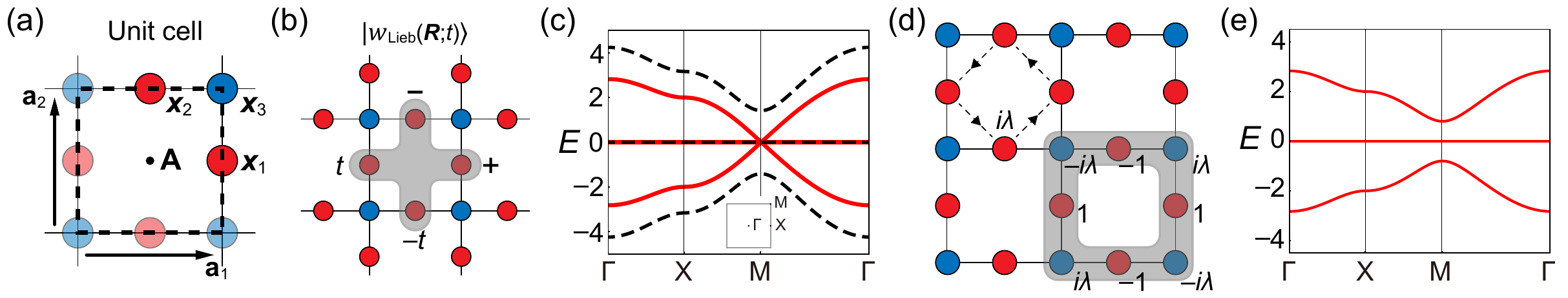}
\caption{
(a) Unit cell of the Lieb lattice.
(b) Description of CLS $\ket{w_{\rm Lieb}(\bR;t)}$.
(c) Band structure of $H_{\rm Lieb}(\bk)$ ($H'_{\rm Lieb}(\bk)$ with $t=2$) is denoted by red lines (black dashed lines).
(d) Description of the gapped FB model $H''_{\rm Lieb}(\bk)$.
Arrows denote the spin-orbit coupling, and the CLS $\ket{w''_{\rm Lieb}(\bR)}$ is drawn in gray region.
(e) Band structure of $H''_{\rm Lieb}(\bk)$ for $\lambda=0.2$.
}
\label{FigS3}
\end{figure*}

The Lieb lattice is illustrated in \fig{FigS3}(a).
We mainly consider three symmetries of the Lieb lattice: $C_4$, $M_x$ and $M_y$.
Their action on the basis atomic orbitals and $\bb Q$ are expressed as
\bg
U_{C_4}(\bk)=\bpm 0 & Q_2 & 0 \\ 1 & 0 & 0 \\ 0 & 0 & Q_2 \epm,
\hat{C}_4: \bb Q \rightarrow (\cm{Q_2},Q_1), \\
U_{M_x}(\bk)=\bpm Q_1 & 0 & 0 \\ 0 & 1 & 0 \\ 0 & 0 & Q_1 \epm,
\hat{M}_x: \bb Q \rightarrow (\cm{Q_1},Q_2), \\
U_{M_y}(\bk)=\bpm 1 & 0 & 0 \\ 0 & Q_2 & 0 \\ 0 & 0 & Q_2 \epm,
\hat{M}_y: \bb Q \rightarrow (Q_1,\cm{Q_2}).
\eg
%

A tight-binding Hamiltonian with the nearest-neighbor hoppings,
\ba
H_{\rm Lieb}(\bk) = \bpm 0 & 0 & 1+Q_2 \\ 0 & 0 & 1+Q_1 \\ c.c. & c.c. & 0 \epm,
\ea
exhibits a FB with a band crossing point at $M=(\pi,\pi)$ [\fig{FigS3}(c)].
The CLS and FT-CLS are given by $\ket{w_{\rm Lieb}(\bR)} = \ket{\bR,1}+\ket{\bR-\bb a_1,1}-\ket{\bR,2}-\ket{\bR-\bb a_2,2}+\ket{\bR,3}$ and $\uef{\rm Lieb}=(1+\cm{Q_1},-1-\cm{Q_2},0)$, respectively.
The CLS $\ket{w_{\rm kgm}(\bR)}$ is centered at $\bb A=(0,0)$.
It has $C_4$, $M_x$ and $M_y$ eigenvalues as $-1$, 1 and 1 respectively.
Importantly, it does not occupy the third sublattice site, \textit{i.e.} $\{\alpha_\varnothing\}=\{3\}$.
The shape of $\ket{w_{\rm Lieb}(\bR)}$ is shown in \fig{FigS2}(b)
[Note that $\ket{w_{\rm Lieb}(\bR)}$ is identical to $\ket{w_{\rm Lieb}(\bR;t=1)}$ in \fig{FigS2}(b)].
The FT-CLS $\uef{\rm Lieb}$ satisfies $U_{C_4}(M) \ket{\hat{u}_{\rm Lieb}(M)} = -\ket{\hat{u}_{\rm Lieb}(M)}$ and $U_{M_{x,y}}(M) \ket{\hat{u}_{\rm Lieb}(M)} = \ket{\hat{u}_{\rm Lieb}(M)}$.
Hence, $\ket{\hat{u}_{\rm Lieb}(M)}=(0,0,c_1)$, $(0,c_2,0)$ and $(c_3,0,0)$ by $C_4$, $M_x$ and $M_y$, respectively, where $c_{1,2,3} \in \mathbb{C}$.
Hence, as long as $\{\alpha_\varnothing\}=0$, \textit{i.e.} $\ket{\hat{u}_{\rm Lieb}(M)}_3=0$, the SR of CLS under $C_4$ alone enforces a band crossing at $M$.
On the other hand, if a CLS with the same SR under $C_4$ and $M_{x,y}$ occupies all sublattices, the singularity at $M$ is enforced by $C_4+M_x$, or $C_4+M_y$, or $M_x+M_y$.

Since the band crossing point at $M$ in the model $H_{\rm Lieb}(\bk)$ is enforced by $C_4$ alone, we expect that the FB becomes a nearly flat Chern band under a gap-opening and $C_4$-preserving perturbation.
Before adding the perturbation, we introduce on-site potential $\ep_3=2$ at the third sublattice for lifting the accidental three-fold degeneracy at $M$ into two-fold one.
Then, we add the next-nearest spin-orbit coupling $\delta H_{\rm Lieb}(\bk)$ with the strength $\lambda=0.1$,
\ba
\delta H_{\rm Lieb}(\bk)=&i \lambda \bpm 0 & g_{\rm soc}(\bk) & 0 \\ -\cm{g_{\rm soc}(\bk)} & 0 & 0 \\ 0 & 0 & 0 \epm,
\label{eq:Lieb_soc}
\ea
where $g_{\rm soc}(\bk)=-(1-\cm{Q_1})(1-Q_2)$.
The relevant band structures are shown in Fig.~\magenta{2}(d) in the main text.
As expected, we obtain a nearly FB with Chern number 1.

Now, we discuss gapped FB in the Lieb lattice.
Recall that $\ket{w_{\rm Lieb}(\bR)}$ must not occupy the third sublattice since otherwise the SR under $C_4$ does not necessarily lead to a SR-enforced band crossing.
Hence, we can obtain gapped FB if (i) $C_4$ symmetry is broken or (ii) the CLS occupies the third sublattice as well.
First, we consider the case (i) where the $C_4$ rotation is broken while the CLS does not occupy the third sublattice.
For this, we deform the CLS $\ket{w_{\rm Lieb}(\bR)}$ as $\ket{w_{\rm Lieb}(\bR;t)}=\ket{\bR,1}+t\ket{\bR-\bb a_1,1}-\ket{\bR,2}-t\ket{\bR-\bb a_2,2}$ with $t\in \mathbb{C}$ [\fig{FigS3}(b)].
Note that $C_4$ is broken for $t \ne 1$.
The corresponding FT-CLS, $\ket{\hat{u}_{\rm Lieb}(\bk;t)}=(1+t\cm{Q_1},-1-t\cm{Q_2},0)$, is nonsingular everywhere in the BZ.
This CLS is realized in a tight-binding model,
\ba
H'_{\rm Lieb}(\bk) = \bpm 0 & 0 & 1+tQ_2 \\ 0 & 0 & 1+tQ_1 \\ c.c. & c.c. & 0 \epm.
\ea
The band structure with $t=2$ is shown in \figs{FigS3}(c).
As expected, the FB is gapped.

Now, we consider the case (ii) where $C_4$ exists while the CLS occupies the third sublattice.
A relevant tight-binding Hamiltonian, also known as the spin-orbit coupled Lieb model~\cite{weeks2010topological}, is expressed as
\ba
H''_{\rm Lieb}(\bk) = H_{\rm Lieb}(\bk)+\delta H_{\rm Lieb}(\bk),
\ea
where $\delta H_{\rm Lieb}(\bk)$ is given as \eq{eq:Lieb_soc}.
The FT-CLS of this model is given by
\ba
\uefL{''}{\rm Lieb} = \bpm 1+\cm{Q_1} \\ -1-\cm{Q_2} \\ i \lambda (1-\cm{Q_1})(1-\cm{Q_2}) \epm,
\label{eq:socLieb_u}
\ea
and the corresponding CLS $\ket{w''_{\rm Lieb}(\bR)}$ is shown in \fig{FigS3}(d).
Since $\uefL{''}{\rm Lieb}$ is nonsingular everywhere in the BZ, hence we have a gapped FB [\fig{FigS3}(e)].

\subsection{Dice lattice}
\begin{figure}[t!]
\centering
\includegraphics[width=0.48\textwidth]{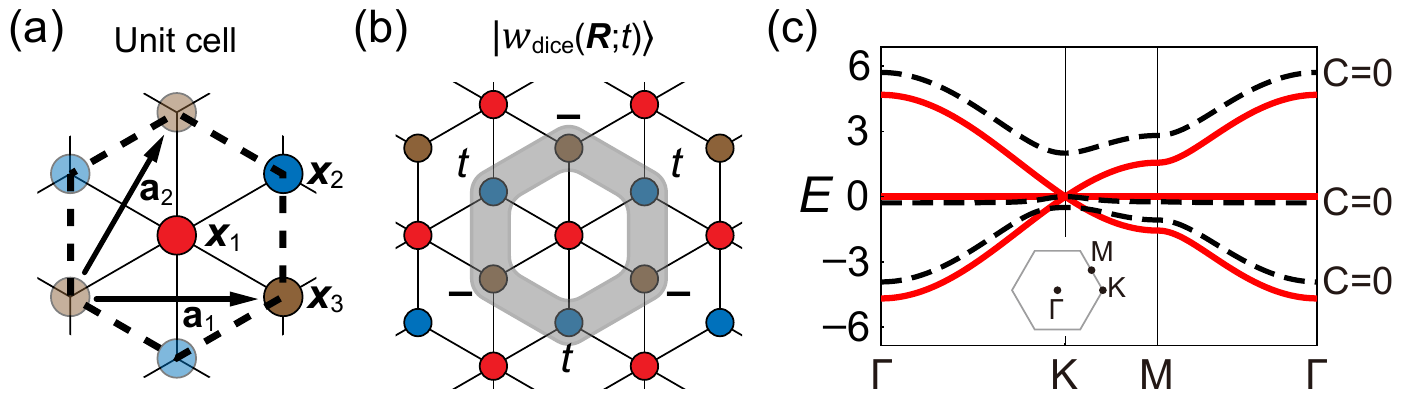}
\caption{
(a) Unit cell of the dice lattice.
(b) Description of CLS $\ket{w_{\rm dice}(\bR;t)}$.
(c) Band structure of $H_{\rm dice}(\bk;t)$ with $t=1.2$ exhibits a FB with two band crossing points $K$ and $-K$ (red lines).
Black dashed lines indicate the band structure when a $C_3$-preserving on-site potential $\delta H_{\rm dice}(\bk)={\rm Diag}(2.0,-0.5,0.0)$ is introduced.
$C$ near each band denotes the corresponding Chern number.
}
\label{FigS4}
\end{figure}

The dice lattice is shown in \fig{FigS4}(a) and has $p6mm$ wallpaper group.
As mentioned in the main text, the tight-binding Hamiltonian $H_{\rm dice}(\bk)$ with only nearest-neighbor hoppings exhibits a FB with two band crossing points at $\pm K$.
These band crossing points are enforced by SR of CLS $\ket{w_{\rm dice}(\bR)}$ under $C_3$.
For this reason, we introduce $C_2$-breaking parameter $t$ and consider a tight-binding Hamiltonian
\ba
H_{\rm dice}(\bk;t) = \bpm 0 & c.c. & c.c. \\ 1+\cm{Q_1}+\cm{Q_2} & 0 & 0 \\ t(1+\cm{Q_1}+\cm{Q_1}Q_2) & 0 & 0 \epm,
\ea
which has $C_3$ and $M_x$ symmetries.
($C_2$ and $M_y$ are symmetries of $H_{\rm dice}(\bk;t)$ only when $t=\pm1$.)
For any $t$, the band crossing points of FB remains the same as $H_{\rm dice}(\bk)$.
The relevant CLS $\ket{w_{\rm dice}(\bR;t)}$ is illustrated in \fig{FigS4}(b), and the FT-CLS is given by
\ba
\ket{\hat{u}_{\rm dice}(\bk;t)} = \bpm 0 \\ t(1+\cm{Q_1}+\cm{Q_2}) \\ -(1+\cm{Q_1}+\cm{Q_1}Q_2) \epm.
\ea
Note that $H_{\rm dice}(\bk)=H_{\rm dice}(\bk;t=1)$ and $\ket{w_{\rm dice}(\bR)}=\ket{w_{\rm dice}(\bR;t=1)}$.
%

Now, let us discuss a singular behavior of $\ket{\hat{u}_{\rm dice}(\bk;t)}$ at $K$.
$K$ is left invariant under $C_3$ and $M_y$.
These symmetries are represented as the direct sum $U_{C_3}(K)=1 \oplus \cm{\om} \oplus \om$ and $U_{M_y}(K)=1 \oplus \tau_1$ where $\om=e^{2\pi i/3}$ and $\tau_{0,1,2,3}$ denote the Pauli matrices.
Since $\ket{w_{\rm dice}(\bR;t)}$ has $C_3$ eigenvalue 1, the FT-CLS becomes $\ket{\hat{u}_{\rm dice}(K;t)}=(c_1,0,0)$ with $c_1 \in \mathbb{C}$.
Hence, if a CLS does not occupies the first sublattice as like $\ket{\hat{u}_{\rm dice}(\bk;t)}$, the SR under $C_3$ enforces a band crossing point of FB at $K$ alone.
Also, the FT-CLS must be singular at $-K$, since it is symmetric under $M_x$, which maps $K$ onto $-K$.
When a FB has multiple SR-enforced band crossing points as in this case, the FB does not have to be a nearly flat Chern band once a gap-opening and symmetry-preserving perturbation is added.
For this, we first add on-site potential $\ep_1=2$ at the first sublattice for lifting the accidental three-fold degeneracy at the SR-enforced band crossing points.
Then, the three-fold degeneracy are lifted to two-fold one while the flat dispersion of the middle band is intact.
The two-fold degeneracies at $K$ and $-K$ can be further lifted and the FB becomes a nearly FB by introducing on-site potential $\ep_2=-0.5$ at the second sublattice [\fig{FigS4}(c)].
However, the nearly FB is trivial and has zero Chern number.

Finally, we comment on the case when a CLS has $C_3$ eigenvalue 1 but it now occupies the first sublattice as well.
In this case, the corresponding FT-CLS is nonsingular.
For instance, consider a CLS $\ket{w'_{\rm dice}(\bR)}=\ket{w_{\rm dice}(\bR;t)}+\ket{\bR,1}$.
Its FT-CLS is equal to $\ket{\hat{u}'_{\rm dice}(\bk)}=\ket{\hat{u}_{\rm dice}(\bk;t)}+(1,0,0)$, which is nonzero everywhere in the BZ.
However, if a CLS is further symmetric under $M_y$ and has $M_y$ eigenvalue $-1$, then the FT-CLS must be singular at $\pm K$.

\subsection{Line graph of Lieb lattice}
\begin{figure}[t!]
\centering
\includegraphics[width=0.48\textwidth]{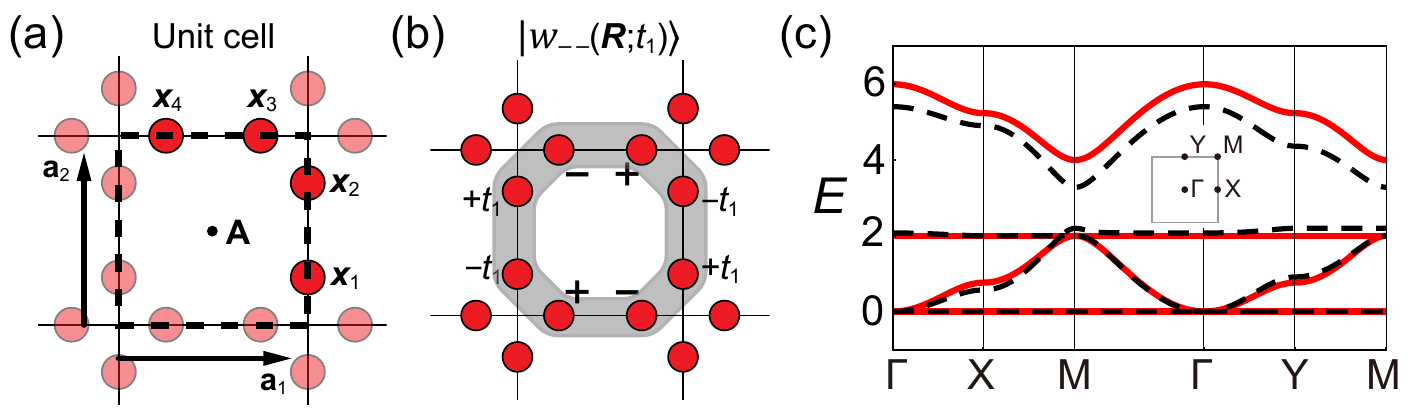}
\caption{
(a) Unit cell of a square lattice with two orthogonal mirrors.
The unit cell has four sublattices.
(b) Description of CLS $\ket{w_{--}(\bR;t_1)}$.
(c) Band structure of $H_{\rm M}(\bk)$ with $t_1=1.0$ and $t_2=1.0$ (with $t_1=0.8$ and $t_2=1.1$) is denoted by red lines (black dashed lines).
In both the band structures, the FB with zero energy has a band crossing point at $\Gamma$.
}
\label{FigS5}
\end{figure}

Consider a square lattice with two orthogonal mirrors $M_x$ and $M_y$ [\fig{FigS5}(a)].
Four sublattices in the unit cell are located at $\bb x_1=\frac{1}{2} \bb a_1-\frac{1}{4} \bb a_2$, $\bb x_2=O_{M_y} \bb x_1$, $\bb x_3=\frac{1}{4} \bb a_1+\frac{1}{2} \bb a_2$ and $\bb x_4=O_{M_x} \bb x_3$.
The mirrors are represented as $U_{M_x}(\bk) = (Q_1 \tau_0) \oplus \tau_1$, $U_{M_y}(\bk) = \tau_1 \oplus (Q_2 \tau_0)$, $M_x:\bb Q \rightarrow (\cm{Q_1},Q_2)$ and $M_y:\bb Q \rightarrow (Q_1,\cm{Q_2})$.
Now, we consider a tight-binding Hamiltonian,
\bg
H_{\rm M}(\bk)=\bpm 1+t_2 & t_2+Q_2 & t_1 Q_2 & t_1 \cm{Q_1} Q_2 \\ c.c. & 1+t_2 & t_1 & t_1 \cm{Q_1} \\ c.c. & c.c. & 1+|t_1|^2 & 1+|t_1|^2 \cm{Q_1} \\ c.c. & c.c. & c.c. & 1+|t_1|^2 \epm,
\eg
with two parameters $t_1 \in \mathbb{C}$ and $t_2 \in \mathbb{R}$.
Note that $H_{\rm M}(\bk)$ with $t_1=t_2=1$ corresponds to the FB model $H_{\rm LG}(\bk)$ in the line graph of the Lieb lattice, which is discussed in the main text.
%

For generic $t_1$ and $t_2$, $H_{\rm M}(\bk)$ has a FB with a band crossing point at $\Gamma$, as shown in \fig{FigS5}(c).
The relevant CLS is shown in \fig{FigS5}(b).
It is centered at the $\bb A=(0,0)$, and has eigenvalue 1 for both $M_x$ and $M_y$.
Accordingly, the FT-CLS is transforms as $U_{M_{x,y}}(\bk) \ket{\hat{u}_{\rm --}(\bk;t_1)} = -\ket{\hat{u}_{\rm --}(O_{M_{x,y}}\bk;t_1)}$.
Note that the explicit form of FT-CLS is $\ket{\hat{u}_{\rm --}(\bk;t_1)}=(t_1-t_1 \cm{Q_1},t_1 \cm{Q_1}-t_1,1-\cm{Q_2},\cm{Q_2}-1)$.
The singularity of $\ket{\hat{u}_{\rm --}(\bk;t_1)}$ at $\Gamma$ originates from the SR under $M_{x,y}$.
That is, $U_{M_{x,y}}(\Gamma) \ket{\hat{u}_{\rm --}(\Gamma;t_1)} = -\ket{\hat{u}_{\rm --}(\Gamma;t_1)}$ can be satisfied only if $\ket{\hat{u}_{\rm --}(\Gamma;t_1)}=(0,0,0,0)$.
Furthermore, it can be straightforwardly shown that when a CLS is centered at $\bb A$ and has $M_x$ and $M_y$ eigenvalues as $(1,-1)$, $(-1,1)$ and $(1,1)$, the corresponding FB must have a band crossing point at $X$, $Y$ and $M$, respectively.

\subsection{Split and line graphs of square and hexagonal lattices}
It has been known that the split and line graph lattices exhibit nondegenerate FB with a band crossing point~\cite{ma2020spin}.
The Lieb lattice corresponds to the split graph ($S$) of square lattice ($\Square$), hence we denote it as $S(\Square)$.
On the other hand, the kagome lattice $L(\varhexagon)$ is the line graph ($L$) of hexagonal lattice ($\varhexagon$).
For the details on the line and split graphs, see Refs.~\onlinecite{kollar2019line,ma2020spin,chiu2020fragile}.

We confirmed that the band crossing points of FBs in the split and line graphs of square and hexagonal lattices can be understood by SR under $M_x+M_y$ and $C_3+M_x$ respectively.
Specifically, in $S(\Square)$, $L(\Square)$, $L(S(\Square))$ and $S(S(\Square))$, a band crossing point of FB is enforced at $\Gamma$ and $M$ if the CLS has $M_x$ and $M_y$ eigenvalues as $(-1,-1)$ and $(1,1)$ respectively.
And, a FB in $S(\varhexagon)$, $L(\varhexagon)$, $L(S(\varhexagon))$ and $S(S(\varhexagon))$ must have a band crossing point at $\Gamma$ when the CLS has $C_3$ and $M_x$ eigenvalues as $1$ and $-1$.
Note that we use the same primitive lattice vectors and coordinate systems as the Lieb and kagome lattices for the square and hexagonal lattices, respectively.

\begin{figure*}[t!]
\centering
\includegraphics[width=0.85\textwidth]{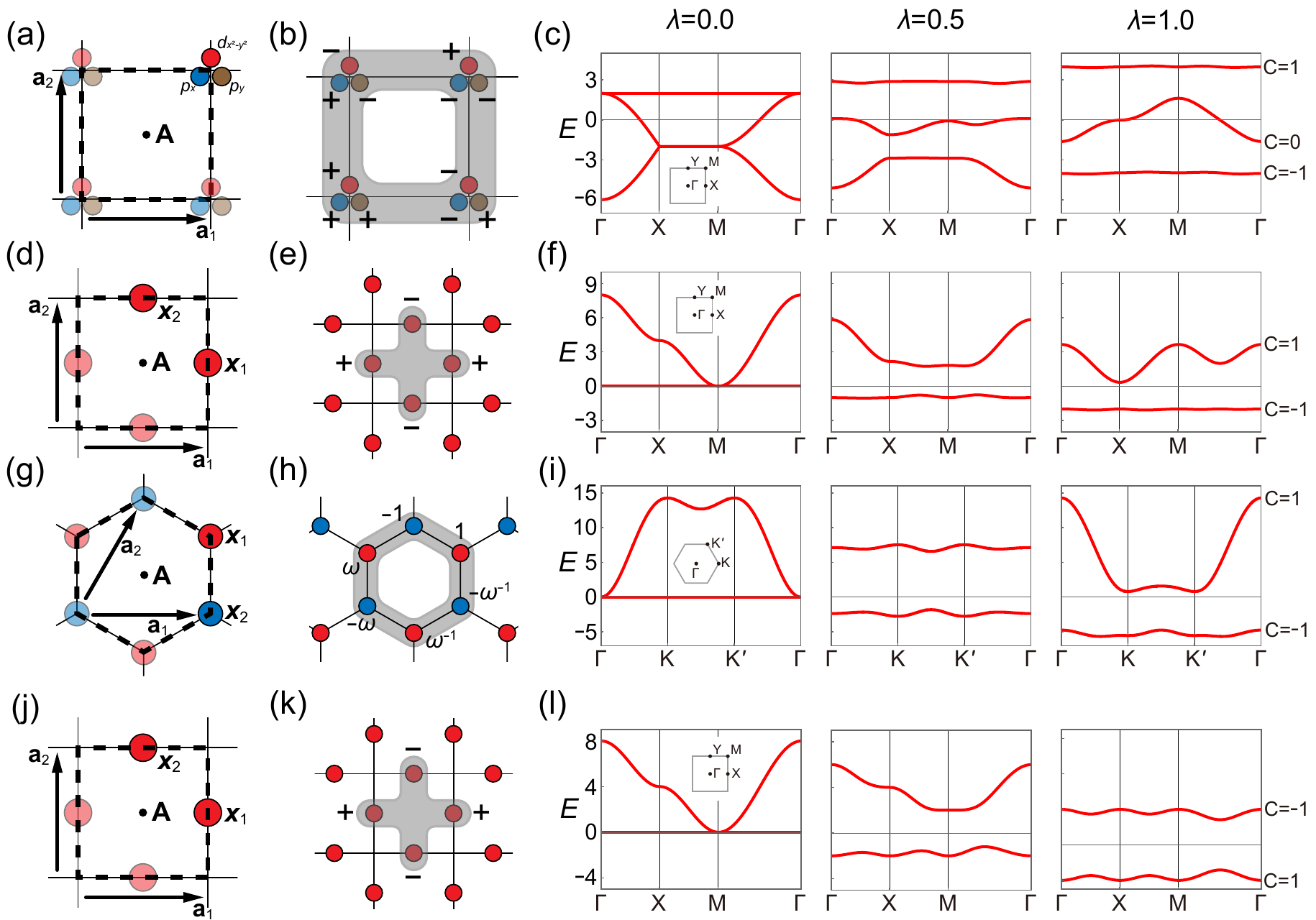}
\caption{
The nearly FB models.
(a)-(c) Description of $H_{\rm I}(\bk)$ and $\mc{H}_{\rm I}(\bk)$.
(a) The unit cell, (b) the CLS of $H_{\rm I}(\bk)$ and (c) the band structures of $H_{\rm I}(\bk;\lambda)$ with $\lambda=0.0,0.5,1.0$.
$C$ near each band denotes its Chern number.
(d)-(f) Description of $H_{\rm II}(\bk)$ and $\mc{H}_{\rm II}(\bk)$.
(g)-(i) Description of $H_{\rm III}(\bk)$ and $\mc{H}_{\rm III}(\bk)$.
(j)-(l) Description of $H_{\rm IV}(\bk)$ and $\mc{H}_{\rm IV}(\bk)$.
}
\label{FigS6}
\end{figure*}

\section{Early models of nearly flat Chern bands \label{app:NFBs}}
In this section, we demonstrate that the early models of the nearly flat Chern bands in Refs.~\onlinecite{sun2011nearly,neupert2011fractional} can be obtained from FB models with SR-enforced band crossings.
These nearly FB models will be denoted as $\mc{H}_a(\bk)$ ($a=\rm I,II,III,IV$).
Now, we show that for each $\mc{H}_a(\bk)$ one can find a \textit{generating} FB model $H_a(\bk)$ in which a FB has a SR-enforced band crossing because of rotation symmetry.
To this end, we introduce a Hamiltonian
\ba
H_a(\bk;\lambda)=(1-\lambda)H_a(\bk)+\lambda \mc{H}_a(\bk),
\ea
which connects $H_a(\bk)$ at $\lambda=0$ and $\mc{H}_a(\bk)$ at $\lambda=1$.
For $\lambda>0$, the FB with a band crossing point becomes the nearly flat Chern band.
As a gap between the nearly FB and other dispersive bands does not close until $\lambda=1$, the nearly FB in $\mc{H}_a(\bk)$ has nonzero Chern number.
These results are summarized in \fig{FigS6}.

The first model in \Rf{sun2011nearly} is defined in a square lattice with three basis orbitals $d_{x^2-y^2}$, $p_x$ and $p_y$ [\figs{FigS6}(a)-(c)].
All three orbitals are located at $\frac{1}{2}\bb a_1+\frac{1}{2}\bb a_2$.
The nearly FB model $\mc{H}_I(\bk)$ and the generating FB model $H_I(\bk)$ are described by $H(\bk;\Delta)$ with $\Delta=2.8$ and $0.0$ respectively.
Here, $H(\bk;\Delta)$ is defined as
\bg
H(\bk;\Delta)=h_0(\bk)+{\rm Diag}(f_1(\bk),f_2(\bk),f_3(\bk)), \\
h_0(\bk)=\bpm 0 & \cm{Q_2}-Q_2 & \cm{Q_1}-Q_1 \\ c.c. & 0 & -i \Delta \\ c.c. & c.c. & 0 \epm,
\eg
with $f_1(\bk)=-2-Q_1-\cm{Q_1}-Q_2-\cm{Q_2}+\Delta- \frac{2\Delta}{\Delta+4}$, $f_2(\bk)=-\frac{\Delta}{\Delta+4}(Q_1+\cm{Q_1})$, $f_3(\bk)=-\frac{\Delta}{\Delta+4}(Q_2+\cm{Q_2})$.
In the generating FB model $H_I(\bk)$, the CLS is centered at the unit cell center, and has $C_4$ eigenvalue 1 [\fig{FigS6}(b)].
Note that symmetry operator for $C_4$ is given by $U_{C_4}(\bk)=(-Q_2) \oplus (-iQ_2\tau_2)$.
The FT-CLS is given by \bg
\uef{\rm I}= \bpm (1-\cm{Q_1})(1-\cm{Q_2}) \\ (\cm{Q_1}-1)(1+\cm{Q_2}) \\ (1+\cm{Q_1})(\cm{Q_2}-1) \epm,
\eg
The singularity of $\uef{\rm I}$ at $\Gamma$ is enforced by SR according to $[U_{C_4}(\Gamma)-\mathds{1}_3] \ket{\hat{u}_{\rm I}(\Gamma)}=0$ and ${\rm Det} [U_{C_4}(\Gamma)-\mathds{1}_3]=-4$.
The band structure of $H_{\rm I}(\bk;\lambda)$ is shown in \fig{FigS6}(c).

The second model in \Rf{sun2011nearly} is defined in the checkerboard lattice [\figs{FigS6}(d)-(f)].
In the unit cell, two basis orbitals are located at $\bb x_1=\frac{1}{2}\bb a_1$ and $\bb x_2=\frac{1}{2}\bb a_2$, respectively.
The nearly FB model is expressed as $\mc{H}_{\rm II}(\bk)=\sum_{i=0}^3 d_{{\rm II},i}(\bk) \tau_i$ with
$d_{{\rm II},0}(\bk)=t'(Q_1Q_2+Q_1\cm{Q_2}+\cm{Q_1}Q_2+\cm{Q_1}\cm{Q_2})$, $d_{{\rm II},3}(\bk)=t_1(-Q_1-\cm{Q_1}+Q_2+\cm{Q_2})$ and $d_{{\rm II},1}(\bk)+id_{{\rm II},2}(\bk)=e^{-\pi i/4}(1+iQ_1+i\cm{Q_2}+Q_1\cm{Q_2})$ with $t_1=1/(2+\sqrt{2})$ and $t'=1/(2+2\sqrt{2})$.
We find the generating FB model,
\bg
H_{\rm II}(\bk) = \bpm 2+Q_2+\cm{Q_2} & (1+\cm{Q_1})(1+Q_2) \\ c.c. & 2+Q_1+\cm{Q_1} \epm,
\eg
whose the FT-CLS is given by $\uef{\rm II}= (1+\cm{Q_1},-1-\cm{Q_2})$.
It is singular at $M$ since the CLS is centered at the unit cell center and has $C_4$ eigenvalue $-1$, \textit{i.e.} $U_{C_4}(\bk) \ket{\hat{u}_{\rm II}(M)} = -\ket{\hat{u}_{\rm II}(M)}$ and ${\rm Det} [U_{C_4}(M)+\mathds{1}_2]=2$, as shown in \fig{FigS6}(e).
Note that $U_{C_4}(\bk)=\bpm 0 & Q_2 \\ 1 & 0 \epm$.

Now, let us discuss the two nearly FB models in \Rf{neupert2011fractional}.
The first model is defined in the honeycomb lattice:
\bg
\mc{H}_{\rm III}(\bk)=\bpm h_1(\bk,\Phi) & t_1(1+\cm{Q_2}+Q_1\cm{Q_2}) \\ c.c. & h_1(\bk,-\Phi) \epm, \\
h_1(\bk,\Phi)=t_2 e^{i\Phi}(Q_1+\cm{Q_2}+\cm{Q_1}Q_2) + c.c.
\eg
with $t_1=12\sqrt{3/43}$, $t_2=1$ and $\Phi=\cos^{-1}(3\sqrt{3/43})$ [\figs{FigS6}(g)-(i)].
The generating FB model is described by
\bg
H_{\rm III}(\bk)=\frac{t_1}{2} \bpm v_2(\bk) \cm{v_2(\bk)} & -v_1(\bk) \cm{v_2(\bk)} \\ -\cm{v_1(\bk)} v_2(\bk) & v_1(\bk) \cm{v_1(\bk)} \epm,
\eg
with $v_1(\bk)=1+\om \cm{Q_1}+\cm{\om Q_2}$ and $v_2(\bk)=-\cm{\om}-\om \cm{Q_1}-\cm{Q_1}Q_2$.
The CLS is centered at the unit cell center and has $C_3$ eigenvalue $\cm{\om}$ [\fig{FigS6}(h)].
Note that symmetry operator of $C_3$ is defined as $U_{C_3}(\bk)={\rm Diag}(Q_2,Q_1)$.
Hence, $U_{C_3}(\bk) \ket{\hat{u}_{\rm III}(\Gamma)} = \cm{\om} \ket{\hat{u}_{\rm III}(\Gamma)}$ and ${\rm Det} [U_{C_3}(\Gamma)-\cm{\om} \mathds{1}_2]=-3\cm{\om}$.
This implies that the FT-CLS, $\uef{\rm III}= (v_1(\bk),v_2(\bk))$, is singular at $\Gamma$.

Finally, let us discuss the second model $\mc{H}_{\rm IV}(\bk)$ in \Rf{neupert2011fractional} [\figs{FigS6}(j)-(l)].
This model is defined in the same lattice as $\mc{H}_{\rm II}(\bk)$.
The relevant Hamiltonian is given by $\mc{H}_{\rm IV}(\bk)=\sum_{i=1}^3 d_{{\rm IV},i}(\bk) \tau_i$ with $d_{{\rm IV},1}(\bk)+id_{{\rm IV},2}(\bk)=e^{-\pi i/4}(1+iQ_1+i\cm{Q_2}+Q_1\cm{Q_2})$ and $d_{{\rm IV},3}(\bk)=\sqrt{2} e^{-\pi i/4}(1+i\cm{Q_1}+iQ_2+\cm{Q_1}Q_2)$.
Since the generating FB model for $\mc{H}_{\rm IV}(\bk)$ is the identical to $H_{\rm II}(\bk)$, the nearly flat Chern band can be explained in the same way as $\mc{H}_{\rm II}(\bk)$.


\begin{thebibliography}{92}%
\makeatletter
\providecommand \@ifxundefined [1]{%
\@ifx{#1\undefined}
}%
\providecommand \@ifnum [1]{%
\ifnum #1\expandafter \@firstoftwo
\else \expandafter \@secondoftwo
\fi
}%
\providecommand \@ifx [1]{%
\ifx #1\expandafter \@firstoftwo
\else \expandafter \@secondoftwo
\fi
}%
\providecommand \natexlab [1]{#1}%
\providecommand \enquote  [1]{``#1''}%
\providecommand \bibnamefont  [1]{#1}%
\providecommand \bibfnamefont [1]{#1}%
\providecommand \citenamefont [1]{#1}%
\providecommand \href@noop [0]{\@secondoftwo}%
\providecommand \href [0]{\begingroup \@sanitize@url \@href}%
\providecommand \@href[1]{\@@startlink{#1}\@@href}%
\providecommand \@@href[1]{\endgroup#1\@@endlink}%
\providecommand \@sanitize@url [0]{\catcode `\\12\catcode `\$12\catcode
`\&12\catcode `\#12\catcode `\^12\catcode `\_12\catcode `\%12\relax}%
\providecommand \@@startlink[1]{}%
\providecommand \@@endlink[0]{}%
\providecommand \url  [0]{\begingroup\@sanitize@url \@url }%
\providecommand \@url [1]{\endgroup\@href {#1}{\urlprefix }}%
\providecommand \urlprefix  [0]{URL }%
\providecommand \Eprint [0]{\href }%
\providecommand \doibase [0]{http://dx.doi.org/}%
\providecommand \selectlanguage [0]{\@gobble}%
\providecommand \bibinfo  [0]{\@secondoftwo}%
\providecommand \bibfield  [0]{\@secondoftwo}%
\providecommand \translation [1]{[#1]}%
\providecommand \BibitemOpen [0]{}%
\providecommand \bibitemStop [0]{}%
\providecommand \bibitemNoStop [0]{.\EOS\space}%
\providecommand \EOS [0]{\spacefactor3000\relax}%
\providecommand \BibitemShut  [1]{\csname bibitem#1\endcsname}%
\let\auto@bib@innerbib\@empty
\bibitem [{\citenamefont {Lieb}(1989)}]{lieb1989two}%
\BibitemOpen
\bibfield  {author} {\bibinfo {author} {\bibfnamefont {Elliott~H.}\
\bibnamefont {Lieb}},\ }\bibfield  {title} {\enquote {\bibinfo {title} {Two
theorems on the {Hubbard} model},}\ }\href {\doibase
10.1103/PhysRevLett.62.1201} {\bibfield  {journal} {\bibinfo  {journal}
{Physical Review Letters}\ }\textbf {\bibinfo {volume} {62}},\ \bibinfo
{pages} {1201} (\bibinfo {year} {1989})}\BibitemShut {NoStop}%
\bibitem [{\citenamefont {Aoki}\ \emph {et~al.}(1996)\citenamefont {Aoki},
\citenamefont {Ando},\ and\ \citenamefont {Matsumura}}]{aoki1996hofstadter}%
\BibitemOpen
\bibfield  {author} {\bibinfo {author} {\bibfnamefont {Hideo}\ \bibnamefont
{Aoki}}, \bibinfo {author} {\bibfnamefont {Masato}\ \bibnamefont {Ando}}, \
and\ \bibinfo {author} {\bibfnamefont {Hajime}\ \bibnamefont {Matsumura}},\
}\bibfield  {title} {\enquote {\bibinfo {title} {Hofstadter butterflies for
flat bands},}\ }\href {\doibase 10.1103/PhysRevB.54.R17296} {\bibfield
{journal} {\bibinfo  {journal} {Physical Review B}\ }\textbf {\bibinfo
{volume} {54}},\ \bibinfo {pages} {R17296} (\bibinfo {year}
{1996})}\BibitemShut {NoStop}%
\bibitem [{\citenamefont {Huber}\ and\ \citenamefont
{Altman}(2010)}]{huber2010bose}%
\BibitemOpen
\bibfield  {author} {\bibinfo {author} {\bibfnamefont {Sebastian~D.}\
\bibnamefont {Huber}}\ and\ \bibinfo {author} {\bibfnamefont {Ehud}\
\bibnamefont {Altman}},\ }\bibfield  {title} {\enquote {\bibinfo {title}
{Bose condensation in flat bands},}\ }\href {\doibase
10.1103/PhysRevB.82.184502} {\bibfield  {journal} {\bibinfo  {journal}
{Physical Review B}\ }\textbf {\bibinfo {volume} {82}},\ \bibinfo {pages}
{184502} (\bibinfo {year} {2010})}\BibitemShut {NoStop}%
\bibitem [{\citenamefont {Weeks}\ and\ \citenamefont
{Franz}(2012)}]{weeks2012flat}%
\BibitemOpen
\bibfield  {author} {\bibinfo {author} {\bibfnamefont {C.}~\bibnamefont
{Weeks}}\ and\ \bibinfo {author} {\bibfnamefont {M.}~\bibnamefont {Franz}},\
}\bibfield  {title} {\enquote {\bibinfo {title} {Flat bands with nontrivial
topology in three dimensions},}\ }\href {\doibase 10.1103/PhysRevB.85.041104}
{\bibfield  {journal} {\bibinfo  {journal} {Physical Review B}\ }\textbf
{\bibinfo {volume} {85}},\ \bibinfo {pages} {041104(R)} (\bibinfo {year}
{2012})}\BibitemShut {NoStop}%
\bibitem [{\citenamefont {Peotta}\ and\ \citenamefont
{T{\"o}rm{\"a}}(2015)}]{peotta2015superfluidity}%
\BibitemOpen
\bibfield  {author} {\bibinfo {author} {\bibfnamefont {Sebastiano}\
\bibnamefont {Peotta}}\ and\ \bibinfo {author} {\bibfnamefont {P{\"a}ivi}\
\bibnamefont {T{\"o}rm{\"a}}},\ }\bibfield  {title} {\enquote {\bibinfo
{title} {Superfluidity in topologically nontrivial flat bands},}\ }\href
{\doibase 10.1038/ncomms9944} {\bibfield  {journal} {\bibinfo  {journal}
{Nature Communications}\ }\textbf {\bibinfo {volume} {6}},\ \bibinfo {pages}
{8944} (\bibinfo {year} {2015})}\BibitemShut {NoStop}%
\bibitem [{\citenamefont {Julku}\ \emph {et~al.}(2016)\citenamefont {Julku},
\citenamefont {Peotta}, \citenamefont {Vanhala}, \citenamefont {Kim},\ and\
\citenamefont {T\"orm\"a}}]{julku2016geometric}%
\BibitemOpen
\bibfield  {author} {\bibinfo {author} {\bibfnamefont {Aleksi}\ \bibnamefont
{Julku}}, \bibinfo {author} {\bibfnamefont {Sebastiano}\ \bibnamefont
{Peotta}}, \bibinfo {author} {\bibfnamefont {Tuomas~I.}\ \bibnamefont
{Vanhala}}, \bibinfo {author} {\bibfnamefont {Dong-Hee}\ \bibnamefont {Kim}},
\ and\ \bibinfo {author} {\bibfnamefont {P\"aivi}\ \bibnamefont
{T\"orm\"a}},\ }\bibfield  {title} {\enquote {\bibinfo {title} {Geometric
origin of superfluidity in the {Lieb}-lattice flat band},}\ }\href {\doibase
10.1103/PhysRevLett.117.045303} {\bibfield  {journal} {\bibinfo  {journal}
{Physical Review Letters}\ }\textbf {\bibinfo {volume} {117}},\ \bibinfo
{pages} {045303} (\bibinfo {year} {2016})}\BibitemShut {NoStop}%
\bibitem [{\citenamefont {Ramachandran}\ \emph {et~al.}(2017)\citenamefont
{Ramachandran}, \citenamefont {Andreanov},\ and\ \citenamefont
{Flach}}]{ramachandran2017chiral}%
\BibitemOpen
\bibfield  {author} {\bibinfo {author} {\bibfnamefont {Ajith}\ \bibnamefont
{Ramachandran}}, \bibinfo {author} {\bibfnamefont {Alexei}\ \bibnamefont
{Andreanov}}, \ and\ \bibinfo {author} {\bibfnamefont {Sergej}\ \bibnamefont
{Flach}},\ }\bibfield  {title} {\enquote {\bibinfo {title} {Chiral flat
bands: {Existence}, engineering, and stability},}\ }\href {\doibase
10.1103/PhysRevB.96.161104} {\bibfield  {journal} {\bibinfo  {journal}
{Physical Review B}\ }\textbf {\bibinfo {volume} {96}},\ \bibinfo {pages}
{161104(R)} (\bibinfo {year} {2017})}\BibitemShut {NoStop}%
\bibitem [{\citenamefont {Misumi}\ and\ \citenamefont
{Aoki}(2017)}]{misumi2017new}%
\BibitemOpen
\bibfield  {author} {\bibinfo {author} {\bibfnamefont {Tatsuhiro}\
\bibnamefont {Misumi}}\ and\ \bibinfo {author} {\bibfnamefont {Hideo}\
\bibnamefont {Aoki}},\ }\bibfield  {title} {\enquote {\bibinfo {title} {New
class of flat-band models on tetragonal and hexagonal lattices: {Gapped}
versus crossing flat bands},}\ }\href {\doibase 10.1103/PhysRevB.96.155137}
{\bibfield  {journal} {\bibinfo  {journal} {Physical Review B}\ }\textbf
{\bibinfo {volume} {96}},\ \bibinfo {pages} {155137} (\bibinfo {year}
{2017})}\BibitemShut {NoStop}%
\bibitem [{\citenamefont {Pal}\ and\ \citenamefont {Saha}(2018)}]{pal2018flat}%
\BibitemOpen
\bibfield  {author} {\bibinfo {author} {\bibfnamefont {Biplab}\ \bibnamefont
{Pal}}\ and\ \bibinfo {author} {\bibfnamefont {Kush}\ \bibnamefont {Saha}},\
}\bibfield  {title} {\enquote {\bibinfo {title} {Flat bands in fractal-like
geometry},}\ }\href {\doibase 10.1103/PhysRevB.97.195101} {\bibfield
{journal} {\bibinfo  {journal} {Physical Review B}\ }\textbf {\bibinfo
{volume} {97}},\ \bibinfo {pages} {195101} (\bibinfo {year}
{2018})}\BibitemShut {NoStop}%
\bibitem [{\citenamefont {Bilitewski}\ and\ \citenamefont
{Moessner}(2018)}]{bilitewski2018disordered}%
\BibitemOpen
\bibfield  {author} {\bibinfo {author} {\bibfnamefont {Thomas}\ \bibnamefont
{Bilitewski}}\ and\ \bibinfo {author} {\bibfnamefont {Roderich}\ \bibnamefont
{Moessner}},\ }\bibfield  {title} {\enquote {\bibinfo {title} {Disordered
flat bands on the kagome lattice},}\ }\href {\doibase
10.1103/PhysRevB.98.235109} {\bibfield  {journal} {\bibinfo  {journal}
{Physical Review B}\ }\textbf {\bibinfo {volume} {98}},\ \bibinfo {pages}
{235109} (\bibinfo {year} {2018})}\BibitemShut {NoStop}%
\bibitem [{\citenamefont {Mizoguchi}\ and\ \citenamefont
{Udagawa}(2019)}]{mizoguchi2019flat}%
\BibitemOpen
\bibfield  {author} {\bibinfo {author} {\bibfnamefont {Tomonari}\
\bibnamefont {Mizoguchi}}\ and\ \bibinfo {author} {\bibfnamefont {Masafumi}\
\bibnamefont {Udagawa}},\ }\bibfield  {title} {\enquote {\bibinfo {title}
{Flat-band engineering in tight-binding models: {Beyond} the nearest-neighbor
hopping},}\ }\href {\doibase 10.1103/PhysRevB.99.235118} {\bibfield
{journal} {\bibinfo  {journal} {Physical Review B}\ }\textbf {\bibinfo
{volume} {99}},\ \bibinfo {pages} {235118} (\bibinfo {year}
{2019})}\BibitemShut {NoStop}%
\bibitem [{\citenamefont {Mizoguchi}\ and\ \citenamefont
{Hatsugai}(2019)}]{mizoguchi2019molecular}%
\BibitemOpen
\bibfield  {author} {\bibinfo {author} {\bibfnamefont {Tomonari}\
\bibnamefont {Mizoguchi}}\ and\ \bibinfo {author} {\bibfnamefont {Yasuhiro}\
\bibnamefont {Hatsugai}},\ }\bibfield  {title} {\enquote {\bibinfo {title}
{Molecular-orbital representation of generic flat-band models},}\ }\href
{\doibase 10.1209/0295-5075/127/47001} {\bibfield  {journal} {\bibinfo
{journal} {EPL (Europhysics Letters)}\ }\textbf {\bibinfo {volume} {127}},\
\bibinfo {pages} {47001} (\bibinfo {year} {2019})}\BibitemShut {NoStop}%
\bibitem [{\citenamefont {Mizoguchi}\ and\ \citenamefont
{Hatsugai}(2020{\natexlab{a}})}]{mizoguchi2020systematic}%
\BibitemOpen
\bibfield  {author} {\bibinfo {author} {\bibfnamefont {Tomonari}\
\bibnamefont {Mizoguchi}}\ and\ \bibinfo {author} {\bibfnamefont {Yasuhiro}\
\bibnamefont {Hatsugai}},\ }\bibfield  {title} {\enquote {\bibinfo {title}
{Systematic construction of topological flat-band models by molecular-orbital
representation},}\ }\href {\doibase 10.1103/PhysRevB.101.235125} {\bibfield
{journal} {\bibinfo  {journal} {Physical Review B}\ }\textbf {\bibinfo
{volume} {101}},\ \bibinfo {pages} {235125} (\bibinfo {year}
{2020}{\natexlab{a}})}\BibitemShut {NoStop}%
\bibitem [{\citenamefont {Mizoguchi}\ and\ \citenamefont
{Hatsugai}(2020{\natexlab{b}})}]{mizoguchi2020type}%
\BibitemOpen
\bibfield  {author} {\bibinfo {author} {\bibfnamefont {Tomonari}\
\bibnamefont {Mizoguchi}}\ and\ \bibinfo {author} {\bibfnamefont {Yasuhiro}\
\bibnamefont {Hatsugai}},\ }\bibfield  {title} {\enquote {\bibinfo {title}
{Type-{III} {Dirac} cones from degenerate directionally flat bands: Viewpoint
from molecular-orbital representation},}\ }\href {\doibase
10.7566/JPSJ.89.103704} {\bibfield  {journal} {\bibinfo  {journal} {Journal
of the Physical Society of Japan}\ }\textbf {\bibinfo {volume} {89}},\
\bibinfo {pages} {103704} (\bibinfo {year} {2020}{\natexlab{b}})}\BibitemShut
{NoStop}%
\bibitem [{\citenamefont {Chiu}\ \emph {et~al.}(2020)\citenamefont {Chiu},
\citenamefont {Ma}, \citenamefont {Song}, \citenamefont {Bernevig},\ and\
\citenamefont {Houck}}]{chiu2020fragile}%
\BibitemOpen
\bibfield  {author} {\bibinfo {author} {\bibfnamefont {C.S.}\ \bibnamefont
{Chiu}}, \bibinfo {author} {\bibfnamefont {Da-Shuai}\ \bibnamefont {Ma}},
\bibinfo {author} {\bibfnamefont {Zhi-Da}\ \bibnamefont {Song}}, \bibinfo
{author} {\bibfnamefont {B.A.}\ \bibnamefont {Bernevig}}, \ and\ \bibinfo
{author} {\bibfnamefont {A.A.}\ \bibnamefont {Houck}},\ }\bibfield  {title}
{\enquote {\bibinfo {title} {Fragile topology in line-graph lattices with
two, three, or four gapped flat bands},}\ }\href {\doibase
10.1103/PhysRevResearch.2.043414} {\bibfield  {journal} {\bibinfo  {journal}
{Physical Review Research}\ }\textbf {\bibinfo {volume} {2}},\ \bibinfo
{pages} {043414} (\bibinfo {year} {2020})}\BibitemShut {NoStop}%
\bibitem [{\citenamefont {Hwang}\ \emph {et~al.}(2020)\citenamefont {Hwang},
\citenamefont {Rhim},\ and\ \citenamefont {Yang}}]{hwang2020geometric}%
\BibitemOpen
\bibfield  {author} {\bibinfo {author} {\bibfnamefont {Yoonseok}\
\bibnamefont {Hwang}}, \bibinfo {author} {\bibfnamefont {Jun-Won}\
\bibnamefont {Rhim}}, \ and\ \bibinfo {author} {\bibfnamefont {Bohm-Jung}\
\bibnamefont {Yang}},\ }\bibfield  {title} {\enquote {\bibinfo {title}
{Geometric characterization of anomalous {Landau} levels of isolated flat
bands},}\ }\href {https://arxiv.org/abs/2012.15132} {\bibfield  {journal}
{\bibinfo  {journal} {arXiv:2012.15132}\ } (\bibinfo {year}
{2020})}\BibitemShut {NoStop}%
\bibitem [{\citenamefont {Kuno}\ \emph {et~al.}(2020)\citenamefont {Kuno},
\citenamefont {Mizoguchi},\ and\ \citenamefont {Hatsugai}}]{kuno2020flat}%
\BibitemOpen
\bibfield  {author} {\bibinfo {author} {\bibfnamefont {Yoshihito}\
\bibnamefont {Kuno}}, \bibinfo {author} {\bibfnamefont {Tomonari}\
\bibnamefont {Mizoguchi}}, \ and\ \bibinfo {author} {\bibfnamefont
{Yasuhiro}\ \bibnamefont {Hatsugai}},\ }\bibfield  {title} {\enquote
{\bibinfo {title} {Flat band quantum scar},}\ }\href {\doibase
10.1103/PhysRevB.102.241115} {\bibfield  {journal} {\bibinfo  {journal}
{Physical Review B}\ }\textbf {\bibinfo {volume} {102}},\ \bibinfo {pages}
{241115(R)} (\bibinfo {year} {2020})}\BibitemShut {NoStop}%
\bibitem [{\citenamefont {Xie}\ \emph {et~al.}(2020)\citenamefont {Xie},
\citenamefont {Song}, \citenamefont {Lian},\ and\ \citenamefont
{Bernevig}}]{xie2020topology}%
\BibitemOpen
\bibfield  {author} {\bibinfo {author} {\bibfnamefont {Fang}\ \bibnamefont
{Xie}}, \bibinfo {author} {\bibfnamefont {Zhida}\ \bibnamefont {Song}},
\bibinfo {author} {\bibfnamefont {Biao}\ \bibnamefont {Lian}}, \ and\
\bibinfo {author} {\bibfnamefont {B.A.}\ \bibnamefont {Bernevig}},\
}\bibfield  {title} {\enquote {\bibinfo {title} {Topology-bounded superfluid
weight in twisted bilayer graphene},}\ }\href {\doibase
10.1103/PhysRevLett.124.167002} {\bibfield  {journal} {\bibinfo  {journal}
{Physical Review Letters}\ }\textbf {\bibinfo {volume} {124}},\ \bibinfo
{pages} {167002} (\bibinfo {year} {2020})}\BibitemShut {NoStop}%
\bibitem [{\citenamefont {Lin}(2020)}]{lin2020chiral}%
\BibitemOpen
\bibfield  {author} {\bibinfo {author} {\bibfnamefont {Yu-Ping}\ \bibnamefont
{Lin}},\ }\bibfield  {title} {\enquote {\bibinfo {title} {Chiral flat band
superconductivity from symmetry-protected three-band crossings},}\ }\href
{\doibase 10.1103/PhysRevResearch.2.043209} {\bibfield  {journal} {\bibinfo
{journal} {Physical Review Research}\ }\textbf {\bibinfo {volume} {2}},\
\bibinfo {pages} {043209} (\bibinfo {year} {2020})}\BibitemShut {NoStop}%
\bibitem [{\citenamefont {Kheirkhah}\ \emph {et~al.}(2020)\citenamefont
{Kheirkhah}, \citenamefont {Nagai}, \citenamefont {Chen},\ and\ \citenamefont
{Marsiglio}}]{kheirkhah2020majorana}%
\BibitemOpen
\bibfield  {author} {\bibinfo {author} {\bibfnamefont {Majid}\ \bibnamefont
{Kheirkhah}}, \bibinfo {author} {\bibfnamefont {Yuki}\ \bibnamefont {Nagai}},
\bibinfo {author} {\bibfnamefont {Chun}\ \bibnamefont {Chen}}, \ and\
\bibinfo {author} {\bibfnamefont {Frank}\ \bibnamefont {Marsiglio}},\
}\bibfield  {title} {\enquote {\bibinfo {title} {{Majorana} corner flat bands
in two-dimensional second-order topological superconductors},}\ }\href
{\doibase 10.1103/PhysRevB.101.104502} {\bibfield  {journal} {\bibinfo
{journal} {Physical Review B}\ }\textbf {\bibinfo {volume} {101}},\ \bibinfo
{pages} {104502} (\bibinfo {year} {2020})}\BibitemShut {NoStop}%
\bibitem [{\citenamefont {Morfonios}\ \emph {et~al.}(2021)\citenamefont
{Morfonios}, \citenamefont {R\"ontgen}, \citenamefont {Pyzh},\ and\
\citenamefont {Schmelcher}}]{morfonios2021flat}%
\BibitemOpen
\bibfield  {author} {\bibinfo {author} {\bibfnamefont {C.~V.}\ \bibnamefont
{Morfonios}}, \bibinfo {author} {\bibfnamefont {M.}~\bibnamefont
{R\"ontgen}}, \bibinfo {author} {\bibfnamefont {M.}~\bibnamefont {Pyzh}}, \
and\ \bibinfo {author} {\bibfnamefont {P.}~\bibnamefont {Schmelcher}},\
}\bibfield  {title} {\enquote {\bibinfo {title} {Flat bands by latent
symmetry},}\ }\href {\doibase 10.1103/PhysRevB.104.035105} {\bibfield
{journal} {\bibinfo  {journal} {Physical Review B}\ }\textbf {\bibinfo
{volume} {104}},\ \bibinfo {pages} {035105} (\bibinfo {year}
{2021})}\BibitemShut {NoStop}%
\bibitem [{\citenamefont {Peri}\ \emph {et~al.}(2021)\citenamefont {Peri},
\citenamefont {Song}, \citenamefont {Bernevig},\ and\ \citenamefont
{Huber}}]{peri2021fragile}%
\BibitemOpen
\bibfield  {author} {\bibinfo {author} {\bibfnamefont {Valerio}\ \bibnamefont
{Peri}}, \bibinfo {author} {\bibfnamefont {Zhi-Da}\ \bibnamefont {Song}},
\bibinfo {author} {\bibfnamefont {B.A.}\ \bibnamefont {Bernevig}}, \ and\
\bibinfo {author} {\bibfnamefont {S.D.}\ \bibnamefont {Huber}},\ }\bibfield
{title} {\enquote {\bibinfo {title} {Fragile topology and flat-band
superconductivity in the strong-coupling regime},}\ }\href {\doibase
10.1103/PhysRevLett.126.027002} {\bibfield  {journal} {\bibinfo  {journal}
{Physical Review Letters}\ }\textbf {\bibinfo {volume} {126}},\ \bibinfo
{pages} {027002} (\bibinfo {year} {2021})}\BibitemShut {NoStop}%
\bibitem [{\citenamefont {Maimaiti}\ \emph {et~al.}(2021)\citenamefont
{Maimaiti}, \citenamefont {Andreanov},\ and\ \citenamefont
{Flach}}]{maimaiti2021flat}%
\BibitemOpen
\bibfield  {author} {\bibinfo {author} {\bibfnamefont {Wulayimu}\
\bibnamefont {Maimaiti}}, \bibinfo {author} {\bibfnamefont {Alexei}\
\bibnamefont {Andreanov}}, \ and\ \bibinfo {author} {\bibfnamefont {Sergej}\
\bibnamefont {Flach}},\ }\bibfield  {title} {\enquote {\bibinfo {title}
{Flat-band generator in two dimensions},}\ }\href {\doibase
10.1103/PhysRevB.103.165116} {\bibfield  {journal} {\bibinfo  {journal}
{Physical Review B}\ }\textbf {\bibinfo {volume} {103}},\ \bibinfo {pages}
{165116} (\bibinfo {year} {2021})}\BibitemShut {NoStop}%
\bibitem [{\citenamefont {Bistritzer}\ and\ \citenamefont
{MacDonald}(2011)}]{bistritzer2011moire}%
\BibitemOpen
\bibfield  {author} {\bibinfo {author} {\bibfnamefont {Rafi}\ \bibnamefont
{Bistritzer}}\ and\ \bibinfo {author} {\bibfnamefont {Allan~H.}\ \bibnamefont
{MacDonald}},\ }\bibfield  {title} {\enquote {\bibinfo {title} {Moir{\'e}
bands in twisted double-layer graphene},}\ }\href {\doibase
10.1073/pnas.1108174108} {\bibfield  {journal} {\bibinfo  {journal}
{Proceedings of the National Academy of Sciences}\ }\textbf {\bibinfo
{volume} {108}},\ \bibinfo {pages} {12233--12237} (\bibinfo {year}
{2011})}\BibitemShut {NoStop}%
\bibitem [{\citenamefont {Cao}\ \emph {et~al.}(2018{\natexlab{a}})\citenamefont
{Cao}, \citenamefont {Fatemi}, \citenamefont {Demir}, \citenamefont {Fang},
\citenamefont {Tomarken}, \citenamefont {Luo}, \citenamefont
{Sanchez-Yamagishi}, \citenamefont {Watanabe}, \citenamefont {Taniguchi},
\citenamefont {Kaxiras} \emph {et~al.}}]{cao2018correlated}%
\BibitemOpen
\bibfield  {author} {\bibinfo {author} {\bibfnamefont {Yuan}\ \bibnamefont
{Cao}}, \bibinfo {author} {\bibfnamefont {Valla}\ \bibnamefont {Fatemi}},
\bibinfo {author} {\bibfnamefont {Ahmet}\ \bibnamefont {Demir}}, \bibinfo
{author} {\bibfnamefont {Shiang}\ \bibnamefont {Fang}}, \bibinfo {author}
{\bibfnamefont {Spencer~L}\ \bibnamefont {Tomarken}}, \bibinfo {author}
{\bibfnamefont {Jason~Y}\ \bibnamefont {Luo}}, \bibinfo {author}
{\bibfnamefont {Javier~D}\ \bibnamefont {Sanchez-Yamagishi}}, \bibinfo
{author} {\bibfnamefont {Kenji}\ \bibnamefont {Watanabe}}, \bibinfo {author}
{\bibfnamefont {Takashi}\ \bibnamefont {Taniguchi}}, \bibinfo {author}
{\bibfnamefont {Efthimios}\ \bibnamefont {Kaxiras}},  \emph {et~al.},\
}\bibfield  {title} {\enquote {\bibinfo {title} {Correlated insulator
behaviour at half-filling in magic-angle graphene superlattices},}\ }\href
{\doibase 10.1038/nature26154} {\bibfield  {journal} {\bibinfo  {journal}
{Nature}\ }\textbf {\bibinfo {volume} {556}},\ \bibinfo {pages} {80--84}
(\bibinfo {year} {2018}{\natexlab{a}})}\BibitemShut {NoStop}%
\bibitem [{\citenamefont {Cao}\ \emph {et~al.}(2018{\natexlab{b}})\citenamefont
{Cao}, \citenamefont {Fatemi}, \citenamefont {Fang}, \citenamefont
{Watanabe}, \citenamefont {Taniguchi}, \citenamefont {Kaxiras},\ and\
\citenamefont {Jarillo-Herrero}}]{cao2018unconventional}%
\BibitemOpen
\bibfield  {author} {\bibinfo {author} {\bibfnamefont {Yuan}\ \bibnamefont
{Cao}}, \bibinfo {author} {\bibfnamefont {Valla}\ \bibnamefont {Fatemi}},
\bibinfo {author} {\bibfnamefont {Shiang}\ \bibnamefont {Fang}}, \bibinfo
{author} {\bibfnamefont {Kenji}\ \bibnamefont {Watanabe}}, \bibinfo {author}
{\bibfnamefont {Takashi}\ \bibnamefont {Taniguchi}}, \bibinfo {author}
{\bibfnamefont {Efthimios}\ \bibnamefont {Kaxiras}}, \ and\ \bibinfo {author}
{\bibfnamefont {Pablo}\ \bibnamefont {Jarillo-Herrero}},\ }\bibfield  {title}
{\enquote {\bibinfo {title} {Unconventional superconductivity in magic-angle
graphene superlattices},}\ }\href {\doibase 10.1038/nature26160} {\bibfield
{journal} {\bibinfo  {journal} {Nature}\ }\textbf {\bibinfo {volume} {556}},\
\bibinfo {pages} {43--50} (\bibinfo {year} {2018}{\natexlab{b}})}\BibitemShut
{NoStop}%
\bibitem [{\citenamefont {Ye}\ \emph {et~al.}(2018)\citenamefont {Ye},
\citenamefont {Kang}, \citenamefont {Liu}, \citenamefont {Von~Cube},
\citenamefont {Wicker}, \citenamefont {Suzuki}, \citenamefont {Jozwiak},
\citenamefont {Bostwick}, \citenamefont {Rotenberg}, \citenamefont {Bell}
\emph {et~al.}}]{ye2018massive}%
\BibitemOpen
\bibfield  {author} {\bibinfo {author} {\bibfnamefont {Linda}\ \bibnamefont
{Ye}}, \bibinfo {author} {\bibfnamefont {Mingu}\ \bibnamefont {Kang}},
\bibinfo {author} {\bibfnamefont {Junwei}\ \bibnamefont {Liu}}, \bibinfo
{author} {\bibfnamefont {Felix}\ \bibnamefont {Von~Cube}}, \bibinfo {author}
{\bibfnamefont {Christina~R}\ \bibnamefont {Wicker}}, \bibinfo {author}
{\bibfnamefont {Takehito}\ \bibnamefont {Suzuki}}, \bibinfo {author}
{\bibfnamefont {Chris}\ \bibnamefont {Jozwiak}}, \bibinfo {author}
{\bibfnamefont {Aaron}\ \bibnamefont {Bostwick}}, \bibinfo {author}
{\bibfnamefont {Eli}\ \bibnamefont {Rotenberg}}, \bibinfo {author}
{\bibfnamefont {David~C}\ \bibnamefont {Bell}},  \emph {et~al.},\ }\bibfield
{title} {\enquote {\bibinfo {title} {Massive {Dirac} fermions in a
ferromagnetic kagome metal},}\ }\href {\doibase 10.1038/nature25987}
{\bibfield  {journal} {\bibinfo  {journal} {Nature}\ }\textbf {\bibinfo
{volume} {555}},\ \bibinfo {pages} {638--642} (\bibinfo {year}
{2018})}\BibitemShut {NoStop}%
\bibitem [{\citenamefont {Li}\ \emph {et~al.}(2018)\citenamefont {Li},
\citenamefont {Zhuang}, \citenamefont {Wang}, \citenamefont {Feng},
\citenamefont {Gao}, \citenamefont {Xu}, \citenamefont {Hao}, \citenamefont
{Wang}, \citenamefont {Zhang}, \citenamefont {Wu} \emph
{et~al.}}]{li2018realization}%
\BibitemOpen
\bibfield  {author} {\bibinfo {author} {\bibfnamefont {Zhi}\ \bibnamefont
{Li}}, \bibinfo {author} {\bibfnamefont {Jincheng}\ \bibnamefont {Zhuang}},
\bibinfo {author} {\bibfnamefont {Li}~\bibnamefont {Wang}}, \bibinfo {author}
{\bibfnamefont {Haifeng}\ \bibnamefont {Feng}}, \bibinfo {author}
{\bibfnamefont {Qian}\ \bibnamefont {Gao}}, \bibinfo {author} {\bibfnamefont
{Xun}\ \bibnamefont {Xu}}, \bibinfo {author} {\bibfnamefont {Weichang}\
\bibnamefont {Hao}}, \bibinfo {author} {\bibfnamefont {Xiaolin}\ \bibnamefont
{Wang}}, \bibinfo {author} {\bibfnamefont {Chao}\ \bibnamefont {Zhang}},
\bibinfo {author} {\bibfnamefont {Kehui}\ \bibnamefont {Wu}},  \emph
{et~al.},\ }\bibfield  {title} {\enquote {\bibinfo {title} {Realization of
flat band with possible nontrivial topology in electronic {Kagome}
lattice},}\ }\href {\doibase 10.1126/sciadv.aau4511} {\bibfield  {journal}
{\bibinfo  {journal} {Science advances}\ }\textbf {\bibinfo {volume} {4}},\
\bibinfo {pages} {eaau4511} (\bibinfo {year} {2018})}\BibitemShut {NoStop}%
\bibitem [{\citenamefont {Kang}\ \emph {et~al.}(2020)\citenamefont {Kang},
\citenamefont {Fang}, \citenamefont {Ye}, \citenamefont {Po}, \citenamefont
{Denlinger}, \citenamefont {Jozwiak}, \citenamefont {Bostwick}, \citenamefont
{Rotenberg}, \citenamefont {Kaxiras}, \citenamefont {Checkelsky} \emph
{et~al.}}]{kang2020topological}%
\BibitemOpen
\bibfield  {author} {\bibinfo {author} {\bibfnamefont {Mingu}\ \bibnamefont
{Kang}}, \bibinfo {author} {\bibfnamefont {Shiang}\ \bibnamefont {Fang}},
\bibinfo {author} {\bibfnamefont {Linda}\ \bibnamefont {Ye}}, \bibinfo
{author} {\bibfnamefont {Hoi~Chun}\ \bibnamefont {Po}}, \bibinfo {author}
{\bibfnamefont {Jonathan}\ \bibnamefont {Denlinger}}, \bibinfo {author}
{\bibfnamefont {Chris}\ \bibnamefont {Jozwiak}}, \bibinfo {author}
{\bibfnamefont {Aaron}\ \bibnamefont {Bostwick}}, \bibinfo {author}
{\bibfnamefont {Eli}\ \bibnamefont {Rotenberg}}, \bibinfo {author}
{\bibfnamefont {Efthimios}\ \bibnamefont {Kaxiras}}, \bibinfo {author}
{\bibfnamefont {Joseph~G}\ \bibnamefont {Checkelsky}},  \emph {et~al.},\
}\bibfield  {title} {\enquote {\bibinfo {title} {Topological flat bands in
frustrated kagome lattice {CoSn}},}\ }\href {\doibase
10.1038/s41467-020-17465-1} {\bibfield  {journal} {\bibinfo  {journal}
{Nature Communications}\ }\textbf {\bibinfo {volume} {11}},\ \bibinfo {pages}
{4004} (\bibinfo {year} {2020})}\BibitemShut {NoStop}%
\bibitem [{\citenamefont {Tang}\ \emph {et~al.}(2011)\citenamefont {Tang},
\citenamefont {Mei},\ and\ \citenamefont {Wen}}]{tang2011high}%
\BibitemOpen
\bibfield  {author} {\bibinfo {author} {\bibfnamefont {Evelyn}\ \bibnamefont
{Tang}}, \bibinfo {author} {\bibfnamefont {Jia-Wei}\ \bibnamefont {Mei}}, \
and\ \bibinfo {author} {\bibfnamefont {Xiao-Gang}\ \bibnamefont {Wen}},\
}\bibfield  {title} {\enquote {\bibinfo {title} {High-temperature fractional
quantum {Hall} states},}\ }\href {\doibase 10.1103/PhysRevLett.106.236802}
{\bibfield  {journal} {\bibinfo  {journal} {Physical Review Letters}\
}\textbf {\bibinfo {volume} {106}},\ \bibinfo {pages} {236802} (\bibinfo
{year} {2011})}\BibitemShut {NoStop}%
\bibitem [{\citenamefont {Sun}\ \emph {et~al.}(2011)\citenamefont {Sun},
\citenamefont {Gu}, \citenamefont {Katsura},\ and\ \citenamefont
{Das~Sarma}}]{sun2011nearly}%
\BibitemOpen
\bibfield  {author} {\bibinfo {author} {\bibfnamefont {Kai}\ \bibnamefont
{Sun}}, \bibinfo {author} {\bibfnamefont {Zhengcheng}\ \bibnamefont {Gu}},
\bibinfo {author} {\bibfnamefont {Hosho}\ \bibnamefont {Katsura}}, \ and\
\bibinfo {author} {\bibfnamefont {S.}~\bibnamefont {Das~Sarma}},\ }\bibfield
{title} {\enquote {\bibinfo {title} {Nearly flatbands with nontrivial
topology},}\ }\href {\doibase 10.1103/PhysRevLett.106.236803} {\bibfield
{journal} {\bibinfo  {journal} {Physical Review Letters}\ }\textbf {\bibinfo
{volume} {106}},\ \bibinfo {pages} {236803} (\bibinfo {year}
{2011})}\BibitemShut {NoStop}%
\bibitem [{\citenamefont {Neupert}\ \emph {et~al.}(2011)\citenamefont
{Neupert}, \citenamefont {Santos}, \citenamefont {Chamon},\ and\
\citenamefont {Mudry}}]{neupert2011fractional}%
\BibitemOpen
\bibfield  {author} {\bibinfo {author} {\bibfnamefont {Titus}\ \bibnamefont
{Neupert}}, \bibinfo {author} {\bibfnamefont {Luiz}\ \bibnamefont {Santos}},
\bibinfo {author} {\bibfnamefont {Claudio}\ \bibnamefont {Chamon}}, \ and\
\bibinfo {author} {\bibfnamefont {Christopher}\ \bibnamefont {Mudry}},\
}\bibfield  {title} {\enquote {\bibinfo {title} {Fractional quantum {Hall}
states at zero magnetic field},}\ }\href {\doibase
10.1103/PhysRevLett.106.236804} {\bibfield  {journal} {\bibinfo  {journal}
{Physical Review Letters}\ }\textbf {\bibinfo {volume} {106}},\ \bibinfo
{pages} {236804} (\bibinfo {year} {2011})}\BibitemShut {NoStop}%
\bibitem [{\citenamefont {Song}\ \emph {et~al.}(2019)\citenamefont {Song},
\citenamefont {Wang}, \citenamefont {Shi}, \citenamefont {Li}, \citenamefont
{Fang},\ and\ \citenamefont {Bernevig}}]{song2019all}%
\BibitemOpen
\bibfield  {author} {\bibinfo {author} {\bibfnamefont {Zhida}\ \bibnamefont
{Song}}, \bibinfo {author} {\bibfnamefont {Zhijun}\ \bibnamefont {Wang}},
\bibinfo {author} {\bibfnamefont {Wujun}\ \bibnamefont {Shi}}, \bibinfo
{author} {\bibfnamefont {Gang}\ \bibnamefont {Li}}, \bibinfo {author}
{\bibfnamefont {Chen}\ \bibnamefont {Fang}}, \ and\ \bibinfo {author}
{\bibfnamefont {B.A.}\ \bibnamefont {Bernevig}},\ }\bibfield  {title}
{\enquote {\bibinfo {title} {All magic angles in twisted bilayer graphene are
topological},}\ }\href {\doibase 10.1103/PhysRevLett.123.036401} {\bibfield
{journal} {\bibinfo  {journal} {Physical Review Letters}\ }\textbf {\bibinfo
{volume} {123}},\ \bibinfo {pages} {036401} (\bibinfo {year}
{2019})}\BibitemShut {NoStop}%
\bibitem [{\citenamefont {Po}\ \emph {et~al.}(2019)\citenamefont {Po},
\citenamefont {Zou}, \citenamefont {Senthil},\ and\ \citenamefont
{Vishwanath}}]{po2019faithful}%
\BibitemOpen
\bibfield  {author} {\bibinfo {author} {\bibfnamefont {Hoi~Chun}\
\bibnamefont {Po}}, \bibinfo {author} {\bibfnamefont {Liujun}\ \bibnamefont
{Zou}}, \bibinfo {author} {\bibfnamefont {T.}~\bibnamefont {Senthil}}, \ and\
\bibinfo {author} {\bibfnamefont {Ashvin}\ \bibnamefont {Vishwanath}},\
}\bibfield  {title} {\enquote {\bibinfo {title} {Faithful tight-binding
models and fragile topology of magic-angle bilayer graphene},}\ }\href
{\doibase 10.1103/PhysRevB.99.195455} {\bibfield  {journal} {\bibinfo
{journal} {Physical Review B}\ }\textbf {\bibinfo {volume} {99}},\ \bibinfo
{pages} {195455} (\bibinfo {year} {2019})}\BibitemShut {NoStop}%
\bibitem [{\citenamefont {Ahn}\ \emph {et~al.}(2019)\citenamefont {Ahn},
\citenamefont {Park},\ and\ \citenamefont {Yang}}]{ahn2019failure}%
\BibitemOpen
\bibfield  {author} {\bibinfo {author} {\bibfnamefont {Junyeong}\
\bibnamefont {Ahn}}, \bibinfo {author} {\bibfnamefont {Sungjoon}\
\bibnamefont {Park}}, \ and\ \bibinfo {author} {\bibfnamefont {Bohm-Jung}\
\bibnamefont {Yang}},\ }\bibfield  {title} {\enquote {\bibinfo {title}
{Failure of {Nielsen}-{Ninomiya} theorem and fragile topology in
two-dimensional systems with space-time inversion symmetry: {Application} to
twisted bilayer graphene at magic angle},}\ }\href {\doibase
10.1103/PhysRevX.9.021013} {\bibfield  {journal} {\bibinfo  {journal}
{Physical Review X}\ }\textbf {\bibinfo {volume} {9}},\ \bibinfo {pages}
{021013} (\bibinfo {year} {2019})}\BibitemShut {NoStop}%
\bibitem [{\citenamefont {Bergman}\ \emph {et~al.}(2008)\citenamefont
{Bergman}, \citenamefont {Wu},\ and\ \citenamefont
{Balents}}]{bergman2008band}%
\BibitemOpen
\bibfield  {author} {\bibinfo {author} {\bibfnamefont {Doron~L.}\
\bibnamefont {Bergman}}, \bibinfo {author} {\bibfnamefont {Congjun}\
\bibnamefont {Wu}}, \ and\ \bibinfo {author} {\bibfnamefont {Leon}\
\bibnamefont {Balents}},\ }\bibfield  {title} {\enquote {\bibinfo {title}
{Band touching from real-space topology in frustrated hopping models},}\
}\href {\doibase 10.1103/PhysRevB.78.125104} {\bibfield  {journal} {\bibinfo
{journal} {Physical Review B}\ }\textbf {\bibinfo {volume} {78}},\ \bibinfo
{pages} {125104} (\bibinfo {year} {2008})}\BibitemShut {NoStop}%
\bibitem [{\citenamefont {Ma}\ \emph {et~al.}(2020{\natexlab{a}})\citenamefont
{Ma}, \citenamefont {Xu}, \citenamefont {Chiu}, \citenamefont {Regnault},
\citenamefont {Houck}, \citenamefont {Song},\ and\ \citenamefont
{Bernevig}}]{ma2020spin}%
\BibitemOpen
\bibfield  {author} {\bibinfo {author} {\bibfnamefont {Da-Shuai}\
\bibnamefont {Ma}}, \bibinfo {author} {\bibfnamefont {Yuanfeng}\ \bibnamefont
{Xu}}, \bibinfo {author} {\bibfnamefont {C.S.}\ \bibnamefont {Chiu}},
\bibinfo {author} {\bibfnamefont {Nicolas}\ \bibnamefont {Regnault}},
\bibinfo {author} {\bibfnamefont {A.A.}\ \bibnamefont {Houck}}, \bibinfo
{author} {\bibfnamefont {Zhida}\ \bibnamefont {Song}}, \ and\ \bibinfo
{author} {\bibfnamefont {B.A.}\ \bibnamefont {Bernevig}},\ }\bibfield
{title} {\enquote {\bibinfo {title} {Spin-orbit-induced topological flat
bands in line and split graphs of bipartite lattices},}\ }\href {\doibase
10.1103/PhysRevLett.125.266403} {\bibfield  {journal} {\bibinfo  {journal}
{Physical Review Letters}\ }\textbf {\bibinfo {volume} {125}},\ \bibinfo
{pages} {266403} (\bibinfo {year} {2020}{\natexlab{a}})}\BibitemShut
{NoStop}%
\bibitem [{\citenamefont {Rhim}\ and\ \citenamefont
{Yang}(2019)}]{rhim2019classification}%
\BibitemOpen
\bibfield  {author} {\bibinfo {author} {\bibfnamefont {Jun-Won}\ \bibnamefont
{Rhim}}\ and\ \bibinfo {author} {\bibfnamefont {Bohm-Jung}\ \bibnamefont
{Yang}},\ }\bibfield  {title} {\enquote {\bibinfo {title} {Classification of
flat bands according to the band-crossing singularity of {Bloch} wave
functions},}\ }\href {\doibase 10.1103/PhysRevB.99.045107} {\bibfield
{journal} {\bibinfo  {journal} {Physical Review B}\ }\textbf {\bibinfo
{volume} {99}},\ \bibinfo {pages} {045107} (\bibinfo {year}
{2019})}\BibitemShut {NoStop}%
\bibitem [{\citenamefont {Rhim}\ and\ \citenamefont
{Yang}(2021)}]{rhim2021singular}%
\BibitemOpen
\bibfield  {author} {\bibinfo {author} {\bibfnamefont {Jun-Won}\ \bibnamefont
{Rhim}}\ and\ \bibinfo {author} {\bibfnamefont {Bohm-Jung}\ \bibnamefont
{Yang}},\ }\bibfield  {title} {\enquote {\bibinfo {title} {Singular flat
bands},}\ }\href {\doibase 10.1080/23746149.2021.1901606} {\bibfield
{journal} {\bibinfo  {journal} {Advances in Physics: X}\ }\textbf {\bibinfo
{volume} {6}},\ \bibinfo {pages} {1901606} (\bibinfo {year}
{2021})}\BibitemShut {NoStop}%
\bibitem [{\citenamefont {Rhim}\ \emph {et~al.}(2020)\citenamefont {Rhim},
\citenamefont {Kim},\ and\ \citenamefont {Yang}}]{rhim2020quantum}%
\BibitemOpen
\bibfield  {author} {\bibinfo {author} {\bibfnamefont {Jun-Won}\ \bibnamefont
{Rhim}}, \bibinfo {author} {\bibfnamefont {Kyoo}\ \bibnamefont {Kim}}, \ and\
\bibinfo {author} {\bibfnamefont {Bohm-Jung}\ \bibnamefont {Yang}},\
}\bibfield  {title} {\enquote {\bibinfo {title} {Quantum distance and
anomalous {Landau} levels of flat bands},}\ }\href {\doibase
10.1038/s41586-020-2540-1} {\bibfield  {journal} {\bibinfo  {journal}
{Nature}\ }\textbf {\bibinfo {volume} {584}},\ \bibinfo {pages} {59--63}
(\bibinfo {year} {2020})}\BibitemShut {NoStop}%
\bibitem [{\citenamefont {Hwang}\ \emph {et~al.}(2021)\citenamefont {Hwang},
\citenamefont {Jung}, \citenamefont {Rhim},\ and\ \citenamefont
{Yang}}]{hwang2021wave}%
\BibitemOpen
\bibfield  {author} {\bibinfo {author} {\bibfnamefont {Yoonseok}\
\bibnamefont {Hwang}}, \bibinfo {author} {\bibfnamefont {Junseo}\
\bibnamefont {Jung}}, \bibinfo {author} {\bibfnamefont {Jun-Won}\
\bibnamefont {Rhim}}, \ and\ \bibinfo {author} {\bibfnamefont {Bohm-Jung}\
\bibnamefont {Yang}},\ }\bibfield  {title} {\enquote {\bibinfo {title}
{Wave-function geometry of band crossing points in two dimensions},}\ }\href
{\doibase 10.1103/PhysRevB.103.L241102} {\bibfield  {journal} {\bibinfo
{journal} {Physical Review B}\ }\textbf {\bibinfo {volume} {103}},\ \bibinfo
{pages} {L241102} (\bibinfo {year} {2021})}\BibitemShut {NoStop}%
\bibitem [{\citenamefont {Sutherland}(1986)}]{sutherland1986localization}%
\BibitemOpen
\bibfield  {author} {\bibinfo {author} {\bibfnamefont {Bill}\ \bibnamefont
{Sutherland}},\ }\bibfield  {title} {\enquote {\bibinfo {title} {Localization
of electronic wave functions due to local topology},}\ }\href {\doibase
10.1103/PhysRevB.34.5208} {\bibfield  {journal} {\bibinfo  {journal}
{Physical Review B}\ }\textbf {\bibinfo {volume} {34}},\ \bibinfo {pages}
{5208} (\bibinfo {year} {1986})}\BibitemShut {NoStop}%
\bibitem [{\citenamefont {Vidal}\ \emph {et~al.}(1998)\citenamefont {Vidal},
\citenamefont {Mosseri},\ and\ \citenamefont {Dou\ifmmode~\mbox{\c{c}}\else
\c{c}\fi{}ot}}]{vidal1998aharonov}%
\BibitemOpen
\bibfield  {author} {\bibinfo {author} {\bibfnamefont {Julien}\ \bibnamefont
{Vidal}}, \bibinfo {author} {\bibfnamefont {R\'emy}\ \bibnamefont {Mosseri}},
\ and\ \bibinfo {author} {\bibfnamefont {Benoit}\ \bibnamefont
{Dou\ifmmode~\mbox{\c{c}}\else \c{c}\fi{}ot}},\ }\bibfield  {title} {\enquote
{\bibinfo {title} {{Aharonov}-{Bohm} cages in two-dimensional structures},}\
}\href {\doibase 10.1103/PhysRevLett.81.5888} {\bibfield  {journal} {\bibinfo
{journal} {Physical Review Letters}\ }\textbf {\bibinfo {volume} {81}},\
\bibinfo {pages} {5888} (\bibinfo {year} {1998})}\BibitemShut {NoStop}%
\bibitem [{\citenamefont {Vidal}\ \emph {et~al.}(2001)\citenamefont {Vidal},
\citenamefont {Butaud}, \citenamefont {Dou\ifmmode~\mbox{\c{c}}\else
\c{c}\fi{}ot},\ and\ \citenamefont {Mosseri}}]{vidal2001disorder}%
\BibitemOpen
\bibfield  {author} {\bibinfo {author} {\bibfnamefont {Julien}\ \bibnamefont
{Vidal}}, \bibinfo {author} {\bibfnamefont {Patrick}\ \bibnamefont {Butaud}},
\bibinfo {author} {\bibfnamefont {Benoit}\ \bibnamefont
{Dou\ifmmode~\mbox{\c{c}}\else \c{c}\fi{}ot}}, \ and\ \bibinfo {author}
{\bibfnamefont {R\'emy}\ \bibnamefont {Mosseri}},\ }\bibfield  {title}
{\enquote {\bibinfo {title} {Disorder and interactions in {Aharonov}-{Bohm}
cages},}\ }\href {\doibase 10.1103/PhysRevB.64.155306} {\bibfield  {journal}
{\bibinfo  {journal} {Physical Review B}\ }\textbf {\bibinfo {volume} {64}},\
\bibinfo {pages} {155306} (\bibinfo {year} {2001})}\BibitemShut {NoStop}%
\bibitem [{\citenamefont {Mukherjee}\ \emph {et~al.}(2015)\citenamefont
{Mukherjee}, \citenamefont {Spracklen}, \citenamefont {Choudhury},
\citenamefont {Goldman}, \citenamefont {\"Ohberg}, \citenamefont
{Andersson},\ and\ \citenamefont {Thomson}}]{mukherjee2015observation}%
\BibitemOpen
\bibfield  {author} {\bibinfo {author} {\bibfnamefont {Sebabrata}\
\bibnamefont {Mukherjee}}, \bibinfo {author} {\bibfnamefont {Alexander}\
\bibnamefont {Spracklen}}, \bibinfo {author} {\bibfnamefont {Debaditya}\
\bibnamefont {Choudhury}}, \bibinfo {author} {\bibfnamefont {Nathan}\
\bibnamefont {Goldman}}, \bibinfo {author} {\bibfnamefont {Patrik}\
\bibnamefont {\"Ohberg}}, \bibinfo {author} {\bibfnamefont {Erika}\
\bibnamefont {Andersson}}, \ and\ \bibinfo {author} {\bibfnamefont
{Robert~R.}\ \bibnamefont {Thomson}},\ }\bibfield  {title} {\enquote
{\bibinfo {title} {Observation of a localized flat-band state in a photonic
{Lieb} lattice},}\ }\href {\doibase 10.1103/PhysRevLett.114.245504}
{\bibfield  {journal} {\bibinfo  {journal} {Physical Review Letters}\
}\textbf {\bibinfo {volume} {114}},\ \bibinfo {pages} {245504} (\bibinfo
{year} {2015})}\BibitemShut {NoStop}%
\bibitem [{\citenamefont {Read}(2017)}]{read2017compactly}%
\BibitemOpen
\bibfield  {author} {\bibinfo {author} {\bibfnamefont {N.}~\bibnamefont
{Read}},\ }\bibfield  {title} {\enquote {\bibinfo {title} {Compactly
supported {Wannier} functions and algebraic {K}-theory},}\ }\href {\doibase
10.1103/PhysRevB.95.115309} {\bibfield  {journal} {\bibinfo  {journal}
{Physical Review B}\ }\textbf {\bibinfo {volume} {95}},\ \bibinfo {pages}
{115309} (\bibinfo {year} {2017})}\BibitemShut {NoStop}%
\bibitem [{\citenamefont {Maimaiti}\ \emph {et~al.}(2017)\citenamefont
{Maimaiti}, \citenamefont {Andreanov}, \citenamefont {Park}, \citenamefont
{Gendelman},\ and\ \citenamefont {Flach}}]{maimaiti2017compact}%
\BibitemOpen
\bibfield  {author} {\bibinfo {author} {\bibfnamefont {Wulayimu}\
\bibnamefont {Maimaiti}}, \bibinfo {author} {\bibfnamefont {Alexei}\
\bibnamefont {Andreanov}}, \bibinfo {author} {\bibfnamefont {Hee~Chul}\
\bibnamefont {Park}}, \bibinfo {author} {\bibfnamefont {Oleg}\ \bibnamefont
{Gendelman}}, \ and\ \bibinfo {author} {\bibfnamefont {Sergej}\ \bibnamefont
{Flach}},\ }\bibfield  {title} {\enquote {\bibinfo {title} {Compact localized
states and flat-band generators in one dimension},}\ }\href {\doibase
10.1103/PhysRevB.95.115135} {\bibfield  {journal} {\bibinfo  {journal}
{Physical Review B}\ }\textbf {\bibinfo {volume} {95}},\ \bibinfo {pages}
{115135} (\bibinfo {year} {2017})}\BibitemShut {NoStop}%
\bibitem [{\citenamefont {Maimaiti}\ \emph {et~al.}(2019)\citenamefont
{Maimaiti}, \citenamefont {Flach},\ and\ \citenamefont
{Andreanov}}]{maimaiti2019universal}%
\BibitemOpen
\bibfield  {author} {\bibinfo {author} {\bibfnamefont {Wulayimu}\
\bibnamefont {Maimaiti}}, \bibinfo {author} {\bibfnamefont {Sergej}\
\bibnamefont {Flach}}, \ and\ \bibinfo {author} {\bibfnamefont {Alexei}\
\bibnamefont {Andreanov}},\ }\bibfield  {title} {\enquote {\bibinfo {title}
{Universal {$d=1$} flat band generator from compact localized states},}\
}\href {\doibase 10.1103/PhysRevB.99.125129} {\bibfield  {journal} {\bibinfo
{journal} {Physical Review B}\ }\textbf {\bibinfo {volume} {99}},\ \bibinfo
{pages} {125129} (\bibinfo {year} {2019})}\BibitemShut {NoStop}%
\bibitem [{\citenamefont {R\"ontgen}\ \emph {et~al.}(2018)\citenamefont
{R\"ontgen}, \citenamefont {Morfonios},\ and\ \citenamefont
{Schmelcher}}]{rontgen2018compact}%
\BibitemOpen
\bibfield  {author} {\bibinfo {author} {\bibfnamefont {M.}~\bibnamefont
{R\"ontgen}}, \bibinfo {author} {\bibfnamefont {C.~V.}\ \bibnamefont
{Morfonios}}, \ and\ \bibinfo {author} {\bibfnamefont {P.}~\bibnamefont
{Schmelcher}},\ }\bibfield  {title} {\enquote {\bibinfo {title} {Compact
localized states and flat bands from local symmetry partitioning},}\ }\href
{\doibase 10.1103/PhysRevB.97.035161} {\bibfield  {journal} {\bibinfo
{journal} {Physical Review B}\ }\textbf {\bibinfo {volume} {97}},\ \bibinfo
{pages} {035161} (\bibinfo {year} {2018})}\BibitemShut {NoStop}%
\bibitem [{\citenamefont {Ma}\ \emph {et~al.}(2020{\natexlab{b}})\citenamefont
{Ma}, \citenamefont {Rhim}, \citenamefont {Tang}, \citenamefont {Xia},
\citenamefont {Wang}, \citenamefont {Zheng}, \citenamefont {Xia},
\citenamefont {Song}, \citenamefont {Hu}, \citenamefont {Li}, \citenamefont
{Yang}, \citenamefont {Leykam},\ and\ \citenamefont {Chen}}]{ma2020direct}%
\BibitemOpen
\bibfield  {author} {\bibinfo {author} {\bibfnamefont {Jina}\ \bibnamefont
{Ma}}, \bibinfo {author} {\bibfnamefont {Jun-Won}\ \bibnamefont {Rhim}},
\bibinfo {author} {\bibfnamefont {Liqin}\ \bibnamefont {Tang}}, \bibinfo
{author} {\bibfnamefont {Shiqi}\ \bibnamefont {Xia}}, \bibinfo {author}
{\bibfnamefont {Haiping}\ \bibnamefont {Wang}}, \bibinfo {author}
{\bibfnamefont {Xiuyan}\ \bibnamefont {Zheng}}, \bibinfo {author}
{\bibfnamefont {Shiqiang}\ \bibnamefont {Xia}}, \bibinfo {author}
{\bibfnamefont {Daohong}\ \bibnamefont {Song}}, \bibinfo {author}
{\bibfnamefont {Yi}~\bibnamefont {Hu}}, \bibinfo {author} {\bibfnamefont
{Yigang}\ \bibnamefont {Li}}, \bibinfo {author} {\bibfnamefont {Bohm-Jung}\
\bibnamefont {Yang}}, \bibinfo {author} {\bibfnamefont {Daniel}\ \bibnamefont
{Leykam}}, \ and\ \bibinfo {author} {\bibfnamefont {Zhigang}\ \bibnamefont
{Chen}},\ }\bibfield  {title} {\enquote {\bibinfo {title} {Direct observation
of flatband loop states arising from nontrivial real-space topology},}\
}\href {\doibase 10.1103/PhysRevLett.124.183901} {\bibfield  {journal}
{\bibinfo  {journal} {Physical Review Letters}\ }\textbf {\bibinfo {volume}
{124}},\ \bibinfo {pages} {183901} (\bibinfo {year}
{2020}{\natexlab{b}})}\BibitemShut {NoStop}%
\bibitem [{\citenamefont
{Mielke}(1991{\natexlab{a}})}]{mielke1991ferromagnetic}%
\BibitemOpen
\bibfield  {author} {\bibinfo {author} {\bibfnamefont {Andreas}\ \bibnamefont
{Mielke}},\ }\bibfield  {title} {\enquote {\bibinfo {title} {Ferromagnetic
ground states for the {Hubbard} model on line graphs},}\ }\href {\doibase
10.1088/0305-4470/24/2/005} {\bibfield  {journal} {\bibinfo  {journal}
{Journal of Physics A: Mathematical and General}\ }\textbf {\bibinfo {volume}
{24}},\ \bibinfo {pages} {L73} (\bibinfo {year}
{1991}{\natexlab{a}})}\BibitemShut {NoStop}%
\bibitem [{\citenamefont
{Mielke}(1991{\natexlab{b}})}]{mielke1991ferromagnetism}%
\BibitemOpen
\bibfield  {author} {\bibinfo {author} {\bibfnamefont {A}~\bibnamefont
{Mielke}},\ }\bibfield  {title} {\enquote {\bibinfo {title} {Ferromagnetism
in the {Hubbard} model on line graphs and further considerations},}\ }\href
{\doibase 10.1088/0305-4470/24/14/018} {\bibfield  {journal} {\bibinfo
{journal} {Journal of Physics A: Mathematical and General}\ }\textbf
{\bibinfo {volume} {24}},\ \bibinfo {pages} {3311} (\bibinfo {year}
{1991}{\natexlab{b}})}\BibitemShut {NoStop}%
\bibitem [{\citenamefont {Mielke}\ and\ \citenamefont
{Tasaki}(1993)}]{mielke1993ferromagnetism}%
\BibitemOpen
\bibfield  {author} {\bibinfo {author} {\bibfnamefont {Andreas}\ \bibnamefont
{Mielke}}\ and\ \bibinfo {author} {\bibfnamefont {Hal}\ \bibnamefont
{Tasaki}},\ }\bibfield  {title} {\enquote {\bibinfo {title} {Ferromagnetism
in the {Hubbard} model},}\ }\href {\doibase 10.1007/BF02108079} {\bibfield
{journal} {\bibinfo  {journal} {Communications in mathematical physics}\
}\textbf {\bibinfo {volume} {158}},\ \bibinfo {pages} {341--371} (\bibinfo
{year} {1993})}\BibitemShut {NoStop}%
\bibitem [{\citenamefont {Koll{\'a}r}\ \emph {et~al.}(2020)\citenamefont
{Koll{\'a}r}, \citenamefont {Fitzpatrick}, \citenamefont {Sarnak},\ and\
\citenamefont {Houck}}]{kollar2019line}%
\BibitemOpen
\bibfield  {author} {\bibinfo {author} {\bibfnamefont {Alicia~J}\
\bibnamefont {Koll{\'a}r}}, \bibinfo {author} {\bibfnamefont {Mattias}\
\bibnamefont {Fitzpatrick}}, \bibinfo {author} {\bibfnamefont {Peter}\
\bibnamefont {Sarnak}}, \ and\ \bibinfo {author} {\bibfnamefont {Andrew~A}\
\bibnamefont {Houck}},\ }\bibfield  {title} {\enquote {\bibinfo {title}
{Line-graph lattices: {Euclidean} and non-{Euclidean} flat bands, and
implementations in circuit quantum electrodynamics},}\ }\href {\doibase
10.1007/s00220-019-03645-8} {\bibfield  {journal} {\bibinfo  {journal}
{Communications in Mathematical Physics}\ }\textbf {\bibinfo {volume}
{376}},\ \bibinfo {pages} {1909} (\bibinfo {year} {2020})}\BibitemShut
{NoStop}%
\bibitem [{\citenamefont {Wang}\ and\ \citenamefont
{Ran}(2011)}]{wang2011nearly}%
\BibitemOpen
\bibfield  {author} {\bibinfo {author} {\bibfnamefont {Fa}~\bibnamefont
{Wang}}\ and\ \bibinfo {author} {\bibfnamefont {Ying}\ \bibnamefont {Ran}},\
}\bibfield  {title} {\enquote {\bibinfo {title} {Nearly flat band with
{Chern} number {$C=2$} on the dice lattice},}\ }\href {\doibase
10.1103/PhysRevB.84.241103} {\bibfield  {journal} {\bibinfo  {journal}
{Physical Review B}\ }\textbf {\bibinfo {volume} {84}},\ \bibinfo {pages}
{241103(R)} (\bibinfo {year} {2011})}\BibitemShut {NoStop}%
\bibitem [{\citenamefont {Trescher}\ and\ \citenamefont
{Bergholtz}(2012)}]{trescher2012flat}%
\BibitemOpen
\bibfield  {author} {\bibinfo {author} {\bibfnamefont {Maximilian}\
\bibnamefont {Trescher}}\ and\ \bibinfo {author} {\bibfnamefont {Emil~J.}\
\bibnamefont {Bergholtz}},\ }\bibfield  {title} {\enquote {\bibinfo {title}
{Flat bands with higher {Chern} number in pyrochlore slabs},}\ }\href
{\doibase 10.1103/PhysRevB.86.241111} {\bibfield  {journal} {\bibinfo
{journal} {Physical Review B}\ }\textbf {\bibinfo {volume} {86}},\ \bibinfo
{pages} {241111(R)} (\bibinfo {year} {2012})}\BibitemShut {NoStop}%
\bibitem [{\citenamefont {Yang}\ \emph {et~al.}(2012)\citenamefont {Yang},
\citenamefont {Gu}, \citenamefont {Sun},\ and\ \citenamefont
{Das~Sarma}}]{yang2012topological}%
\BibitemOpen
\bibfield  {author} {\bibinfo {author} {\bibfnamefont {Shuo}\ \bibnamefont
{Yang}}, \bibinfo {author} {\bibfnamefont {Zheng-Cheng}\ \bibnamefont {Gu}},
\bibinfo {author} {\bibfnamefont {Kai}\ \bibnamefont {Sun}}, \ and\ \bibinfo
{author} {\bibfnamefont {S.}~\bibnamefont {Das~Sarma}},\ }\bibfield  {title}
{\enquote {\bibinfo {title} {Topological flat band models with arbitrary
{Chern} numbers},}\ }\href {\doibase 10.1103/PhysRevB.86.241112} {\bibfield
{journal} {\bibinfo  {journal} {Physical Review B}\ }\textbf {\bibinfo
{volume} {86}},\ \bibinfo {pages} {241112(R)} (\bibinfo {year}
{2012})}\BibitemShut {NoStop}%
\bibitem [{\citenamefont {Kalesaki}\ \emph {et~al.}(2014)\citenamefont
{Kalesaki}, \citenamefont {Delerue}, \citenamefont {Morais~Smith},
\citenamefont {Beugeling}, \citenamefont {Allan},\ and\ \citenamefont
{Vanmaekelbergh}}]{kalesaki2014dirac}%
\BibitemOpen
\bibfield  {author} {\bibinfo {author} {\bibfnamefont {E.}~\bibnamefont
{Kalesaki}}, \bibinfo {author} {\bibfnamefont {C.}~\bibnamefont {Delerue}},
\bibinfo {author} {\bibfnamefont {C.}~\bibnamefont {Morais~Smith}}, \bibinfo
{author} {\bibfnamefont {W.}~\bibnamefont {Beugeling}}, \bibinfo {author}
{\bibfnamefont {G.}~\bibnamefont {Allan}}, \ and\ \bibinfo {author}
{\bibfnamefont {D.}~\bibnamefont {Vanmaekelbergh}},\ }\bibfield  {title}
{\enquote {\bibinfo {title} {Dirac cones, topological edge states, and
nontrivial flat bands in two-dimensional semiconductors with a honeycomb
nanogeometry},}\ }\href {\doibase 10.1103/PhysRevX.4.011010} {\bibfield
{journal} {\bibinfo  {journal} {Physical Review X}\ }\textbf {\bibinfo
{volume} {4}},\ \bibinfo {pages} {011010} (\bibinfo {year}
{2014})}\BibitemShut {NoStop}%
\bibitem [{\citenamefont {Pal}(2018)}]{pal2018nontrivial}%
\BibitemOpen
\bibfield  {author} {\bibinfo {author} {\bibfnamefont {Biplab}\ \bibnamefont
{Pal}},\ }\bibfield  {title} {\enquote {\bibinfo {title} {Nontrivial
topological flat bands in a diamond-octagon lattice geometry},}\ }\href
{\doibase 10.1103/PhysRevB.98.245116} {\bibfield  {journal} {\bibinfo
{journal} {Physical Review B}\ }\textbf {\bibinfo {volume} {98}},\ \bibinfo
{pages} {245116} (\bibinfo {year} {2018})}\BibitemShut {NoStop}%
\bibitem [{\citenamefont {Bhattacharya}\ and\ \citenamefont
{Pal}(2019)}]{bhattacharya2019flat}%
\BibitemOpen
\bibfield  {author} {\bibinfo {author} {\bibfnamefont {Ankita}\ \bibnamefont
{Bhattacharya}}\ and\ \bibinfo {author} {\bibfnamefont {Biplab}\ \bibnamefont
{Pal}},\ }\bibfield  {title} {\enquote {\bibinfo {title} {Flat bands and
nontrivial topological properties in an extended {Lieb} lattice},}\ }\href
{\doibase 10.1103/PhysRevB.100.235145} {\bibfield  {journal} {\bibinfo
{journal} {Physical Review B}\ }\textbf {\bibinfo {volume} {100}},\ \bibinfo
{pages} {235145} (\bibinfo {year} {2019})}\BibitemShut {NoStop}%
\bibitem [{sup()}]{supple}%
\BibitemOpen
\bibinfo {note} {See the Supplemental Material, which includes
Refs.~\cite{shiozaki2017topological,dummit2004abstract,zak1980symmetry,zak1981band,bradlyn2017topological,cano2018building,alexandradinata2018no,holler2018topological,alexandradinata2020crystallographic,Chen2014impossibility,elcoro2020magnetic,po2018fragile,cano2018topology,bradlyn2019disconnected,bouhon2019wilson,else2019fragile,wieder2018axion,liu2019shift,hwang2019fragile,bouhon2020geometric,song2020fragile,song2020twisted,peri2020experimental,zhang2021tunable,weeks2010topological},
for more details about the definition and properties of CLS and FT-CLS, the
relation between band crossing points of flat band and singular points of
FT-CLS, the symmetry representation of flat band and SR-enforced band
crossing. The detailed description of flat-band models is also
included.}\BibitemShut {Stop}%
\bibitem [{\citenamefont {Fang}\ \emph {et~al.}(2012)\citenamefont {Fang},
\citenamefont {Gilbert},\ and\ \citenamefont {Bernevig}}]{fang2012bulk}%
\BibitemOpen
\bibfield  {author} {\bibinfo {author} {\bibfnamefont {Chen}\ \bibnamefont
{Fang}}, \bibinfo {author} {\bibfnamefont {M.J.}\ \bibnamefont {Gilbert}}, \
and\ \bibinfo {author} {\bibfnamefont {B.A.}\ \bibnamefont {Bernevig}},\
}\bibfield  {title} {\enquote {\bibinfo {title} {Bulk topological invariants
in noninteracting point group symmetric insulators},}\ }\href {\doibase
10.1103/PhysRevB.86.115112} {\bibfield  {journal} {\bibinfo  {journal}
{Physical Review B}\ }\textbf {\bibinfo {volume} {86}},\ \bibinfo {pages}
{115112} (\bibinfo {year} {2012})}\BibitemShut {NoStop}%
\bibitem [{\citenamefont {Beugeling}\ \emph {et~al.}(2012)\citenamefont
{Beugeling}, \citenamefont {Everts},\ and\ \citenamefont
{Morais~Smith}}]{beugeling2012topological}%
\BibitemOpen
\bibfield  {author} {\bibinfo {author} {\bibfnamefont {W.}~\bibnamefont
{Beugeling}}, \bibinfo {author} {\bibfnamefont {J.~C.}\ \bibnamefont
{Everts}}, \ and\ \bibinfo {author} {\bibfnamefont {C.}~\bibnamefont
{Morais~Smith}},\ }\bibfield  {title} {\enquote {\bibinfo {title}
{Topological phase transitions driven by next-nearest-neighbor hopping in
two-dimensional lattices},}\ }\href {\doibase 10.1103/PhysRevB.86.195129}
{\bibfield  {journal} {\bibinfo  {journal} {Physical Review B}\ }\textbf
{\bibinfo {volume} {86}},\ \bibinfo {pages} {195129} (\bibinfo {year}
{2012})}\BibitemShut {NoStop}%
\bibitem [{\citenamefont {Zhang}\ \emph {et~al.}(2013)\citenamefont {Zhang},
\citenamefont {Ren}, \citenamefont {Wang},\ and\ \citenamefont
{Li}}]{zhang2013topological}%
\BibitemOpen
\bibfield  {author} {\bibinfo {author} {\bibfnamefont {Lifa}\ \bibnamefont
{Zhang}}, \bibinfo {author} {\bibfnamefont {Jie}\ \bibnamefont {Ren}},
\bibinfo {author} {\bibfnamefont {Jian-Sheng}\ \bibnamefont {Wang}}, \ and\
\bibinfo {author} {\bibfnamefont {Baowen}\ \bibnamefont {Li}},\ }\bibfield
{title} {\enquote {\bibinfo {title} {Topological magnon insulator in
insulating ferromagnet},}\ }\href {\doibase 10.1103/PhysRevB.87.144101}
{\bibfield  {journal} {\bibinfo  {journal} {Physical Review B}\ }\textbf
{\bibinfo {volume} {87}},\ \bibinfo {pages} {144101} (\bibinfo {year}
{2013})}\BibitemShut {NoStop}%
\bibitem [{\citenamefont {Mook}\ \emph {et~al.}(2014)\citenamefont {Mook},
\citenamefont {Henk},\ and\ \citenamefont {Mertig}}]{mook2014magnon}%
\BibitemOpen
\bibfield  {author} {\bibinfo {author} {\bibfnamefont {Alexander}\
\bibnamefont {Mook}}, \bibinfo {author} {\bibfnamefont {J\"urgen}\
\bibnamefont {Henk}}, \ and\ \bibinfo {author} {\bibfnamefont {Ingrid}\
\bibnamefont {Mertig}},\ }\bibfield  {title} {\enquote {\bibinfo {title}
{Magnon {Hall} effect and topology in kagome lattices: {A} theoretical
investigation},}\ }\href {\doibase 10.1103/PhysRevB.89.134409} {\bibfield
{journal} {\bibinfo  {journal} {Physical Review B}\ }\textbf {\bibinfo
{volume} {89}},\ \bibinfo {pages} {134409} (\bibinfo {year}
{2014})}\BibitemShut {NoStop}%
\bibitem [{\citenamefont {Chisnell}\ \emph {et~al.}(2015)\citenamefont
{Chisnell}, \citenamefont {Helton}, \citenamefont {Freedman}, \citenamefont
{Singh}, \citenamefont {Bewley}, \citenamefont {Nocera},\ and\ \citenamefont
{Lee}}]{chisnell2015topological}%
\BibitemOpen
\bibfield  {author} {\bibinfo {author} {\bibfnamefont {R.}~\bibnamefont
{Chisnell}}, \bibinfo {author} {\bibfnamefont {J.~S.}\ \bibnamefont
{Helton}}, \bibinfo {author} {\bibfnamefont {D.~E.}\ \bibnamefont
{Freedman}}, \bibinfo {author} {\bibfnamefont {D.~K.}\ \bibnamefont {Singh}},
\bibinfo {author} {\bibfnamefont {R.~I.}\ \bibnamefont {Bewley}}, \bibinfo
{author} {\bibfnamefont {D.~G.}\ \bibnamefont {Nocera}}, \ and\ \bibinfo
{author} {\bibfnamefont {Y.~S.}\ \bibnamefont {Lee}},\ }\bibfield  {title}
{\enquote {\bibinfo {title} {Topological magnon bands in a kagome lattice
ferromagnet},}\ }\href {\doibase 10.1103/PhysRevLett.115.147201} {\bibfield
{journal} {\bibinfo  {journal} {Physical Review Letters}\ }\textbf {\bibinfo
{volume} {115}},\ \bibinfo {pages} {147201} (\bibinfo {year}
{2015})}\BibitemShut {NoStop}%
\bibitem [{\citenamefont {Yang}\ \emph {et~al.}(2021)\citenamefont {Yang},
\citenamefont {Nakano},\ and\ \citenamefont {Katsura}}]{yang2021symmetry}%
\BibitemOpen
\bibfield  {author} {\bibinfo {author} {\bibfnamefont {Hong}\ \bibnamefont
{Yang}}, \bibinfo {author} {\bibfnamefont {Hayate}\ \bibnamefont {Nakano}}, \
and\ \bibinfo {author} {\bibfnamefont {Hosho}\ \bibnamefont {Katsura}},\
}\bibfield  {title} {\enquote {\bibinfo {title} {Symmetry-protected
topological phases in spinful bosons with a flat band},}\ }\href {\doibase
10.1103/PhysRevResearch.3.023210} {\bibfield  {journal} {\bibinfo  {journal}
{Physical Review Research}\ }\textbf {\bibinfo {volume} {3}},\ \bibinfo
{pages} {023210} (\bibinfo {year} {2021})}\BibitemShut {NoStop}%
\bibitem [{\citenamefont {Shiozaki}\ \emph {et~al.}(2017)\citenamefont
{Shiozaki}, \citenamefont {Sato},\ and\ \citenamefont
{Gomi}}]{shiozaki2017topological}%
\BibitemOpen
\bibfield  {author} {\bibinfo {author} {\bibfnamefont {Ken}\ \bibnamefont
{Shiozaki}}, \bibinfo {author} {\bibfnamefont {Masatoshi}\ \bibnamefont
{Sato}}, \ and\ \bibinfo {author} {\bibfnamefont {Kiyonori}\ \bibnamefont
{Gomi}},\ }\bibfield  {title} {\enquote {\bibinfo {title} {Topological
crystalline materials: {General} formulation, module structure, and wallpaper
groups},}\ }\href {\doibase 10.1103/PhysRevB.95.235425} {\bibfield  {journal}
{\bibinfo  {journal} {Physical Review B}\ }\textbf {\bibinfo {volume} {95}},\
\bibinfo {pages} {235425} (\bibinfo {year} {2017})}\BibitemShut {NoStop}%
\bibitem [{\citenamefont {Dummit}\ and\ \citenamefont
{Foote}(2004)}]{dummit2004abstract}%
\BibitemOpen
\bibfield  {author} {\bibinfo {author} {\bibfnamefont {David~Steven}\
\bibnamefont {Dummit}}\ and\ \bibinfo {author} {\bibfnamefont {Richard~M}\
\bibnamefont {Foote}},\ }\href@noop {} {\emph {\bibinfo {title} {Abstract
Algebra}}},\ Vol.~\bibinfo {volume} {3}\ (\bibinfo  {publisher} {Wiley
Hoboken},\ \bibinfo {year} {2004})\BibitemShut {NoStop}%
\bibitem [{\citenamefont {Zak}(1980)}]{zak1980symmetry}%
\BibitemOpen
\bibfield  {author} {\bibinfo {author} {\bibfnamefont {J.}~\bibnamefont
{Zak}},\ }\bibfield  {title} {\enquote {\bibinfo {title} {Symmetry
specification of bands in solids},}\ }\href {\doibase
10.1103/PhysRevLett.45.1025} {\bibfield  {journal} {\bibinfo  {journal}
{Physical Review Letters}\ }\textbf {\bibinfo {volume} {45}},\ \bibinfo
{pages} {1025} (\bibinfo {year} {1980})}\BibitemShut {NoStop}%
\bibitem [{\citenamefont {Zak}(1981)}]{zak1981band}%
\BibitemOpen
\bibfield  {author} {\bibinfo {author} {\bibfnamefont {J.}~\bibnamefont
{Zak}},\ }\bibfield  {title} {\enquote {\bibinfo {title} {Band
representations and symmetry types of bands in solids},}\ }\href {\doibase
10.1103/PhysRevB.23.2824} {\bibfield  {journal} {\bibinfo  {journal}
{Physical Review B}\ }\textbf {\bibinfo {volume} {23}},\ \bibinfo {pages}
{2824} (\bibinfo {year} {1981})}\BibitemShut {NoStop}%
\bibitem [{\citenamefont {Bradlyn}\ \emph {et~al.}(2017)\citenamefont
{Bradlyn}, \citenamefont {Elcoro}, \citenamefont {Cano}, \citenamefont
{Vergniory}, \citenamefont {Wang}, \citenamefont {Felser}, \citenamefont
{Aroyo},\ and\ \citenamefont {Bernevig}}]{bradlyn2017topological}%
\BibitemOpen
\bibfield  {author} {\bibinfo {author} {\bibfnamefont {Barry}\ \bibnamefont
{Bradlyn}}, \bibinfo {author} {\bibfnamefont {L}~\bibnamefont {Elcoro}},
\bibinfo {author} {\bibfnamefont {Jennifer}\ \bibnamefont {Cano}}, \bibinfo
{author} {\bibfnamefont {MG}~\bibnamefont {Vergniory}}, \bibinfo {author}
{\bibfnamefont {Zhijun}\ \bibnamefont {Wang}}, \bibinfo {author}
{\bibfnamefont {C}~\bibnamefont {Felser}}, \bibinfo {author} {\bibfnamefont
{MI}~\bibnamefont {Aroyo}}, \ and\ \bibinfo {author} {\bibfnamefont
{B~Andrei}\ \bibnamefont {Bernevig}},\ }\bibfield  {title} {\enquote
{\bibinfo {title} {Topological quantum chemistry},}\ }\href {\doibase
10.1038/nature23268} {\bibfield  {journal} {\bibinfo  {journal} {Nature}\
}\textbf {\bibinfo {volume} {547}},\ \bibinfo {pages} {298--305} (\bibinfo
{year} {2017})}\BibitemShut {NoStop}%
\bibitem [{\citenamefont {Cano}\ \emph
{et~al.}(2018{\natexlab{a}})\citenamefont {Cano}, \citenamefont {Bradlyn},
\citenamefont {Wang}, \citenamefont {Elcoro}, \citenamefont {Vergniory},
\citenamefont {Felser}, \citenamefont {Aroyo},\ and\ \citenamefont
{Bernevig}}]{cano2018building}%
\BibitemOpen
\bibfield  {author} {\bibinfo {author} {\bibfnamefont {Jennifer}\
\bibnamefont {Cano}}, \bibinfo {author} {\bibfnamefont {Barry}\ \bibnamefont
{Bradlyn}}, \bibinfo {author} {\bibfnamefont {Zhijun}\ \bibnamefont {Wang}},
\bibinfo {author} {\bibfnamefont {L.}~\bibnamefont {Elcoro}}, \bibinfo
{author} {\bibfnamefont {M.~G.}\ \bibnamefont {Vergniory}}, \bibinfo {author}
{\bibfnamefont {C.}~\bibnamefont {Felser}}, \bibinfo {author} {\bibfnamefont
{M.~I.}\ \bibnamefont {Aroyo}}, \ and\ \bibinfo {author} {\bibfnamefont
{B.A.}\ \bibnamefont {Bernevig}},\ }\bibfield  {title} {\enquote {\bibinfo
{title} {Building blocks of topological quantum chemistry: {Elementary} band
representations},}\ }\href {\doibase 10.1103/PhysRevB.97.035139} {\bibfield
{journal} {\bibinfo  {journal} {Physical Review B}\ }\textbf {\bibinfo
{volume} {97}},\ \bibinfo {pages} {035139} (\bibinfo {year}
{2018}{\natexlab{a}})}\BibitemShut {NoStop}%
\bibitem [{\citenamefont {Alexandradinata}\ and\ \citenamefont
{H\"oller}(2018)}]{alexandradinata2018no}%
\BibitemOpen
\bibfield  {author} {\bibinfo {author} {\bibfnamefont {A.}~\bibnamefont
{Alexandradinata}}\ and\ \bibinfo {author} {\bibfnamefont {J.}~\bibnamefont
{H\"oller}},\ }\bibfield  {title} {\enquote {\bibinfo {title} {No-go theorem
for topological insulators and high-throughput identification of {Chern}
insulators},}\ }\href {\doibase 10.1103/PhysRevB.98.184305} {\bibfield
{journal} {\bibinfo  {journal} {Physical Review B}\ }\textbf {\bibinfo
{volume} {98}},\ \bibinfo {pages} {184305} (\bibinfo {year}
{2018})}\BibitemShut {NoStop}%
\bibitem [{\citenamefont {H\"oller}\ and\ \citenamefont
{Alexandradinata}(2018)}]{holler2018topological}%
\BibitemOpen
\bibfield  {author} {\bibinfo {author} {\bibfnamefont {J.}~\bibnamefont
{H\"oller}}\ and\ \bibinfo {author} {\bibfnamefont {A.}~\bibnamefont
{Alexandradinata}},\ }\bibfield  {title} {\enquote {\bibinfo {title}
{Topological {Bloch} oscillations},}\ }\href {\doibase
10.1103/PhysRevB.98.024310} {\bibfield  {journal} {\bibinfo  {journal}
{Physical Review B}\ }\textbf {\bibinfo {volume} {98}},\ \bibinfo {pages}
{024310} (\bibinfo {year} {2018})}\BibitemShut {NoStop}%
\bibitem [{\citenamefont {Alexandradinata}\ \emph {et~al.}(2020)\citenamefont
{Alexandradinata}, \citenamefont {H\"oller}, \citenamefont {Wang},
\citenamefont {Cheng},\ and\ \citenamefont
{Lu}}]{alexandradinata2020crystallographic}%
\BibitemOpen
\bibfield  {author} {\bibinfo {author} {\bibfnamefont {A.}~\bibnamefont
{Alexandradinata}}, \bibinfo {author} {\bibfnamefont {J.}~\bibnamefont
{H\"oller}}, \bibinfo {author} {\bibfnamefont {Chong}\ \bibnamefont {Wang}},
\bibinfo {author} {\bibfnamefont {Hengbin}\ \bibnamefont {Cheng}}, \ and\
\bibinfo {author} {\bibfnamefont {Ling}\ \bibnamefont {Lu}},\ }\bibfield
{title} {\enquote {\bibinfo {title} {Crystallographic splitting theorem for
band representations and fragile topological photonic crystals},}\ }\href
{\doibase 10.1103/PhysRevB.102.115117} {\bibfield  {journal} {\bibinfo
{journal} {Physical Review B}\ }\textbf {\bibinfo {volume} {102}},\ \bibinfo
{pages} {115117} (\bibinfo {year} {2020})}\BibitemShut {NoStop}%
\bibitem [{\citenamefont {Chen}\ \emph {et~al.}(2014)\citenamefont {Chen},
\citenamefont {Mazaheri}, \citenamefont {Seidel},\ and\ \citenamefont
{Tang}}]{Chen2014impossibility}%
\BibitemOpen
\bibfield  {author} {\bibinfo {author} {\bibfnamefont {Li}~\bibnamefont
{Chen}}, \bibinfo {author} {\bibfnamefont {Tahereh}\ \bibnamefont
{Mazaheri}}, \bibinfo {author} {\bibfnamefont {Alexander}\ \bibnamefont
{Seidel}}, \ and\ \bibinfo {author} {\bibfnamefont {Xiang}\ \bibnamefont
{Tang}},\ }\bibfield  {title} {\enquote {\bibinfo {title} {The impossibility
of exactly flat non-trivial {Chern} bands in strictly local periodic tight
binding models},}\ }\href {\doibase 10.1088/1751-8113/47/15/152001}
{\bibfield  {journal} {\bibinfo  {journal} {Journal of Physics A:
Mathematical and Theoretical}\ }\textbf {\bibinfo {volume} {47}},\ \bibinfo
{pages} {152001} (\bibinfo {year} {2014})}\BibitemShut {NoStop}%
\bibitem [{\citenamefont {Elcoro}\ \emph {et~al.}(2020)\citenamefont {Elcoro},
\citenamefont {Wieder}, \citenamefont {Song}, \citenamefont {Xu},
\citenamefont {Bradlyn},\ and\ \citenamefont
{Bernevig}}]{elcoro2020magnetic}%
\BibitemOpen
\bibfield  {author} {\bibinfo {author} {\bibfnamefont {Luis}\ \bibnamefont
{Elcoro}}, \bibinfo {author} {\bibfnamefont {Benjamin~J}\ \bibnamefont
{Wieder}}, \bibinfo {author} {\bibfnamefont {Zhida}\ \bibnamefont {Song}},
\bibinfo {author} {\bibfnamefont {Yuanfeng}\ \bibnamefont {Xu}}, \bibinfo
{author} {\bibfnamefont {Barry}\ \bibnamefont {Bradlyn}}, \ and\ \bibinfo
{author} {\bibfnamefont {B~Andrei}\ \bibnamefont {Bernevig}},\ }\bibfield
{title} {\enquote {\bibinfo {title} {Magnetic topological quantum
chemistry},}\ }\href {https://arxiv.org/abs/2010.00598} {\bibfield  {journal}
{\bibinfo  {journal} {arXiv:2010.00598}\ } (\bibinfo {year}
{2020})}\BibitemShut {NoStop}%
\bibitem [{\citenamefont {Po}\ \emph {et~al.}(2018)\citenamefont {Po},
\citenamefont {Watanabe},\ and\ \citenamefont {Vishwanath}}]{po2018fragile}%
\BibitemOpen
\bibfield  {author} {\bibinfo {author} {\bibfnamefont {Hoi~Chun}\
\bibnamefont {Po}}, \bibinfo {author} {\bibfnamefont {Haruki}\ \bibnamefont
{Watanabe}}, \ and\ \bibinfo {author} {\bibfnamefont {Ashvin}\ \bibnamefont
{Vishwanath}},\ }\bibfield  {title} {\enquote {\bibinfo {title} {Fragile
topology and {Wannier} obstructions},}\ }\href {\doibase
10.1103/PhysRevLett.121.126402} {\bibfield  {journal} {\bibinfo  {journal}
{Phys. Rev. Lett.}\ }\textbf {\bibinfo {volume} {121}},\ \bibinfo {pages}
{126402} (\bibinfo {year} {2018})}\BibitemShut {NoStop}%
\bibitem [{\citenamefont {Cano}\ \emph
{et~al.}(2018{\natexlab{b}})\citenamefont {Cano}, \citenamefont {Bradlyn},
\citenamefont {Wang}, \citenamefont {Elcoro}, \citenamefont {Vergniory},
\citenamefont {Felser}, \citenamefont {Aroyo},\ and\ \citenamefont
{Bernevig}}]{cano2018topology}%
\BibitemOpen
\bibfield  {author} {\bibinfo {author} {\bibfnamefont {Jennifer}\
\bibnamefont {Cano}}, \bibinfo {author} {\bibfnamefont {Barry}\ \bibnamefont
{Bradlyn}}, \bibinfo {author} {\bibfnamefont {Zhijun}\ \bibnamefont {Wang}},
\bibinfo {author} {\bibfnamefont {L.}~\bibnamefont {Elcoro}}, \bibinfo
{author} {\bibfnamefont {M.~G.}\ \bibnamefont {Vergniory}}, \bibinfo {author}
{\bibfnamefont {C.}~\bibnamefont {Felser}}, \bibinfo {author} {\bibfnamefont
{M.~I.}\ \bibnamefont {Aroyo}}, \ and\ \bibinfo {author} {\bibfnamefont
{B.A.}\ \bibnamefont {Bernevig}},\ }\bibfield  {title} {\enquote {\bibinfo
{title} {Topology of disconnected elementary band representations},}\ }\href
{\doibase 10.1103/PhysRevLett.120.266401} {\bibfield  {journal} {\bibinfo
{journal} {Physical Review Letters}\ }\textbf {\bibinfo {volume} {120}},\
\bibinfo {pages} {266401} (\bibinfo {year} {2018}{\natexlab{b}})}\BibitemShut
{NoStop}%
\bibitem [{\citenamefont {Bradlyn}\ \emph {et~al.}(2019)\citenamefont
{Bradlyn}, \citenamefont {Wang}, \citenamefont {Cano},\ and\ \citenamefont
{Bernevig}}]{bradlyn2019disconnected}%
\BibitemOpen
\bibfield  {author} {\bibinfo {author} {\bibfnamefont {Barry}\ \bibnamefont
{Bradlyn}}, \bibinfo {author} {\bibfnamefont {Zhijun}\ \bibnamefont {Wang}},
\bibinfo {author} {\bibfnamefont {Jennifer}\ \bibnamefont {Cano}}, \ and\
\bibinfo {author} {\bibfnamefont {B.A.}\ \bibnamefont {Bernevig}},\
}\bibfield  {title} {\enquote {\bibinfo {title} {Disconnected elementary band
representations, fragile topology, and {Wilson} loops as topological indices:
{An} example on the triangular lattice},}\ }\href {\doibase
10.1103/PhysRevB.99.045140} {\bibfield  {journal} {\bibinfo  {journal}
{Physical Review B}\ }\textbf {\bibinfo {volume} {99}},\ \bibinfo {pages}
{045140} (\bibinfo {year} {2019})}\BibitemShut {NoStop}%
\bibitem [{\citenamefont {Bouhon}\ \emph {et~al.}(2019)\citenamefont {Bouhon},
\citenamefont {Black-Schaffer},\ and\ \citenamefont
{Slager}}]{bouhon2019wilson}%
\BibitemOpen
\bibfield  {author} {\bibinfo {author} {\bibfnamefont {Adrien}\ \bibnamefont
{Bouhon}}, \bibinfo {author} {\bibfnamefont {Annica~M.}\ \bibnamefont
{Black-Schaffer}}, \ and\ \bibinfo {author} {\bibfnamefont {Robert-Jan}\
\bibnamefont {Slager}},\ }\bibfield  {title} {\enquote {\bibinfo {title}
{Wilson loop approach to fragile topology of split elementary band
representations and topological crystalline insulators with time-reversal
symmetry},}\ }\href {\doibase 10.1103/PhysRevB.100.195135} {\bibfield
{journal} {\bibinfo  {journal} {Physical Review B}\ }\textbf {\bibinfo
{volume} {100}},\ \bibinfo {pages} {195135} (\bibinfo {year}
{2019})}\BibitemShut {NoStop}%
\bibitem [{\citenamefont {Else}\ \emph {et~al.}(2019)\citenamefont {Else},
\citenamefont {Po},\ and\ \citenamefont {Watanabe}}]{else2019fragile}%
\BibitemOpen
\bibfield  {author} {\bibinfo {author} {\bibfnamefont {Dominic~V.}\
\bibnamefont {Else}}, \bibinfo {author} {\bibfnamefont {Hoi~Chun}\
\bibnamefont {Po}}, \ and\ \bibinfo {author} {\bibfnamefont {Haruki}\
\bibnamefont {Watanabe}},\ }\bibfield  {title} {\enquote {\bibinfo {title}
{Fragile topological phases in interacting systems},}\ }\href {\doibase
10.1103/PhysRevB.99.125122} {\bibfield  {journal} {\bibinfo  {journal}
{Physical Review B}\ }\textbf {\bibinfo {volume} {99}},\ \bibinfo {pages}
{125122} (\bibinfo {year} {2019})}\BibitemShut {NoStop}%
\bibitem [{\citenamefont {Wieder}\ and\ \citenamefont
{Bernevig}(2018)}]{wieder2018axion}%
\BibitemOpen
\bibfield  {author} {\bibinfo {author} {\bibfnamefont {Benjamin~J}\
\bibnamefont {Wieder}}\ and\ \bibinfo {author} {\bibfnamefont {B~Andrei}\
\bibnamefont {Bernevig}},\ }\bibfield  {title} {\enquote {\bibinfo {title}
{The axion insulator as a pump of fragile topology},}\ }\href
{https://arxiv.org/abs/1810.02373} {\bibfield  {journal} {\bibinfo  {journal}
{arXiv:1810.02373}\ } (\bibinfo {year} {2018})}\BibitemShut {NoStop}%
\bibitem [{\citenamefont {Liu}\ \emph {et~al.}(2019)\citenamefont {Liu},
\citenamefont {Vishwanath},\ and\ \citenamefont {Khalaf}}]{liu2019shift}%
\BibitemOpen
\bibfield  {author} {\bibinfo {author} {\bibfnamefont {Shang}\ \bibnamefont
{Liu}}, \bibinfo {author} {\bibfnamefont {Ashvin}\ \bibnamefont
{Vishwanath}}, \ and\ \bibinfo {author} {\bibfnamefont {Eslam}\ \bibnamefont
{Khalaf}},\ }\bibfield  {title} {\enquote {\bibinfo {title} {Shift
insulators: {Rotation}-protected two-dimensional topological crystalline
insulators},}\ }\href {\doibase 10.1103/PhysRevX.9.031003} {\bibfield
{journal} {\bibinfo  {journal} {Physical Review X}\ }\textbf {\bibinfo
{volume} {9}},\ \bibinfo {pages} {031003} (\bibinfo {year}
{2019})}\BibitemShut {NoStop}%
\bibitem [{\citenamefont {Hwang}\ \emph {et~al.}(2019)\citenamefont {Hwang},
\citenamefont {Ahn},\ and\ \citenamefont {Yang}}]{hwang2019fragile}%
\BibitemOpen
\bibfield  {author} {\bibinfo {author} {\bibfnamefont {Yoonseok}\
\bibnamefont {Hwang}}, \bibinfo {author} {\bibfnamefont {Junyeong}\
\bibnamefont {Ahn}}, \ and\ \bibinfo {author} {\bibfnamefont {Bohm-Jung}\
\bibnamefont {Yang}},\ }\bibfield  {title} {\enquote {\bibinfo {title}
{Fragile topology protected by inversion symmetry: {Diagnosis}, bulk-boundary
correspondence, and {Wilson} loop},}\ }\href {\doibase
10.1103/PhysRevB.100.205126} {\bibfield  {journal} {\bibinfo  {journal}
{Physical Review B}\ }\textbf {\bibinfo {volume} {100}},\ \bibinfo {pages}
{205126} (\bibinfo {year} {2019})}\BibitemShut {NoStop}%
\bibitem [{\citenamefont {Bouhon}\ \emph {et~al.}(2020)\citenamefont {Bouhon},
\citenamefont {Bzdu\ifmmode~\check{s}\else \v{s}\fi{}ek},\ and\ \citenamefont
{Slager}}]{bouhon2020geometric}%
\BibitemOpen
\bibfield  {author} {\bibinfo {author} {\bibfnamefont {Adrien}\ \bibnamefont
{Bouhon}}, \bibinfo {author} {\bibfnamefont {Tom\'a\ifmmode
\check{s}\else~\v{s}\fi{}}\ \bibnamefont {Bzdu\ifmmode~\check{s}\else
\v{s}\fi{}ek}}, \ and\ \bibinfo {author} {\bibfnamefont {Robert-Jan}\
\bibnamefont {Slager}},\ }\bibfield  {title} {\enquote {\bibinfo {title}
{Geometric approach to fragile topology beyond symmetry indicators},}\ }\href
{\doibase 10.1103/PhysRevB.102.115135} {\bibfield  {journal} {\bibinfo
{journal} {Physical Review B}\ }\textbf {\bibinfo {volume} {102}},\ \bibinfo
{pages} {115135} (\bibinfo {year} {2020})}\BibitemShut {NoStop}%
\bibitem [{\citenamefont {Song}\ \emph
{et~al.}(2020{\natexlab{a}})\citenamefont {Song}, \citenamefont {Elcoro},
\citenamefont {Xu}, \citenamefont {Regnault},\ and\ \citenamefont
{Bernevig}}]{song2020fragile}%
\BibitemOpen
\bibfield  {author} {\bibinfo {author} {\bibfnamefont {Zhi-Da}\ \bibnamefont
{Song}}, \bibinfo {author} {\bibfnamefont {Luis}\ \bibnamefont {Elcoro}},
\bibinfo {author} {\bibfnamefont {Yuan-Feng}\ \bibnamefont {Xu}}, \bibinfo
{author} {\bibfnamefont {Nicolas}\ \bibnamefont {Regnault}}, \ and\ \bibinfo
{author} {\bibfnamefont {B.A.}\ \bibnamefont {Bernevig}},\ }\bibfield
{title} {\enquote {\bibinfo {title} {Fragile phases as affine monoids:
{Classification} and material examples},}\ }\href {\doibase
10.1103/PhysRevX.10.031001} {\bibfield  {journal} {\bibinfo  {journal}
{Physical Review X}\ }\textbf {\bibinfo {volume} {10}},\ \bibinfo {pages}
{031001} (\bibinfo {year} {2020}{\natexlab{a}})}\BibitemShut {NoStop}%
\bibitem [{\citenamefont {Song}\ \emph
{et~al.}(2020{\natexlab{b}})\citenamefont {Song}, \citenamefont {Elcoro},\
and\ \citenamefont {Bernevig}}]{song2020twisted}%
\BibitemOpen
\bibfield  {author} {\bibinfo {author} {\bibfnamefont {Zhi-Da}\ \bibnamefont
{Song}}, \bibinfo {author} {\bibfnamefont {Luis}\ \bibnamefont {Elcoro}}, \
and\ \bibinfo {author} {\bibfnamefont {B.~Andrei}\ \bibnamefont {Bernevig}},\
}\bibfield  {title} {\enquote {\bibinfo {title} {Twisted bulk-boundary
correspondence of fragile topology},}\ }\href {\doibase
10.1126/science.aaz7650} {\bibfield  {journal} {\bibinfo  {journal}
{Science}\ }\textbf {\bibinfo {volume} {367}},\ \bibinfo {pages} {794--797}
(\bibinfo {year} {2020}{\natexlab{b}})}\BibitemShut {NoStop}%
\bibitem [{\citenamefont {Peri}\ \emph {et~al.}(2020)\citenamefont {Peri},
\citenamefont {Song}, \citenamefont {Serra-Garcia}, \citenamefont {Engeler},
\citenamefont {Queiroz}, \citenamefont {Huang}, \citenamefont {Deng},
\citenamefont {Liu}, \citenamefont {Bernevig},\ and\ \citenamefont
{Huber}}]{peri2020experimental}%
\BibitemOpen
\bibfield  {author} {\bibinfo {author} {\bibfnamefont {Valerio}\ \bibnamefont
{Peri}}, \bibinfo {author} {\bibfnamefont {Zhi-Da}\ \bibnamefont {Song}},
\bibinfo {author} {\bibfnamefont {Marc}\ \bibnamefont {Serra-Garcia}},
\bibinfo {author} {\bibfnamefont {Pascal}\ \bibnamefont {Engeler}}, \bibinfo
{author} {\bibfnamefont {Raquel}\ \bibnamefont {Queiroz}}, \bibinfo {author}
{\bibfnamefont {Xueqin}\ \bibnamefont {Huang}}, \bibinfo {author}
{\bibfnamefont {Weiyin}\ \bibnamefont {Deng}}, \bibinfo {author}
{\bibfnamefont {Zhengyou}\ \bibnamefont {Liu}}, \bibinfo {author}
{\bibfnamefont {B.~Andrei}\ \bibnamefont {Bernevig}}, \ and\ \bibinfo
{author} {\bibfnamefont {Sebastian~D.}\ \bibnamefont {Huber}},\ }\bibfield
{title} {\enquote {\bibinfo {title} {Experimental characterization of fragile
topology in an acoustic metamaterial},}\ }\href {\doibase
10.1126/science.aaz7654} {\bibfield  {journal} {\bibinfo  {journal}
{Science}\ }\textbf {\bibinfo {volume} {367}},\ \bibinfo {pages} {797--800}
(\bibinfo {year} {2020})}\BibitemShut {NoStop}%
\bibitem [{\citenamefont {Zhang}\ and\ \citenamefont
{Yang}(2021)}]{zhang2021tunable}%
\BibitemOpen
\bibfield  {author} {\bibinfo {author} {\bibfnamefont {Rui-Xing}\
\bibnamefont {Zhang}}\ and\ \bibinfo {author} {\bibfnamefont {Zhi-Cheng}\
\bibnamefont {Yang}},\ }\bibfield  {title} {\enquote {\bibinfo {title}
{Tunable fragile topology in floquet systems},}\ }\href {\doibase
10.1103/PhysRevB.103.L121115} {\bibfield  {journal} {\bibinfo  {journal}
{Physical Review B}\ }\textbf {\bibinfo {volume} {103}},\ \bibinfo {pages}
{L121115} (\bibinfo {year} {2021})}\BibitemShut {NoStop}%
\bibitem [{\citenamefont {Weeks}\ and\ \citenamefont
{Franz}(2010)}]{weeks2010topological}%
\BibitemOpen
\bibfield  {author} {\bibinfo {author} {\bibfnamefont {C.}~\bibnamefont
{Weeks}}\ and\ \bibinfo {author} {\bibfnamefont {M.}~\bibnamefont {Franz}},\
}\bibfield  {title} {\enquote {\bibinfo {title} {Topological insulators on
the {Lieb} and perovskite lattices},}\ }\href {\doibase
10.1103/PhysRevB.82.085310} {\bibfield  {journal} {\bibinfo  {journal}
{Physical Review B}\ }\textbf {\bibinfo {volume} {82}},\ \bibinfo {pages}
{085310} (\bibinfo {year} {2010})}\BibitemShut {NoStop}%
\end{thebibliography}
\end{document}